\documentclass[11pt]{article}

\usepackage{latexsym,amssymb,amsfonts,amsmath}
\usepackage{graphicx}
\usepackage{mathrsfs,multirow}
\usepackage[numbers,sort&compress]{natbib}

\usepackage[pdftex,
            pdftitle={Localization of Dirac modes in finite-temperature
  $Z_2$ gauge theory on the lattice}]{hyperref}

\newcommand{\f}[2]{\frac{#1}{#2}}
\newcommand{\tf}[2]{{\textstyle\f{#1}{#2}}}

\newcommand{\la}{\langle}
\newcommand{\ra}{\rangle}
\newcommand{\Po}{{\mathscr P}}
\newcommand{\U}{{\mathscr U}}

\title{Localization of Dirac modes in finite-temperature
  $\mathbb{Z}_2$ gauge theory on the lattice}

\author{Gy{\"o}rgy Baranka\footnote{barankagy@caesar.elte.hu}~ and Matteo Giordano\footnote{giordano@bodri.elte.hu} \vspace{0.15cm} \\
  ELTE E\"otv\"os Lor\'and University,\\ Institute for Theoretical
  Physics,\\ P\'azm\'any P\'eter s\'et\'any 1/A, H-1117, Budapest,
  Hungary}
\date{}

\begin{document}

\maketitle

\begin{abstract}
  We study the localization properties of the eigenmodes of the
  staggered Dirac operator in finite-temperature $\mathbb{Z}_2$ pure
  gauge theory on the lattice in 2+1 dimensions. We find that the low
  modes turn from delocalized to localized as the system crosses over
  from the confined to the deconfined phase in the ``physical'' sector
  (positive average Polyakov loop) selected by external fermionic
  probes, while they remain delocalized in the ``unphysical'' sector
  (negative average Polyakov loop). This confirms that the close
  connection between deconfinement and localization of the low Dirac
  modes in the physical sector, already observed in other models,
  holds also in the simplest gauge theory displaying a deconfinement
  transition. We also observe a clear correlation of localized modes
  with fluctuations of the Polyakov loop away from the ordered value,
  as expected according to the ``sea/islands'' picture of
  localization, and with clusters of negative plaquettes. A novel
  finding is the presence of localized modes at the high end of the
  Dirac spectrum in all phases/sectors of the theory.
\end{abstract}

\section{Introduction}
\label{sec:intro}

In the imaginary-time functional-integral formulation of gauge
theories at finite temperature, the effects of dynamical fermions are
entirely encoded in the Euclidean Dirac operator in a gauge-field
background. The properties of the spectrum of the Euclidean Dirac
operator play an important role in various nonperturbative aspects of
QCD and other gauge theories, including, e.g., the fate of chiral
symmetry~\cite{Banks:1979yr} and of the ${\rm U}(1)_A$
anomaly~\cite{Dick:2015twa,Ding:2020xlj,Kaczmarek:2021ser}, and
anomalous dimensions in conformal gauge
theories~\cite{Patella:2012da}. More generally, eigenvalues and
eigenvectors of the Euclidean Dirac operator probe the background
gauge configuration in a nontrivial way, even in the absence of
dynamical fermions.

In recent years it has become apparent that there is a close relation
between the localization properties of the low-lying Dirac modes and
the confining properties of gauge theories.  It is by now fairly well
established, by means of numerical calculations on the lattice, that
in the high-temperature phase of QCD the low-lying Dirac modes become
localized~\cite{GarciaGarcia:2006gr,Kovacs:2012zq,Giordano:2013taa,
  Cossu:2016scb,Holicki:2018sms} on the scale of the inverse
temperature~\cite{Kovacs:2012zq,Cossu:2016scb}, up to a
temperature-dependent point in the
spectrum~\cite{Kovacs:2012zq,Holicki:2018sms}. The same situation is
also found in the high-temperature phase of other gauge
theories~\cite{Gockeler:2001hr,Gattringer:2001ia,Gavai:2008xe,
  Kovacs:2009zj,Kovacs:2010wx,Bruckmann:2011cc,Giordano:2016nuu,
  Kovacs:2017uiz,Giordano:2019pvc,Vig:2020pgq,Bonati:2020lal} and
gauge-theory related
models~\cite{GarciaGarcia:2005vj,Giordano:2015vla,Giordano:2016cjs,
  Giordano:2016vhx,Bruckmann:2017ywh}. This phenomenon has been put
into a relation with the ordering of the Polyakov loops in the
high-temperature
phase~\cite{Bruckmann:2011cc,Giordano:2015vla,Giordano:2016cjs}, or
more precisely with the presence of disorder in the form of local
fluctuations of the Polyakov loop away from its ordered value. In what
has been named the ``sea/islands'' picture of localization, the
Polyakov loop fluctuations provide ``energetically'' favorable
``islands'' for the modes to localize in the middle of a ``sea'' of
ordered Polyakov loops. Clear numerical evidence of the correlation
between localized modes and Polyakov loop fluctuations has been
presented in the literature, for staggered and overlap spectra in
quenched SU(2) gauge theory~\cite{Bruckmann:2011cc}, and for overlap
spectra in the background of SU(3) gauge fields with dynamical domain
wall fermions~\cite{Cossu:2016scb} and twisted-mass Wilson
fermions~\cite{Holicki:2018sms} (although with heavier-than-physical
pions).

Localization of eigenmodes induced by disorder is a well-studied
phenomenon in condensed matter physics, dating back to the
groundbreaking work of Anderson~\cite{Anderson:1958vr}. In the
language of condensed matter, disorder leads to the spatial
localization of eigenmodes in certain spectral regions, separated by
so-called ``mobility edges'' from regions where modes are
delocalized. Extensive work has led to a thorough understanding of
localization, in particular concerning its aspects at criticality
(see, e.g., Refs.~\cite{thouless1974electrons,lee1985disordered,
  Evers:2008zz,anderson50}). Detailed numerical studies have confirmed
that localization of the low-lying Dirac modes in QCD at high
temperature $T$ shares the same critical features with
three-dimensional Hamiltonians with on-site disorder in the
appropriate symmetry
class~\cite{Giordano:2013taa,Nishigaki:2013uya,Ujfalusi:2015nha}. This
provides evidence in support of the conjectured connection between
localized modes and Polyakov loop fluctuations, that have the right
properties to be the relevant source of disorder.\footnote{In a
  three-dimensional Anderson model with uncorrelated purely
  off-diagonal logarithmically-distributed disorder in the chiral
  unitary class, Garc{\'i}a-Garc{\'i}a and
  Cuevas~\cite{Garcia_Garcia_2006} found that the amount of
  off-diagonal disorder required to achieve localization at the band
  center is much larger than what is available in QCD. This suggests
  that on-site disorder (from the spatial, three-dimensional point of
  view) is needed to explain localization of the low modes in QCD, and
  Polyakov loops match the description.  Interestingly, however,
  similar results were obtained by Takaishi {\it et
    al.}~\cite{takaishi2018localization} for an Anderson model with
  purely off-diagonal but correlated disorder in the three-dimensional
  chiral unitary class, where they found the same critical statistics
  as in QCD.  It is worth mentioning that while the Dirac operator in
  finite-temperature QCD belongs to the three-dimensional {\it chiral}
  unitary class, the critical behavior is expected to match that of
  the non-chiral class when the Anderson transition is away from the
  origin. This is confirmed by results for the correlation-length
  critical exponent~\cite{Giordano:2013taa}, the critical
  statistics~\cite{Giordano:2013taa,Nishigaki:2013uya} and the
  multifractal exponents~\cite{Ujfalusi:2015nha}.  Near the origin,
  instead, Ref.~\cite{Garcia_Garcia_2006} found multifractal exponents
  different from those of the corresponding non-chiral
  class. Recently, the Anderson transition at the origin was studied
  in a chiral unitary model with on-site disorder in
  Ref.~\cite{wang2021universality}, that found a correlation length
  critical exponent different from the one pertaining to the
  non-chiral class.}

Further insight was obtained in Ref.~\cite{Giordano:2016cjs}, using an
explicit representation of the staggered lattice Dirac operator as a
set of coupled Anderson models, with random on-site potential related
to the phases of the untraced Polyakov loops. There it was pointed
out that besides the presence of disorder in the form of Polyakov loop
fluctuations, an important role in the appearance of localized low
modes is played by the presence of strong correlations of gauge fields
in the temporal direction. Temporal correlations in fact lead to the
decoupling of the various Anderson models, and this makes localization
of the low modes possible. While Polyakov loop fluctuations are
present also at low $T$ (although there the spatial average of the
Polyakov loop is close to zero), temporal correlations are induced by
the ordering of the Polyakov loops and so are specific to the high-$T$
phase.

Perhaps the most interesting aspect of the sea/islands picture is its
simplicity: all that it requires is the presence of order in the
Polyakov loop configuration, inducing strong temporal correlations on
the one hand, and providing spatially localized fluctuations away from
the ordered value on the other. This suggests that all that is
required for localized low Dirac modes to appear is the existence of a
deconfined phase of the theory, irrespectively of the dimensionality
of the system\footnote{An exception are 1+1 dimensional systems: only
  localized modes are in fact expected in (spatially) one-dimensional
  disordered systems, for any amount of disorder (see
  Refs.~\cite{thouless1974electrons,lee1985disordered} and references
  therein). This is the case for the 1+1 dimensional CP${}^3$ model at
  finite temperature, also studied in Ref.~\cite{Bruckmann:2017ywh},
  where only localized modes are found in both phases of the model.}
or the nature of the gauge group. This idea has since been
investigated in several models, including pure-gauge SU(3) in
3+1~\cite{Kovacs:2017uiz,Vig:2020pgq} (see also
Refs.~\cite{Gockeler:2001hr,Gattringer:2001ia,GarciaGarcia:2006gr} for
older results) and 2+1~\cite{Giordano:2019pvc} dimensions, a lattice
toy model of QCD with unimproved staggered fermions on $N_t=4$
lattices~\cite{Giordano:2016nuu}, a QCD-inspired spin
model~\cite{Giordano:2016cjs,Giordano:2016vhx}, the CP${}^3$ model in
2+1 dimensions~\cite{Bruckmann:2017ywh}, and recently in
trace-deformed SU(3) gauge theory in 3+1
dimensions~\cite{Bonati:2020lal}. Results were already available in
the case of the SU(2) theory in 3+1
dimensions~\cite{Kovacs:2009zj,Kovacs:2010wx}, two-flavor
QCD~\cite{Gavai:2008xe}, and the Instanton Liquid Model for QCD of
Ref.~\cite{GarciaGarcia:2005vj}. In all these cases localized low
modes have been found in the deconfined phase of the theory. Detailed
studies of SU(3) gauge theory in 3+1~\cite{Kovacs:2017uiz,Vig:2020pgq}
and 2+1 dimensions~\cite{Giordano:2019pvc}, as well as of the QCD toy
model with unimproved fermions~\cite{Giordano:2016nuu}, the
QCD-inspired spin model of Ref.~\cite{Giordano:2016vhx}, and of
trace-deformed SU(3) gauge theory~\cite{Bonati:2020lal}, have found
convincing evidence that localization appears precisely at the
deconfinement transition. All these findings strongly support the
existence of a very close connection between localization of the
low-lying Dirac modes and deconfinement.

The purpose of this work is to push the localization/deconfinement
connection to its limit by studying the simplest lattice gauge theory
displaying a finite-temperature deconfining transition: $\mathbb{Z}_2$
pure gauge theory in 2+1
dimensions~\cite{Wegner:1971ising,Balian:1974ir,Kogut:1979wt}. 
This is the simplest possible gauge group, and the lowest dimension where
a deconfining transition is found. Furthermore, due to the discrete
nature of the gauge group, there are no instanton-related topological
aspects of the gauge configurations to take into account, that have
been shown to strongly affect the low end of the spectrum in SU(3)
gauge theory in 3+1 dimensions~\cite{Vig:2021oyt}.

It is worth at this point clarifying in what sense an Anderson
transition can take place in the spectrum of the Dirac operator (or
more precisely, of a suitable lattice discretization thereof) when one
is dealing with a pure gauge theory. The following discussion focuses
on $\mathbb{Z}_2$ lattice gauge theory, but can be adapted to any
gauge theory with a non-trivial center symmetry (see below Section
\ref{sec:z2recap}).  In the high-temperature phase of these theories
center symmetry is broken, and degenerate but inequivalent ground
states exist with non-vanishing Polyakov-loop expectation value
aligned to one of the center elements. To select a ground state, the
standard procedure is to introduce a small symmetry-breaking
perturbation in the finite-volume system, that lifts the degeneracy,
and in the thermodynamic limit suppresses all but one of the
(would-be) ground states. After this limit is taken the perturbation
is removed, and the desired unperturbed ground state is selected. In
contrast, in the low-temperature unbroken phase the unique ground
state is always recovered after this procedure.

In the system at hand, such a perturbation is provided by very heavy
but otherwise perfectly physical fermion fields, coupled to the gauge
fields via the (suitably discretized) Dirac operator. These modify the
weight of the gauge configurations in the partition function,
preferring configurations with positive spatially-averaged Polyakov
loop, thus selecting the ``physical'' positive center sector, and the
corresponding (slightly perturbed) ground state in the thermodynamic
limit.\footnote{\label{foot:polloop} A positive (resp.\ negative)
  spatially-averaged Polyakov loop correlates with a lower (resp.\
  higher) density of low modes. Since fermions (resp.\ pseudofermions)
  modify the weight that gauge configurations have in the pure gauge
  theory by a factor equal to the determinant (resp.\ inverse of the
  determinant) of the Dirac operator plus the mass, the positive
  (resp.\ negative) center sector is preferred when they are included
  in the dynamics.  This is demonstrated explicitly for SU(2) gauge
  fields, see, e.g., Ref.~\cite{Scheffler:2013jaa}. We will show that
  this is the case also for $\mathbb{Z}_2$ fields when discussing the
  spectral density of the staggered Dirac operator in Section
  \ref{sec:num_spd}.}  The perturbation is then removed by making the
fermions static, i.e., by taking the infinite-mass, quenched limit
resulting in pure gauge theory. Preference for the physical center
sector of course persists if dynamical (non-static) fermions are
instead kept in the theory, in which case center symmetry is broken
explicitly.  Similarly, introducing unphysical pseudofermion fields,
i.e., spin-$\f{1}{2}$ commuting fields again coupled to the gauge
fields through the Dirac operator, configurations with negative
spatially-averaged Polyakov loop are preferred, thus selecting the
``unphysical'' negative center sector.\footnote{Another way to select
  a nontrivial center sector, which can be applied also to SU($N_c$)
  theories, is to introduce a suitable imaginary chemical potential
  for the fermions.}

The static fields discussed above can also be used to probe the system
across the deconfinement transition, by studying the eigenmodes of the
Dirac operator in the background of the relevant gauge configurations.
In general, the transition looks different to different types of
probes, which see the system transition from the confined phase at low
temperature to the deconfined phase in a specific center sector at
high temperature. How the localization properties of the modes change
across the transition, and whether an Anderson transition appears in
the spectrum, can depend on the probe being used, i.e., on the center
sector selected in the high-temperature phase.  According to the
sea/islands picture, such an Anderson transition is expected to take
place in the physical sector selected by fermion probes, but not in
the unphysical sector selected by pseudofermion probes. In fact,
islands of negative Polyakov loops in a sea of positive Polyakov
loops, as found in the physical sector, are needed for localization to
take place, while in the opposite case found in the unphysical sector
one expects an even stronger delocalization on the favorable sea of
negative Polyakov loops.

From the discussion above, it is clear that the sea/islands picture
leads to expect specific correlations between the low modes and the
local properties of the gauge configurations, in particular the local
orientation of the Polyakov loops, which are therefore worth
investigating.  This is particularly important in the light of the
strong connection between low-mode localization and deconfinement
implied by this picture. Indeed, establishing the correctness of the
sea/islands picture of localization would allow to attack the problem
of confinement from a new perspective, possibly leading to a better
understanding of the microscopic mechanism that underlies confinement
and the deconfinement transition.

The plan of the paper is the following. In Section \ref{sec:z2recap}
we briefly review the relevant aspects of $\mathbb{Z}_2$ lattice gauge
theory. In Section \ref{sec:locD} we discuss how to detect
localization of Dirac eigenmodes, and which observables to use in
order to study the correlation between localization and various
properties of the gauge field configuration. In Section
\ref{sec:numerical} we show our numerical results. Finally, in Section
\ref{sec:concl} we present our conclusions and outlook for future
studies. Technical details concerning the treatment of degenerate
modes are discussed in Appendix \ref{sec:appA}.

\section{$\mathbb{Z}_2$ lattice gauge theory}
\label{sec:z2recap}

The dynamical variables of $\mathbb{Z}_2$ gauge theory on a hypercubic
lattice are the link variables $U_\mu(n)=\pm 1$, taking values in
$\mathbb{Z}_2$, associated to the oriented lattice links connecting
site $n$ with its neighbors $n+\hat{\mu}$. In 2+1 dimensions
$n=(n_1,n_2,n_3)=(t,\vec{x})$, where $0\le t=n_1\le N_t-1$,
$0\le x_{1,2}=n_{2,3}\le N_s-1$, and $\hat{\mu}$ is the unit lattice
vector in direction $\mu$, with $\mu=1,2,3$. Periodic boundary
conditions are imposed in all directions. The Wilson action for the
$\mathbb{Z}_2$ gauge theory reads (up to an irrelevant additive
constant)
\begin{equation}
  \label{eq:z2rec1}
  S[U] = -\beta\sum_n \sum_{\substack{\mu,\nu=1\\\mu<\nu}}^3 U_{\mu\nu}(n)\,,
\end{equation}
where the plaquette variables $U_{\mu\nu}$ read simply
$U_{\mu\nu}(n)=U_\mu(n)U_\nu(n+\hat{\mu})U_\mu(n+\hat{\nu})U_\nu(n)$,
and $\beta= 1/(e^2 a)$ with $e$ the coupling constant (of mass
dimension $1/2$) and $a$ the lattice spacing. Expectation values are
defined as
\begin{equation}
  \label{eq:z2rec3bis}
    \la O[U] \ra =  Z^{-1} \sum_{\{U_\mu(n)=\pm 1\}} e^{-S[U]} \,O[U]\,,\qquad
    Z = \sum_{\{U_\mu(n)=\pm 1\}} e^{-S[U]}\,,
\end{equation}
where $Z$ is the partition function and the sum is over all possible
gauge configurations. The finite-temperature case is studied by taking
the thermodynamic limit $N_s\to\infty$ at fixed $N_t$, and varying
$\beta$. In this approach $\beta$ can be identified with the
temperature $T=1/(a N_t)$ of the system, $T/e^2 = \beta /N_t$ (we
ignore scaling violations for simplicity).

The phase diagram of the $\mathbb{Z}_2$ gauge model has been studied
in detail in Ref.~\cite{Caselle:1995wn} exploiting the duality relation
with the three-dimensional Ising model. A second-order phase
transition takes place at a critical $\beta_c=\beta_c(N_t)$,
separating a confined phase at $\beta<\beta_c$ from a deconfined one
at $\beta>\beta_c$. Deconfinement is signaled by the spontaneous
breaking of the center symmetry of the model, i.e., the symmetry under
the transformation
\begin{equation}
  \label{eq:z2rec4_bis}
  U_1(N_t-1,\vec{x})\to
  z U_1(N_t-1,\vec{x}) \quad \forall\vec{x}\,, \quad z\in\mathbb{Z}_2\,,
\end{equation}
with $z$ an element of the group center (which for an Abelian group
coincides with the group itself). The relevant order parameter is the
expectation value, $\la P(\vec{x})\ra$, of the Polyakov loop,
\begin{equation}
  \label{eq:z2rec4}
  P(\vec{x}) \equiv \prod_{t=0}^{N_t-1}U_1(t,\vec{x})\,.
\end{equation}
In the thermodynamic limit, this vanishes in the confined phase, while
it is nonzero in the deconfined phase where the Polyakov loop gets
ordered, taking mostly either the $+1$ or the $-1$ value, depending on
the choice of ground state.  This choice is made in practice already
in a finite volume, by selecting configurations with the desired sign
of the spatially-averaged Polyakov loop,
$\bar{P} \equiv V^{-1}\sum_{\vec{x}} P(\vec{x})$, where $V=N_s^2$ is
the lattice spatial volume. Here and in the following all quantities
are expressed in lattice units, unless explicitly stated otherwise.
We will refer to configurations with positive $\bar{P}$ as belonging
to the physical sector, and to those with negative $\bar{P}$ as
belonging to the unphysical sector. As already explained in the
Introduction, the reason for this nomenclature is that these
configurations are favored respectively by dynamical fermion or
pseudofermion fields, and in the spontaneously broken phase they are
strictly selected after the thermodynamic limit is taken. Restricting
to one sector provides an accurate approximation of the partition
function in the broken phase for large enough volumes, while both
sectors must always be included in the unbroken phase.

In this paper we study the spectrum of the staggered Dirac operator in
the background of quenched $\mathbb{Z}_2$ configurations, thus probing
the gauge fields with external fermion or pseudofermion fields.  In
the case at hand, the staggered operator reads
\begin{equation}
  \label{eq:z2rec5}
  \begin{aligned}
    D^{\rm stag}_{n,n'} &= \f{1}{2}\sum_{\mu=1}^3
    \eta_\mu(n)(U_\mu(n)\delta_{n+\hat{\mu},n'}
    -U_\mu(n-\hat{\mu})\delta_{n-\hat{\mu},n'})\,, \\ \eta_\mu(n) &=
    (-1)^{\sum_{\nu<\mu}n_\nu}\,,
  \end{aligned}
\end{equation}
with periodic boundary conditions in the spatial directions and
antiperiodic boundary conditions in the temporal direction. The
operator $D^{\rm stag}$ is anti-Hermitian and so has purely imaginary
eigenvalues,
\begin{equation}
  \label{eq:z2rec5bis}
  D^{\rm stag} \psi_l = i\lambda_l \psi_l\,, \qquad \lambda_l\in\mathbb{R}\,.
\end{equation}
The staggered operator satisfies furthermore the chiral property
$\{\varepsilon,D^{\rm stag}\}=0$, where
$\varepsilon(n)=(-1)^{\sum_{\mu=1}^3 n_\mu}$, which implies
\begin{equation}
  \label{eq:z2rec5ter}
   D^{\rm stag} \varepsilon\psi_l = -i\lambda_l \varepsilon\psi_l\,,
\end{equation}
so that the spectrum is symmetric about zero. 

In the continuum limit, $D^{\rm stag}$ describes $N_f=4$ degenerate
species of fermions also in 2+1
dimensions~\cite{Burden:1986by}.\footnote{In
  Ref.~\cite{Giordano:2019pvc} it is mistakenly stated $N_f=2$.}
Here we do not attempt any extrapolation to the continuum limit and
work at finite lattice spacing. In this case the staggered fermionic
action in the massless limit,
$S^{\rm stag}=\sum_{n,n'}\bar{\chi}(n) D^{\rm stag}(n,n') \chi(n')$,
still retains an exact $\text{U}(1)_1\times \text{U}(1)_\varepsilon$
chiral symmetry~\cite{Burden:1986by} under the following
transformations of the fermion fields $\chi$, $\bar{\chi}$:
\begin{equation}
  \label{eq:z2rec6}
  \begin{aligned}
    {\rm U}(1)_1:~&   
    \chi(n) \to e^{i\alpha}\chi(n)\,, &&& &\bar{\chi}(n) \to
    \bar{\chi}(n)e^{-i\alpha}\,, 
\\
     {\rm U}(1)_\varepsilon:~&  
        \chi(n) \to e^{i\alpha\varepsilon(n)}\chi(n)\,, &&& &\bar{\chi}(n) \to
        \bar{\chi}(n)e^{i\alpha\varepsilon(n)}\,.
  \end{aligned}
\end{equation}
The ${\rm U}(1)_\varepsilon$ symmetry can break down spontaneously by
formation of a quark-antiquark condensate, signalled by a nonzero
density of near-zero modes~\cite{Banks:1979yr}.  Although we are
working in the opposite, ``quenched'' limit of infinitely massive
fermions, it is nonetheless worth checking whether a nonzero density
of near-zero modes is present in the thermodynamic limit.

\section{Localization of Dirac eigenmodes}
\label{sec:locD}

The simplest way to detect localization is to study the so-called {\it
  inverse participation ratio} (IPR) of the Dirac eigenmodes $\psi_l$,
\begin{equation}
  \label{eq:loc1}
  {\rm IPR}_l \equiv \sum_n|\psi_l(n)|^4\,,
\end{equation}
where it is understood that modes satisfy the normalization condition
$\sum_n|\psi_l(n)|^2=1$. Since
$|\varepsilon(n) \psi_l(n)| = |\psi_l(n)|$, one finds the same
${\rm IPR}$ for the eigenmodes corresponding to $\lambda_l$ and
$-\lambda_l$; similar symmetry considerations hold for any observable
built out of $|\psi_l(n)|$. It then suffices to consider only
$\lambda_l\ge 0$.

For delocalized modes extended throughout the whole lattice, one has
the qualitative behavior $|\psi_{\rm ext}(n)|^2\sim (N_tV)^{-1}$, and
so $ {\rm IPR}_{\rm ext} \sim (N_tV)^{-1}$, which vanishes in the
large-volume limit. For modes localized in a region of spatial volume
$V_0$ one has instead approximately
$|\psi_{\rm loc}(n)|^2\sim (N_tV_0)^{-1}$ within that region and zero
outside, and so ${\rm IPR}_{\rm loc} \sim (N_tV_0)^{-1}$, which
remains finite as the volume increases. Equivalently, one can consider
the {\it participation ratio} (PR) of the modes,
\begin{equation}
  \label{eq:loc2}
  {\rm PR}_l \equiv {\rm IPR}_l^{-1}(N_tV)^{-1} \,,
\end{equation}
which measures the fraction of spacetime that they effectively occupy,
and as $V\to\infty$ approaches a constant for extended modes, and zero
for localized modes; or the ``size'' of the modes, ${\rm PR}_l\cdot
N_t V ={\rm IPR}_l^{-1}$, which diverges like $V$ for extended modes,
and tends to a constant for localized modes.

To identify the spectral regions where modes are localized, one divides
the support of the spectrum in small (ideally infinitesimal) bins, and
averages, e.g., the PR of the modes within each bin and over gauge
configurations,
\begin{equation}
  \label{eq:loc3}
  {\rm PR}(\lambda,N_s)
  \equiv \f{\left\la  \sum_l
      \delta(\lambda-\lambda_l)  {\rm PR}_l \right\ra}{\left\la  \sum_l
      \delta(\lambda-\lambda_l)  \right\ra}\,,
\end{equation}
studying then the behavior of ${\rm PR}(\lambda,N_s)$ as $N_s$ is
changed at fixed $\lambda$. In practice, to estimate this quantity
locally in the spectrum we divided the full spectral range in bins of
small but finite width $\Delta\lambda$, did the averaging within each
bin separately, and assigned the result to the average eigenvalue in
the bin.  The same procedure was used for the other observables
considered in this work. At large $N_s$ one expects the following
scaling behavior of ${\rm PR}(\lambda,N_s)$,
\begin{equation}
  \label{eq:loc4}
  {\rm PR}(\lambda,N_s)
  \simeq c(\lambda)N_s^{\alpha(\lambda)-2} \,,
\end{equation}
with some volume-independent $c(\lambda)$, and where $\alpha(\lambda)$
is the fractal dimension of the eigenmodes in the given spectral
region. For extended, fully delocalized modes one has $\alpha=2$,
while for localized modes $\alpha=0$. The fractal dimension can then
be extracted from pairs of different system sizes $N_{s1,2}$ via
\begin{equation}
  \label{eq:loc5}
  \alpha(\lambda) = 2  + \left.\log\left(\f{{\rm
          PR}(\lambda,N_{s1})}{{\rm
          PR}(\lambda,N_{s2})}\right)\middle/\log\left(\f{N_{s1}}{N_{s2}}\right)\right. \,, 
\end{equation}
for sufficiently large $N_{s1,2}$.

The presence and the nature of Anderson transitions in two-dimensional
systems is a much more delicate issue than in the three-dimensional
case. According to the scaling theory of
localization~\cite{locgangoffour}, Anderson transitions are generally
expected to be absent in two-dimensional disordered systems, and all
modes are expected to be localized, except for the symplectic class
(see Refs.~\cite{lee1985disordered,Evers:2008zz} for details and
references). In practice, however, this issue strongly depends on the
details of the system. The inclusion of topological effects in the
field-theoretical description of localization (see
Ref.~\cite{Evers:2008zz} for references) allows the presence of
Anderson transitions also in certain other symmetry classes (see
Ref.~\cite{Evers:2008zz} and references cited in
Ref.~\cite{xie1998kosterlitz}), in particular in the chiral
classes~\cite{Konig:2012ud}.  Numerical results show that a
second-order Anderson transition is present in the theory of the
integer Quantum Hall Effect~\cite{slevin2009critical}, in the
two-dimensional unitary class. The critical behavior has been
elucidated in terms of a marginally-perturbed conformal
theory~\cite{Zirnbauer:2018ooz}, which is supported by very recent
numerical results~\cite{Dresselhaus:2021fbx}.  An Anderson transition
of BKT (Berezinski\u{\i}-Kosterlitz-Thouless) type is instead present
in the two-dimensional unitary Anderson
model~\cite{xie1998kosterlitz}, which is in the same, non-chiral
unitary class, and in SU(3) gauge theory in 2+1
dimension~\cite{Giordano:2019pvc}, which is in the two-dimensional
chiral unitary class. An Anderson transition is present also in the
CP${}^3$ model in 2+1 dimensions~\cite{Bruckmann:2017ywh}, again in
the two-dimensional chiral unitary class, while for the
two-dimensional model with correlated, purely off-diagonal disorder
studied in Ref.~\cite{takaishi2018localization}, in the same symmetry
class, results are admittedly not conclusive. Finally, a transition of
BKT type has been observed also in the orthogonal class, in a model
for disordered graphene with strong long-range impurities mimicked by
a suitable on-site random potential~\cite{zhang2009localization}.

The random ``Hamiltonian'' of interest in this paper is
$H=-iD^{\rm stag}$, which has purely imaginary matrix elements and
satisfies the chiral property $\{\varepsilon,H\}=0$, and so is
invariant under the anti-unitary transformation $T=\varepsilon K$,
where $K$ denotes complex conjugation,
\begin{equation}
  \label{eq:rmt_class}
  T H T^\dag = \varepsilon H^* \varepsilon  =  -\varepsilon H \varepsilon = H\,.
\end{equation}
Since $T^2=1$ and $[T,\varepsilon]=0$, $H$ belongs to the chiral
orthogonal
class~\cite{mehta2004random,Verbaarschot:2000dy,Evers:2008zz}.
Numerical studies of Anderson models in this class, employing
bipartite lattices and purely off-diagonal disorder, have found that
all non-zero energy modes are localized. The zero-energy point is
singled out by chiral symmetry and critical modes are found
there~\cite{soukoulis1982study,eilmes1998two,xiong2001power,
  schweitzer2012scaling}, that become localized for sufficiently
strong disorder in certain models~\cite{motrunich2002}.  In the case
at hand, however, an argument analogous to that of
Refs.~\cite{Bruckmann:2011cc,Giordano:2015vla,Giordano:2016cjs} shows
how local fluctuations of the Polyakov loop effectively provide a
two-dimensional on-site disorder.  In analogy with what has been found
for the SU(3) gauge theory~\cite{Giordano:2019pvc}, which is in the
chiral unitary class but behaves similarly to the non-chiral unitary
Anderson model~\cite{xie1998kosterlitz}, displaying a BKT-type
Anderson transition at finite energy, we are then led to expect a
behavior similar to that found in the above-mentioned model in the
non-chiral orthogonal class~\cite{zhang2009localization}.  We then
expect to find a true Anderson transition in the spectrum of the Dirac
operator in the physical sector of the deconfined phase of
$\mathbb{Z}_2$ gauge theory in 2+1 dimensions, and we expect it to be
of BKT type.  This means in particular that we do not expect to find a
sudden transition from an ``insulating'' side (fractal dimension 0) to
a ``metallic'' side (fractal dimension 2) when crossing the mobility
edge. All points in the spectrum beyond the mobility edge are in fact
expected to be critical, with nontrivial fractal properties of the
corresponding eigenmodes, intermediate between insulating and
metallic.

It is expected from the sea/islands picture of localization that
localized modes correlate with the fluctuations of the Polyakov loop
away from order.  To study this correlation we have considered the
following observable,
\begin{equation}
  \label{eq:loc6}
  \Po(\lambda) \equiv \f{\la  \sum_l \delta(\lambda-\lambda_l)
      \sum_{t,\vec{x}} P(\vec{x}) |\psi_l(t,\vec{x})|^2
    \ra}{\la  \sum_l \delta(\lambda-\lambda_l)\ra}
  \,,  
\end{equation}
i.e., the spatial average of the Polyakov loop weighted by the
eigenmode density, or the ``Polyakov loop seen by a
mode''~\cite{Bruckmann:2011cc}. For fully delocalized modes one
expects
\begin{equation}
  \label{eq:loc6_pol_deloc}
    \sum_{t,\vec{x}} P(\vec{x}) |\psi_l(t,\vec{x})|^2 \simeq
    \f{1}{N_tV}    \sum_{t,\vec{x}} P(\vec{x}) =
    \f{1}{V}    \sum_{\vec{x}} P(\vec{x}) =
    \bar{P}\,,
\end{equation}
and so approximately $\Po \simeq \la P \ra$ in spectral regions where
modes are extended. In the physical sector, localized modes are
expected to prefer to live close to negative Polyakov loops, and so
$\Po$ is expected to be considerably smaller than $\la P \ra$. Notice
that $\f{1-\Po(\lambda)}{2}$ measures exactly ``how much'' of the mode
lives on sites corresponding to a negative Polyakov loop.

The dynamics of the $\mathbb{Z}_2$ gauge theory can be trivially
reexpressed in terms of clusters of negative plaquettes. As pointed
out in Ref.~\cite{Gliozzi:2002ht}, what is nontrivial is that at the
deconfinement transition (in 2+1 dimensions) the largest such cluster
ceases to scale like the system size, i.e., is not fully
delocalized. It is then interesting to check whether there is
correlation between localized modes and negative plaquettes. To do so
we have used two different observables, namely
\begin{equation}
  \label{eq:loc6_bis}
    \U(\lambda) \equiv \f{\la \sum_l \delta(\lambda-\lambda_l)
        \sum_{n} A(n) |\psi_l(n)|^2 \ra
    }{\la  \sum_l \delta(\lambda-\lambda_l)\ra}  \,, 
\end{equation}
and
\begin{equation}
  \label{eq:loc6_ter}
  \tilde\U(\lambda) \equiv \f{ \la \sum_l \delta(\lambda-\lambda_l)
    \sum_{n, A(n)> 0} |\psi_l(n)|^2 \ra }{\la \sum_l
    \delta(\lambda-\lambda_l)\ra}\,,
\end{equation}
where $A(n)\ge 0$ counts the number of negative plaquettes touching
the lattice site $n$,
\begin{equation}
  \label{eq:loc7}
    A(n) \equiv \f{1}{2}\sum_{\substack{\mu,\nu=1\\\mu<\nu}}^3[ 4 -
    U_{\mu\nu}(n)- U_{\mu\nu}(n-\hat{\mu}) 
    -U_{\mu\nu}(n-\hat{\nu}) - U_{\mu\nu}(n-\hat{\mu}-\hat{\nu})]\,.
\end{equation}
$\U(\lambda)$ then measures the average number of negative plaquettes
touched by the modes, while $\tilde\U(\lambda)$ measures how much of
the modes lives on sites touched by negative plaquettes.  For
delocalized modes one expects
\begin{equation}
  \label{eq:loc6_plaq_deloc}
   \sum_{n} A(n)  |\psi_l(n)|^2 \simeq \f{1}{N_tV} \sum_{n} A(n)\,,
\end{equation}
and so approximately $\U \simeq 6(1- \la U_{\mu\nu}\ra) $ in spectral
regions where modes are extended. Estimating the probability that
$A(n)>0$, needed to estimate $\tilde\U$ for delocalized modes, is
instead not so straightforward.

To study the shape of the eigenmodes in more detail we measured their
inertia tensor in the center-of-mass frame, and identified the
corresponding principal axes and moments of inertia. A slight
complication is due to the fact that the system of interest, i.e., the
``mass'' distribution $m(n) = |\psi_l(n)|^2$ of mode $l$, lives on a
lattice with periodic boundary conditions, i.e., on a discrete
torus. To identify the center of mass we followed
Ref.~\cite{bai2008calculating}. Each direction is embedded in a
two-dimensional plane as a circle of radius $\f{N_\mu}{2\pi}$, where
$N_\mu$ is the linear size in direction $\mu$, with embedding
coordinates $(\xi_\mu,\zeta_\mu)$,
\begin{equation}
  \label{eq:loc11}
  \begin{aligned}
&  \xi_\mu = N_\mu \cos\theta_\mu(n)\,, &&& & \zeta_\mu = N_\mu
  \sin\theta_\mu(n)\,, \\ & \theta_\mu(n) =\f{2\pi
    n_\mu}{N_\mu}\,,&&& & 0\le n_\mu\le N_\mu-1\,.
  \end{aligned}
\end{equation}
Since $\sum_n m(n) =1$, the position of the center-of-mass in the
plane is easily found to be
\begin{equation}
  \label{eq:loc12}
  \begin{aligned}
  \bar{\xi}_\mu &= N_\mu \sum_n m(n)\cos\theta_\mu(n)\,, \\
  \bar{\zeta}_\mu &= N_\mu \sum_n m(n)\sin\theta_\mu(n)\,.
  \end{aligned}
\end{equation}
This is then projected back on the circle via
$\bar{\xi}_\mu = \bar{r} \cos \bar{\theta}_\mu$,
$\bar{\zeta}_\mu = \bar{r}\sin \bar{\theta}_\mu$, with $\bar{\theta}_\mu\in
[0,2\pi)$ and $\bar{r}^2=\bar{\xi}_\mu^{\hspace{0.02cm}2}+\bar{\zeta}_\mu^2$.
Finally, as long as $\bar{r}\neq 0$, the lattice coordinates of the
center of mass are obtained as
\begin{equation}
  \label{eq:loc14}
  \bar{n}_\mu = N_\mu \f{\bar{\theta}_\mu}{2\pi}\,.
\end{equation}
If $\bar{r}=0$, i.e., $\bar{\xi}_\mu=\bar{\zeta}_\mu= 0$, then the mass
distribution is uniform in direction $\mu$, and any point can be
treated as the center of mass. The inertia tensor is now defined as
\begin{equation}
  \label{eq:loc15}
  \Theta_{\mu\nu} \equiv \sum_n m(n) \textstyle \Big[  \delta_{\mu\nu}\sum_\rho
    (n_\rho -\bar{n}_\rho)_P^2   -
    (n_\mu - \bar{n}_\mu)_P(n_\nu - 
    \bar{n}_\nu)_P\Big], 
\end{equation}
where the periodic coordinate difference $(n_\mu - \bar{n}_\mu)_P$ is
defined as\footnote{Since the center of mass is almost never located
  on a lattice site, the case in which one finds equal magnitudes for
  the coordinate differences obtained following the two possible
  routes on the torus is very unlikely to happen, and can be
  ignored. In this case, since neither route is preferred, both should
  contribute equally to the sum in Eq.~\protect\eqref{eq:loc15}, yielding a
  vanishing contribution to off-diagonal terms.}
\begin{equation}
  \label{eq:loc16}
    (n_\mu - \bar{n}_\mu)_P  = \left\{
    \begin{aligned}
      & n_\mu - \bar{n}_\mu &&& &\text{if}~|n_\mu -
      \bar{n}_\mu|<|N_\mu- n_\mu + \bar{n}_\mu|\,,\\
      & N_\mu - n_\mu + \bar{n}_\mu &&& &\text{if}~|n_\mu -
      \bar{n}_\mu|>|N_\mu- n_\mu + \bar{n}_\mu|\,.
    \end{aligned}\right.
\end{equation}
From the eigenvalues $\theta_1\ge \theta_2\ge \theta_3$ of
$\Theta_{\mu\nu}$ one can find out the approximate shape of the
mode. There are three particularly interesting cases: i.)
$\theta_1\approx \theta_2 > \theta_3$: the mode is more extended in
one direction than in the other two, which are instead approximately
equal (elongated, prolate shape); ii.)
$\theta_1> \theta_2 \approx \theta_3$: the mode is less extended in
one direction than in the other two, which are instead approximately
equal (flattened, oblate shape); iii.)
$\theta_1\approx \theta_2\approx \theta_3$: the mode is extended
approximately in the same way in all directions (spherical shape).
Since the temporal direction is singled out, it is also interesting to
check the relative orientation of the principal axes of the mode,
$v_{1,2,3}$, with the temporal direction. For a randomly oriented
axis, one would find for the average of the absolute value of
$\cos\varphi_j = v_j\cdot \hat{1}$ the following result,
\begin{equation}
  \label{eq:loc17}
  \langle |\cos\varphi_j| \rangle = \f{1}{2 \pi} \int_0^{2\pi} d \phi
  \int_0^{\f{\pi}{2}} d \theta\,  \sin 
  \theta \cos\theta = \f{1}{2}\,.
\end{equation}
Values larger than $1/2$ indicate that the corresponding principal
axis prefers to be oriented closer to the temporal direction, while
values smaller than $1/2$ indicate that it prefers to lie closer to
the spatial plane.

\section{Numerical results}
\label{sec:numerical}

\begin{table}[t]
  \centering
  \begin{tabular}{c|cc|cc} 
    \hline\hline     \multirow{2}{*}{$N_s$}                           & 
                                                                        \multirow{2}{*}{$\beta=$} 
    &
      $0.67,0.68,0.69,0.70,$ &
                               \multirow{2}{*}{$\beta=$} 
    &
      $0.71,0.72,0.725,0.73,0.7325,$
    \\

&                           & $0.75,0.76,0.77$
                                                             & & $ 0.735,0.7375,0.74,0.745$
    \\ \hline
    20 & \multicolumn{2}{c|}{2000} & \multicolumn{2}{c}{4000} \\
    24 & \multicolumn{2}{c|}{1000} & \multicolumn{2}{c}{2000} \\
    28 & \multicolumn{2}{c|}{600} & \multicolumn{2}{c}{1500} \\
    32 & \multicolumn{2}{c|}{500} & \multicolumn{2}{c}{1000}
  \end{tabular}
  \caption{Number of configurations (per center sector) for the various
    lattice sizes and values of $\beta$.}
  \label{tab:stat}
\end{table}

We performed numerical simulations in both phases of the
$\mathbb{Z}_2$ gauge theory on periodic $N_t\times N_s^2$ cubic
lattices with $N_t=4$ and $N_s=20,24,28,32$. We used $\beta$ values in
the range $[0.67,0.77]$ on both sides of the critical coupling, that
for $N_t=4$ is $\beta_c(N_t=4)=
0.73107(2)$~\cite{Caselle:1995wn}. Details about the $\beta$ values
and the accumulated statistics can be found in Table
\ref{tab:stat}. Simulations were performed with a standard Metropolis
algorithm. For each configuration we measured the spatially-averaged
Polyakov loop $\bar{P}$, and depending on its sign we assigned the
configuration to the physical sector ($\bar{P}> 0$) or to the
unphysical sector ($\bar{P}<0$). We then obtained a new configuration
with the same Boltzmann weight but belonging to the opposite sector by
changing the sign of all the temporal links on the last time
slice. Such a configuration was stored separately, leaving the flow of
the simulation undisturbed. For both configurations we then obtained
all the eigenvalues and eigenvectors of the staggered Dirac operator,
Eq.~\eqref{eq:z2rec5}, using the LAPACK
library~\cite{anderson1999lapack}. We then studied the properties of
the eigenmodes locally in the spectrum, dividing the full spectral
range in bins of width $\Delta\lambda=0.05$. Statistical errors were
estimated with the jackknife method. The chiral property allowed us to
restrict to $\lambda_l\ge 0$ without any loss of information.

Degenerate eigenvalues show up in the spectrum, especially frequently
(for some reason unknown to us) in the spectral region around
$\lambda_*\equiv \sqrt{3/2}$, which corresponds to the square root of
the average squared eigenvalue of the staggered Dirac
operator.\footnote{In $d$ dimensions, for link variables in an
  $N$-dimensional representation of the gauge group (and for $N_t>2$)
  one has $-(D^{{\rm stag}\,2})_{nn} = \f{d}{2}\mathbf{1}_{N}$, and so
  $(N_t V N)^{-1}\protect\sum_l\lambda_l^2 = \f{d}{2}$.}  This is probably a
finite-size effect, as signalled by the fact that the amount of
degenerate modes tends to decrease as the lattice size is
increased. Nonetheless, it is clearly visible for the lattice sizes
studied in this paper. For degenerate eigenspaces it is more
appropriate to assign a single value of the ${\rm IPR}$ to the whole
eigenspace. This is obtained by averaging over the eigenspace with the
procedure discussed in Appendix \ref{sec:appA}. For observables
measuring the correlation of the local density of the modes,
$|\psi_l(n)|^2$, with local gauge observables, the result of this
averaging procedure coincides with the simple average over any
orthonormal basis of the degenerate eigenspace, and no particular
treatment is therefore required.

The simplicity of the $\mathbb{Z}_2$ gauge model allows one to obtain
full spectra on lattices of moderately large size in a reasonably
short time (e.g., 500 configurations for $N_s=32$ are analyzed in less
than a day on a standard CPU).  The precise determination of the
mobility edge separating regions of delocalized modes from regions of
extended modes, on the other hand, requires the use of large system
sizes for which full diagonalization is impractical.  Full
diagonalization algorithms in fact typically scale with the matrix
size $N$ like $N^3$, so in our case like $N_s^6$, and the time
required for the analysis quickly ceases to be reasonable. In this
work we have preferred to obtain an overview of the localization
properties of the modes throughout the whole spectrum, rather than
focussing on the low end only, postponing the detailed study of the
mobility edge to future work.

For all the observables considered in this paper we have studied the
two center sectors separately. In the deconfined phase, where center
symmetry is spontaneously broken, this provides physical results
corresponding to the two possible choices for the ground state,
formally following from the use of fermion ($\bar{P}>0$) or
pseudofermion ($\bar{P}<0$) fields to probe the system. In the
confined phase the center symmetry is instead unbroken and the two
sectors should be combined together. Furthermore, in this phase
$\bar{P}$ is small on typical configurations, and one expects little
difference between the two sectors. Nevertheless, it is informative to
study the two sectors separately, in particular to check how they
evolve as the critical coupling is approached.

For couplings close to $\beta_c$ one expects large finite-size effects
for the available lattice sizes, and the corresponding results are
therefore not expected to represent accurately the behavior of the
system in the thermodynamic limit. We nonetheless included them in our
plots, and they seem qualitatively in line with the general trend.

\subsection{Participation ratio}
\label{sec:num_pr}

\begin{figure}[t]
   \centering
   \includegraphics[width=0.48\textwidth]{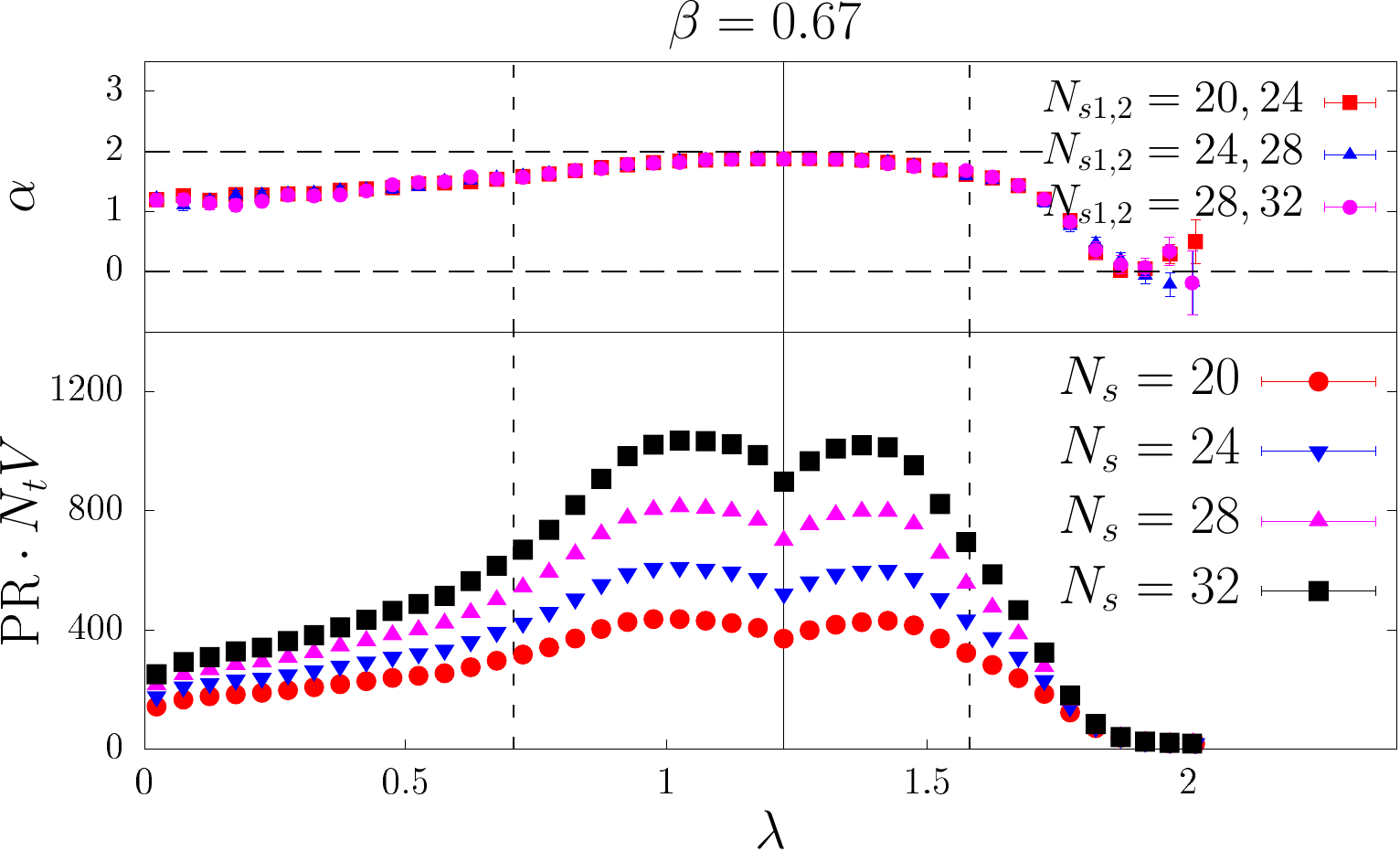}
\hfil   
   \includegraphics[width=0.48\textwidth]{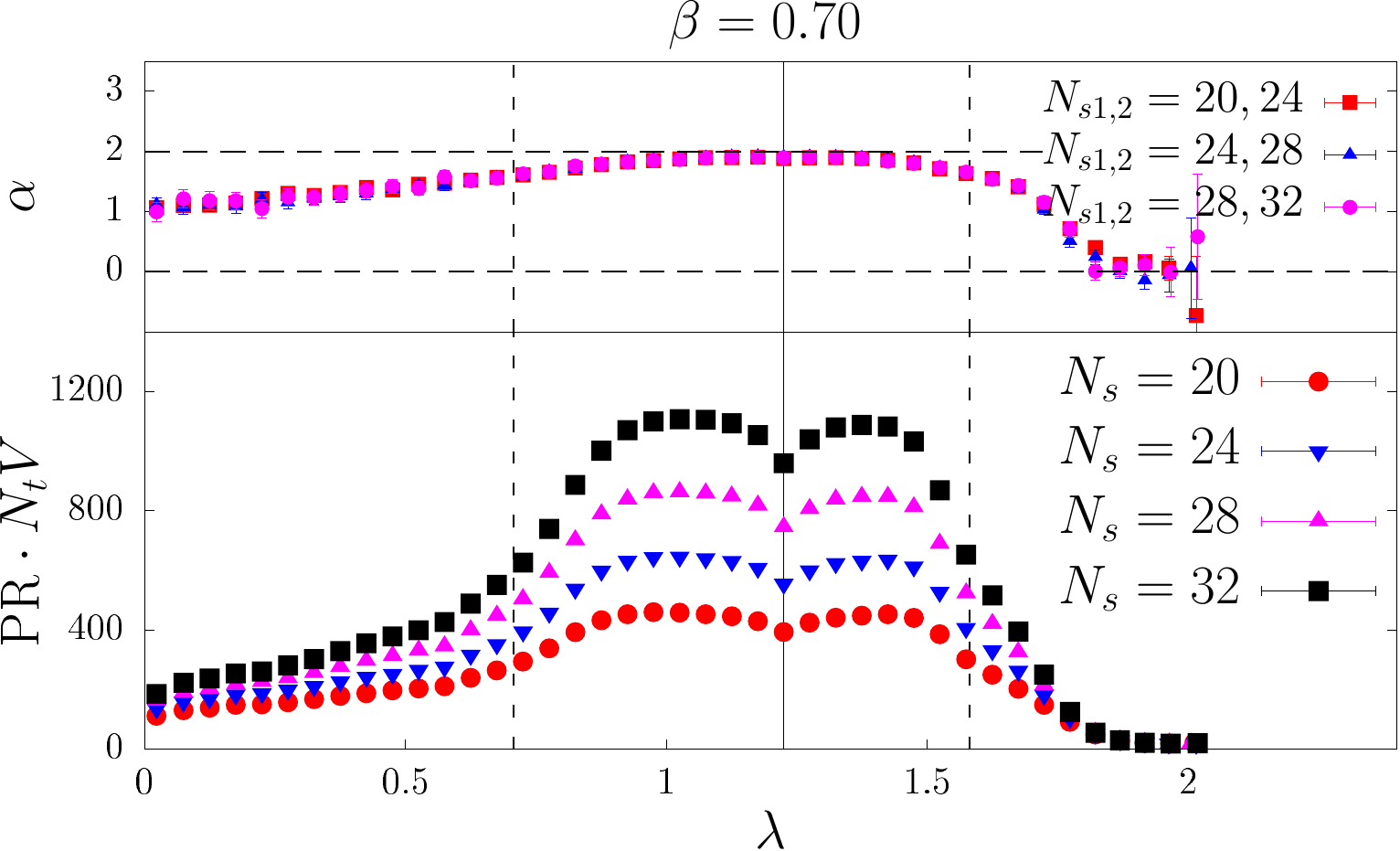}
   
   \vspace{0.1cm}
   \includegraphics[width=0.48\textwidth]{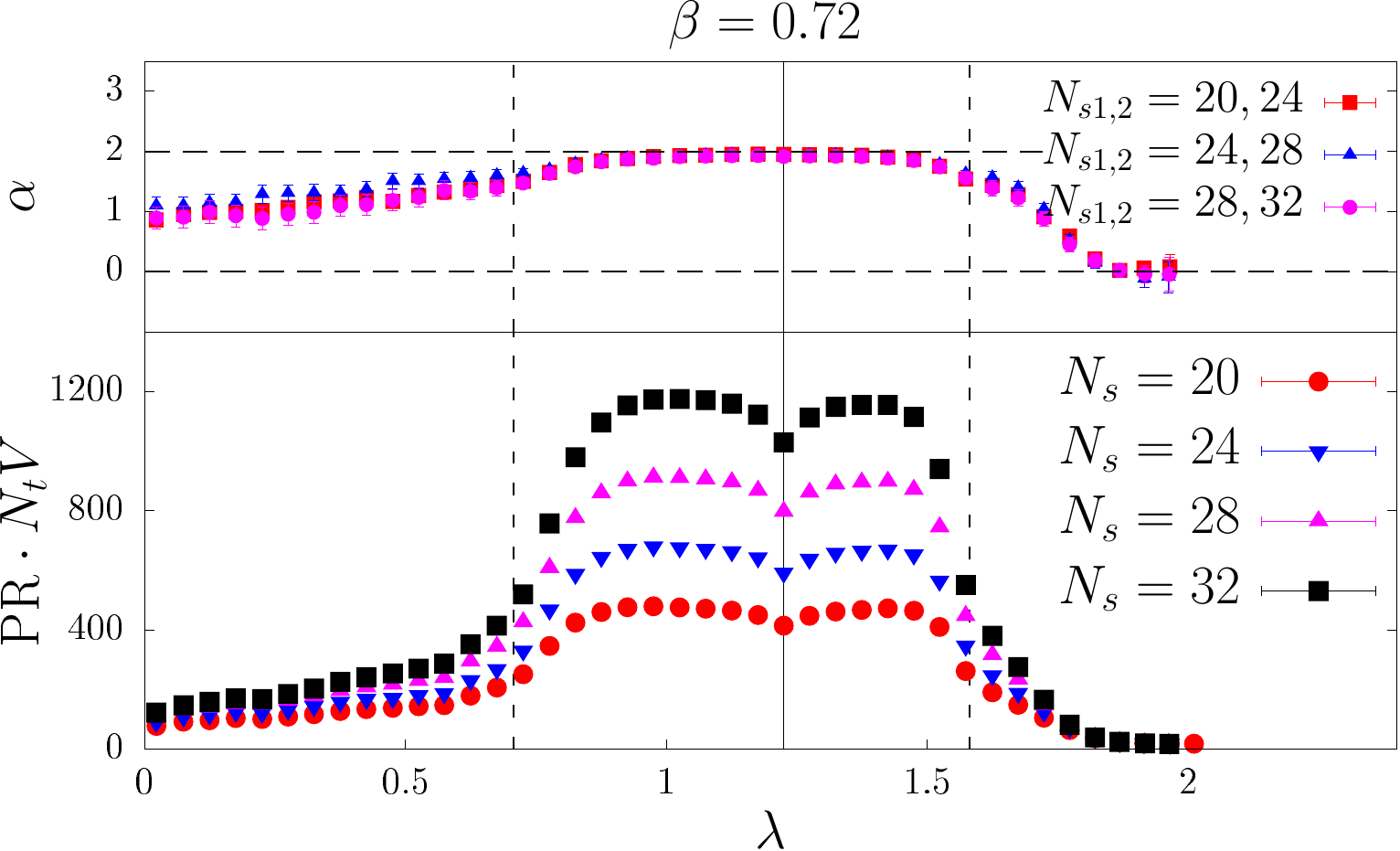}    
   \hfil
   \includegraphics[width=0.48\textwidth]{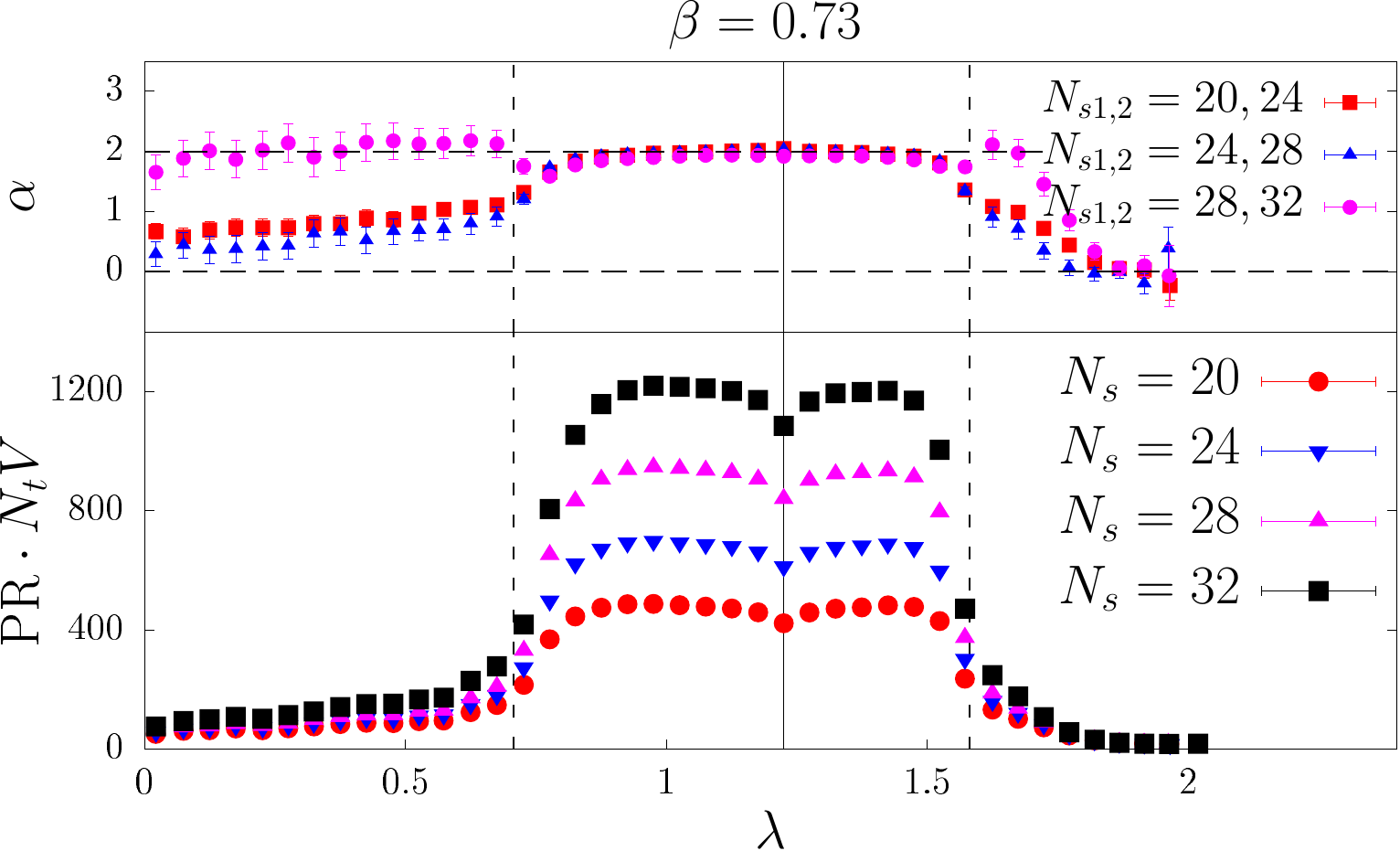}

      \caption{Confined phase, physical sector: mode size
     ${\rm PR} \cdot N_t V = {\rm IPR}^{-1}$ (bottom panels) and
     fractal dimension $\alpha$ (top panels). The points
     $\lambda_{(0)}$, $\lambda_*$ and $\lambda_{(1)}$ (see
     Eq.~\eqref{eq:ns_matsu}) are marked with vertical lines.}

   \label{fig:prVfdim}
\end{figure}

\begin{figure}[t]
   \centering
   \includegraphics[width=0.48\textwidth]{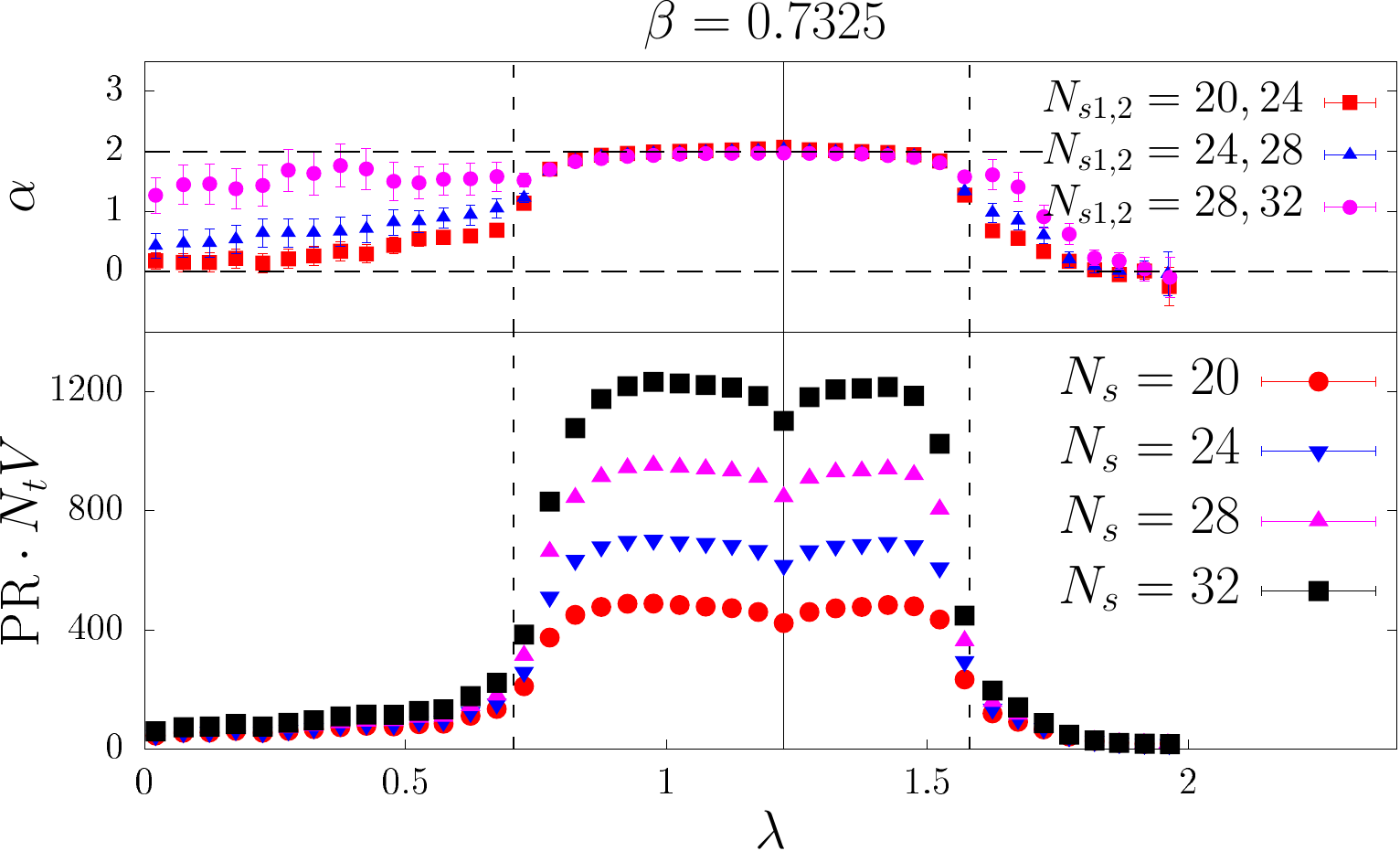}
\hfil      
   \includegraphics[width=0.48\textwidth]{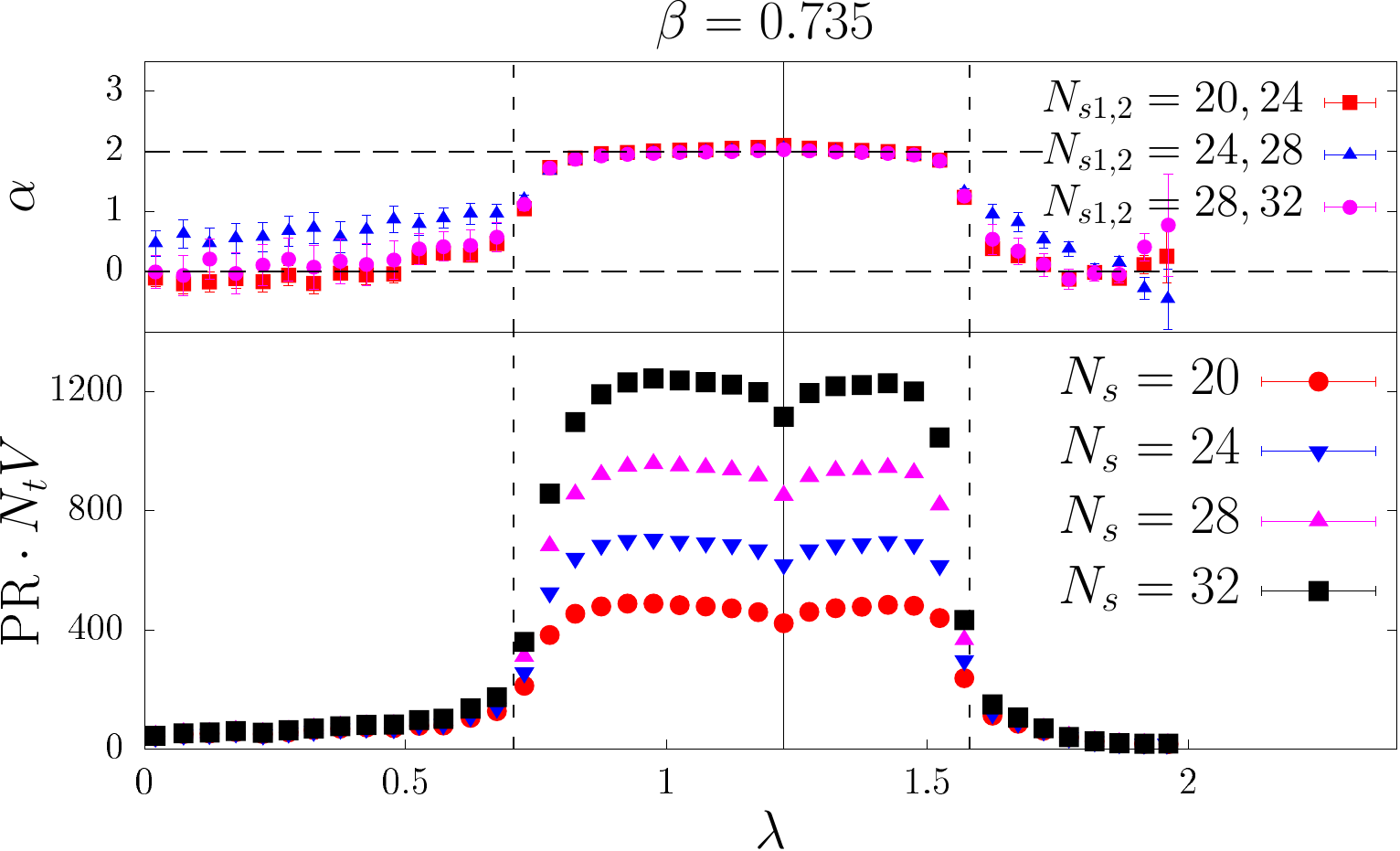}
   
   \vspace{0.1cm}
   \includegraphics[width=0.48\textwidth]{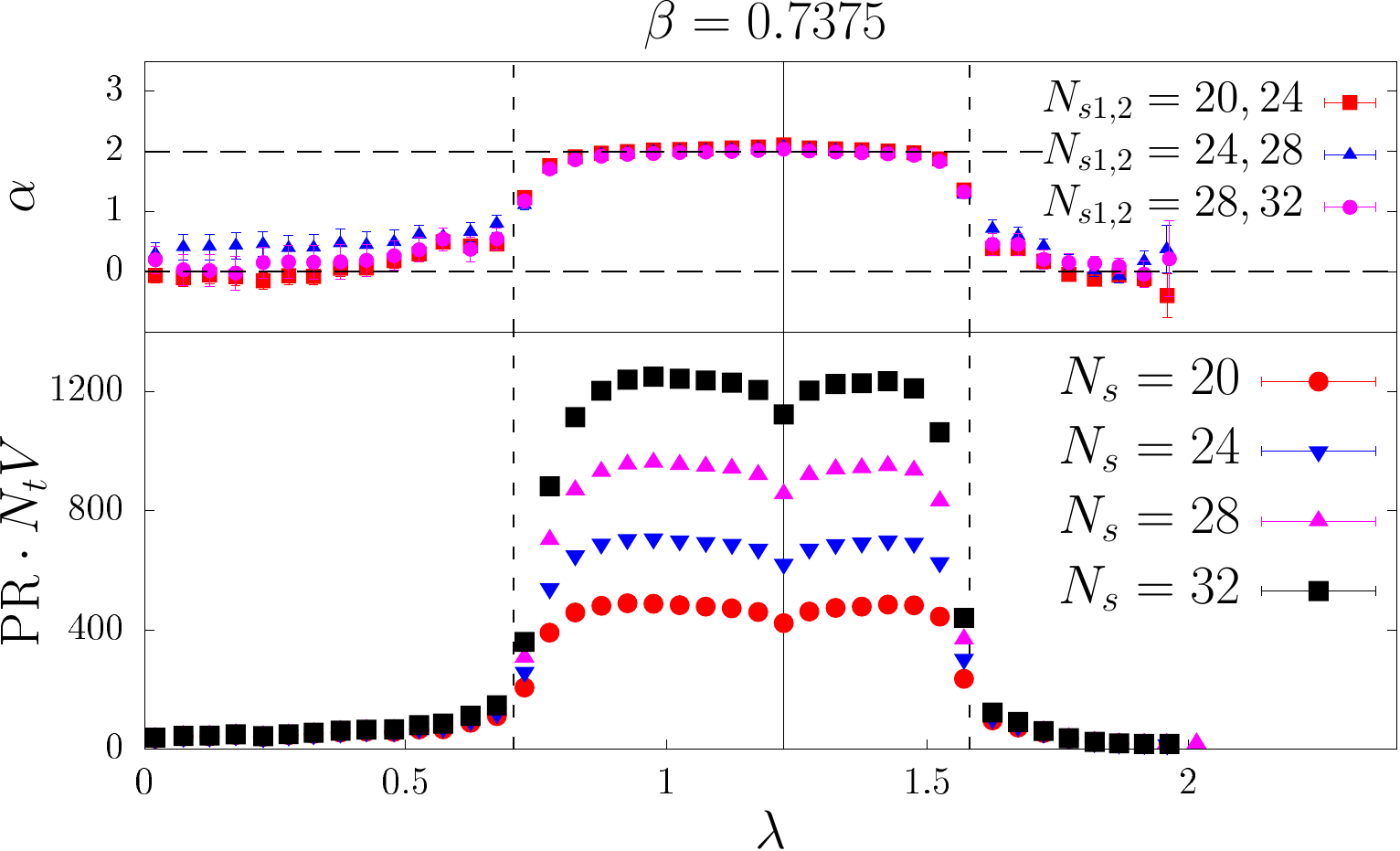}
   \hfil     
   \includegraphics[width=0.48\textwidth]{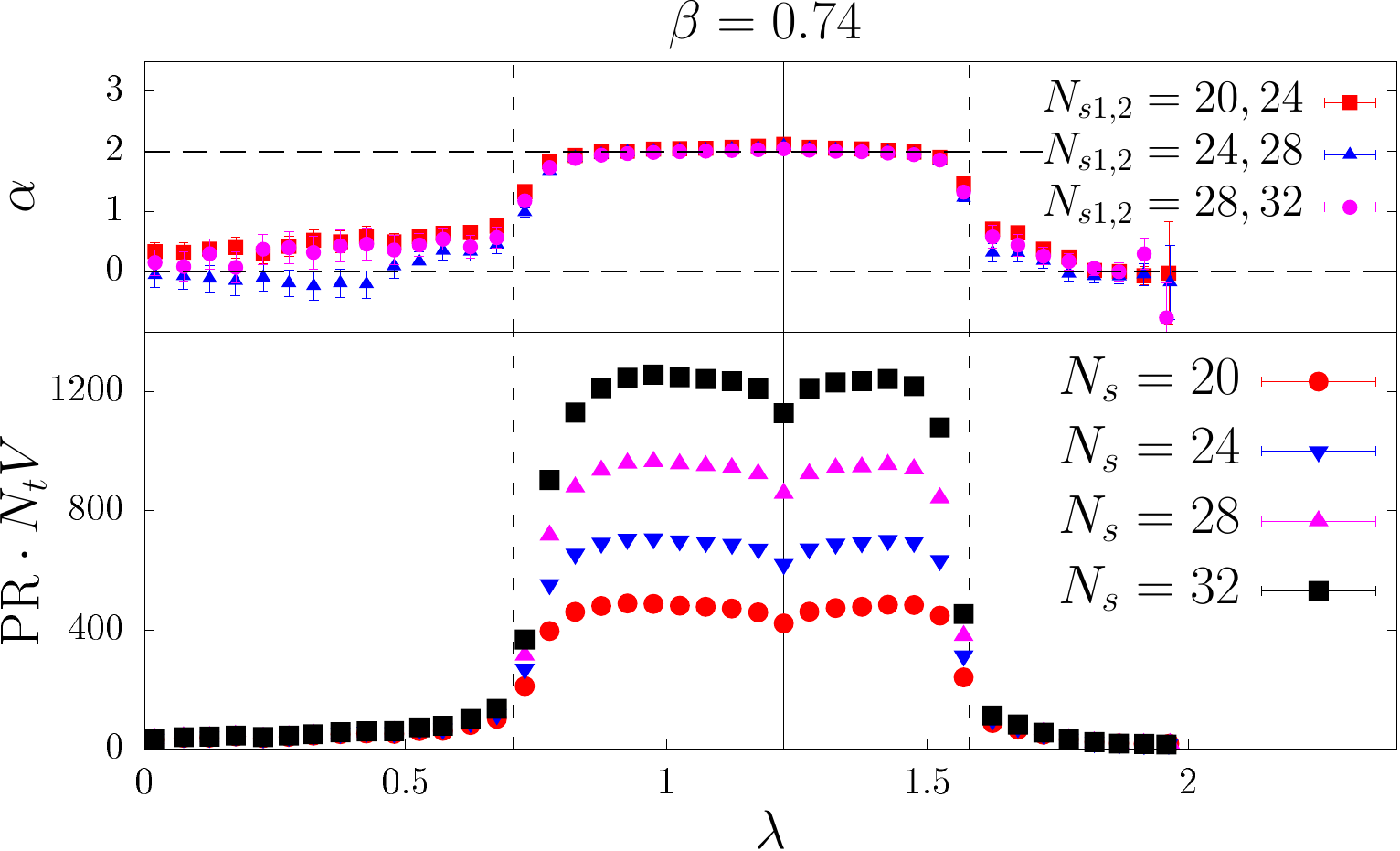}
   
   \caption{Deconfined phase, physical sector: mode size
     ${\rm PR} \cdot N_t V = {\rm IPR}^{-1}$ and fractal dimension
     $\alpha$.}

   \label{fig:prVfdim2}
\end{figure}

\begin{figure}[t]
   \centering
   \includegraphics[width=0.48\textwidth]{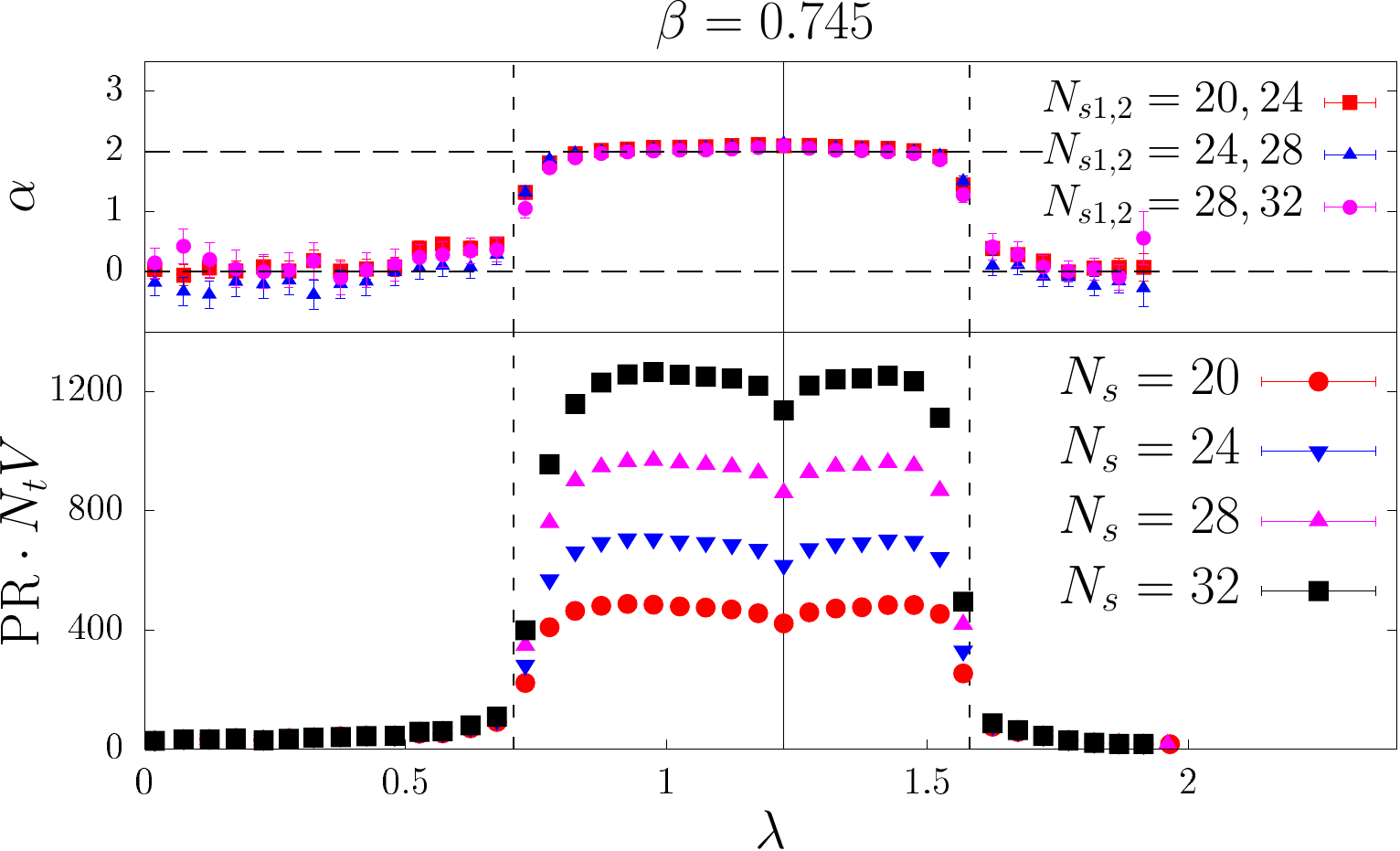}
   \hfil         
   \includegraphics[width=0.48\textwidth]{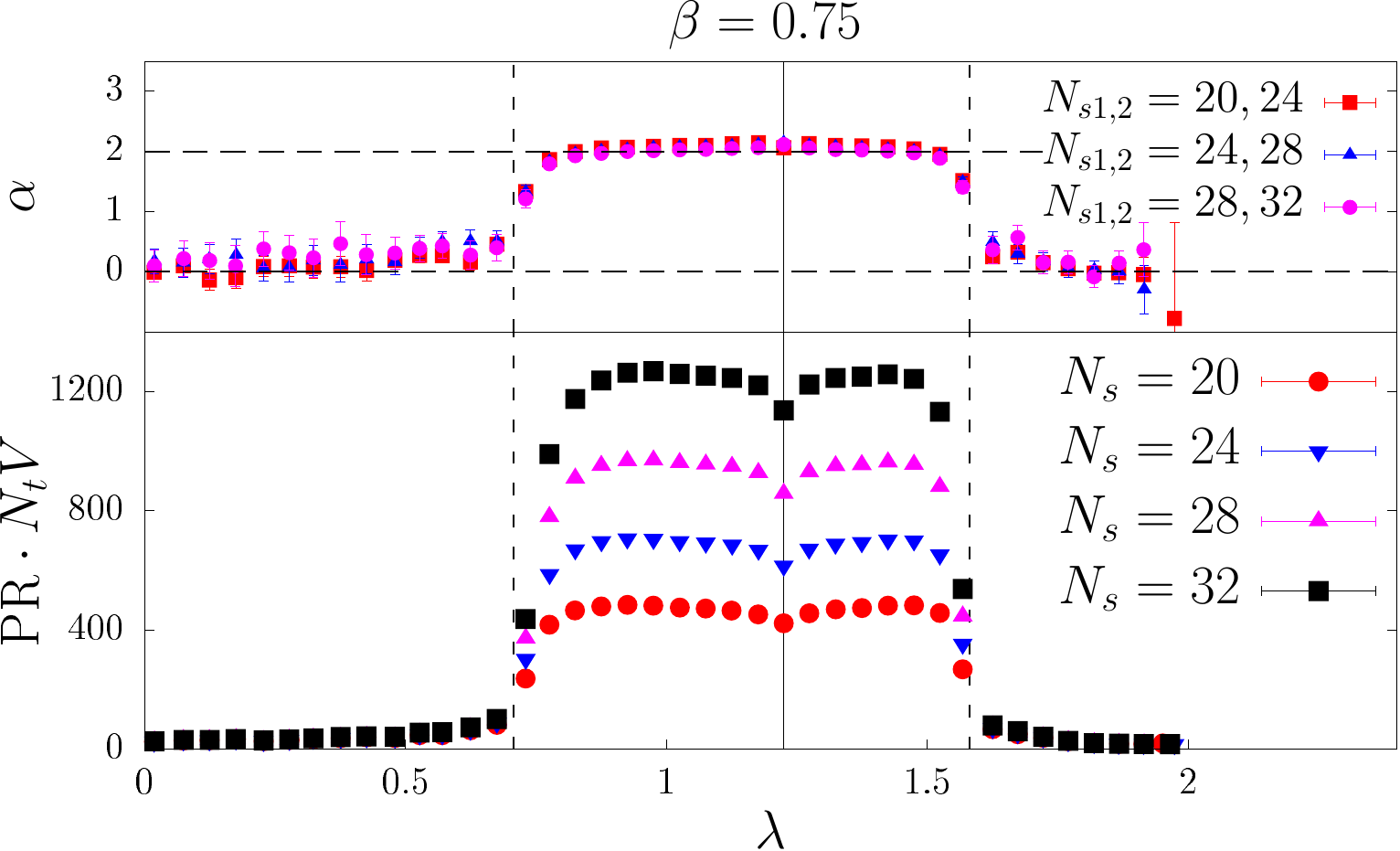}
   
   \vspace{0.1cm}
   \includegraphics[width=0.48\textwidth]{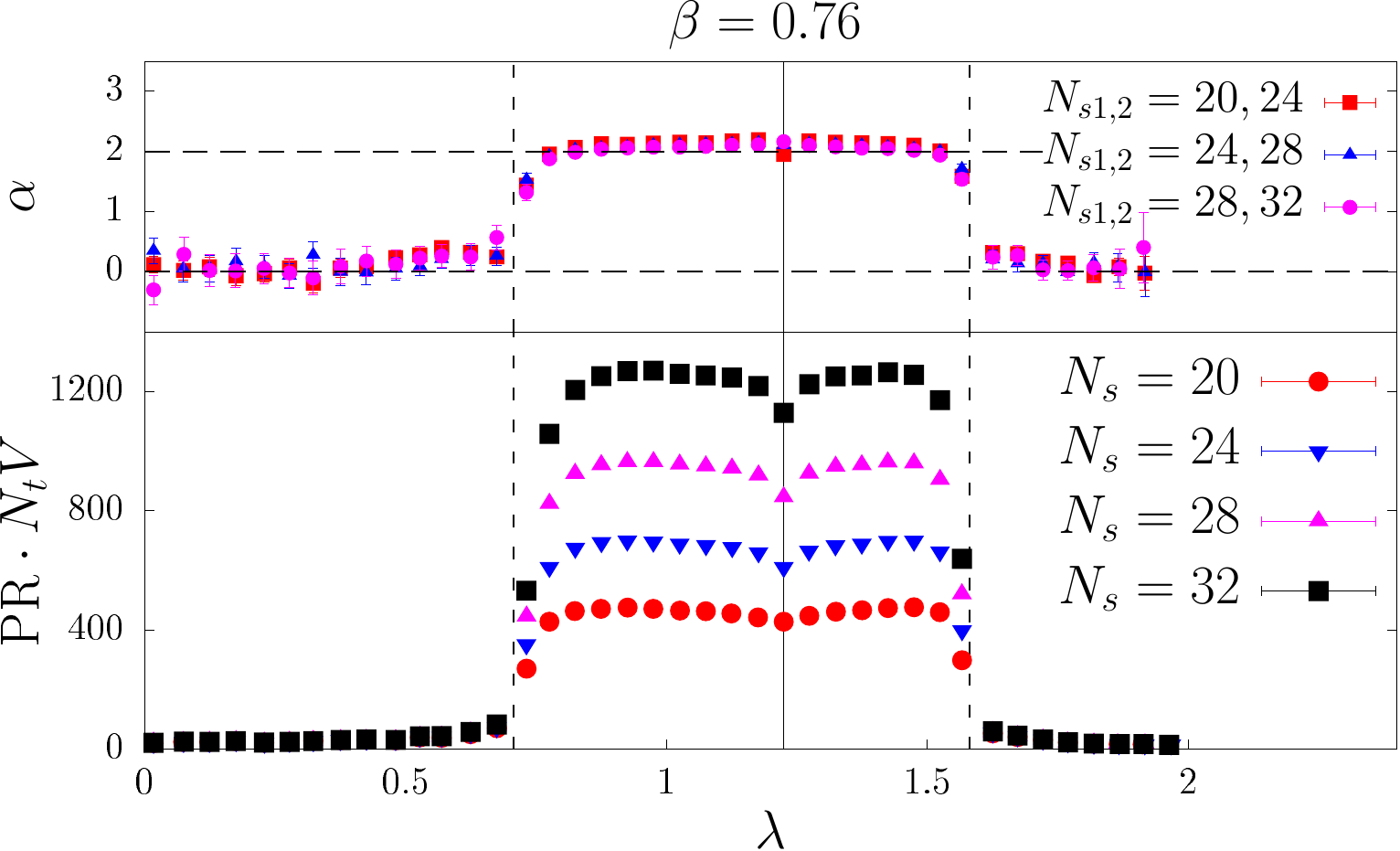}
   \hfil        
   \includegraphics[width=0.48\textwidth]{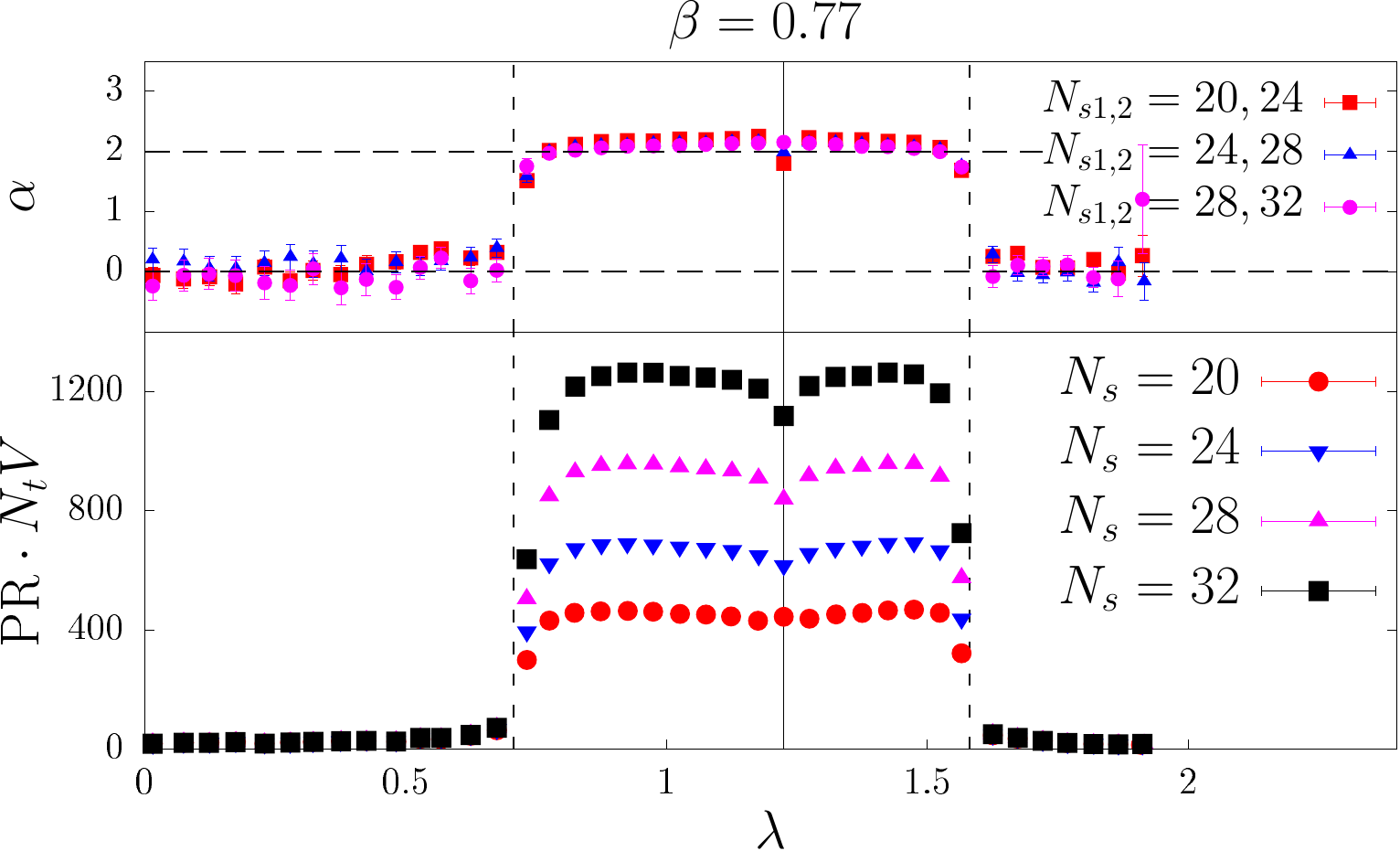}
   
   \caption{Deconfined phase, physical sector: mode size
     ${\rm PR} \cdot N_t V = {\rm IPR}^{-1}$ and fractal dimension
     $\alpha$ (continued).}
   
   \label{fig:prVfdim2bis}
\end{figure}

\begin{figure}[t]
   \centering
   \includegraphics[width=0.48\textwidth]{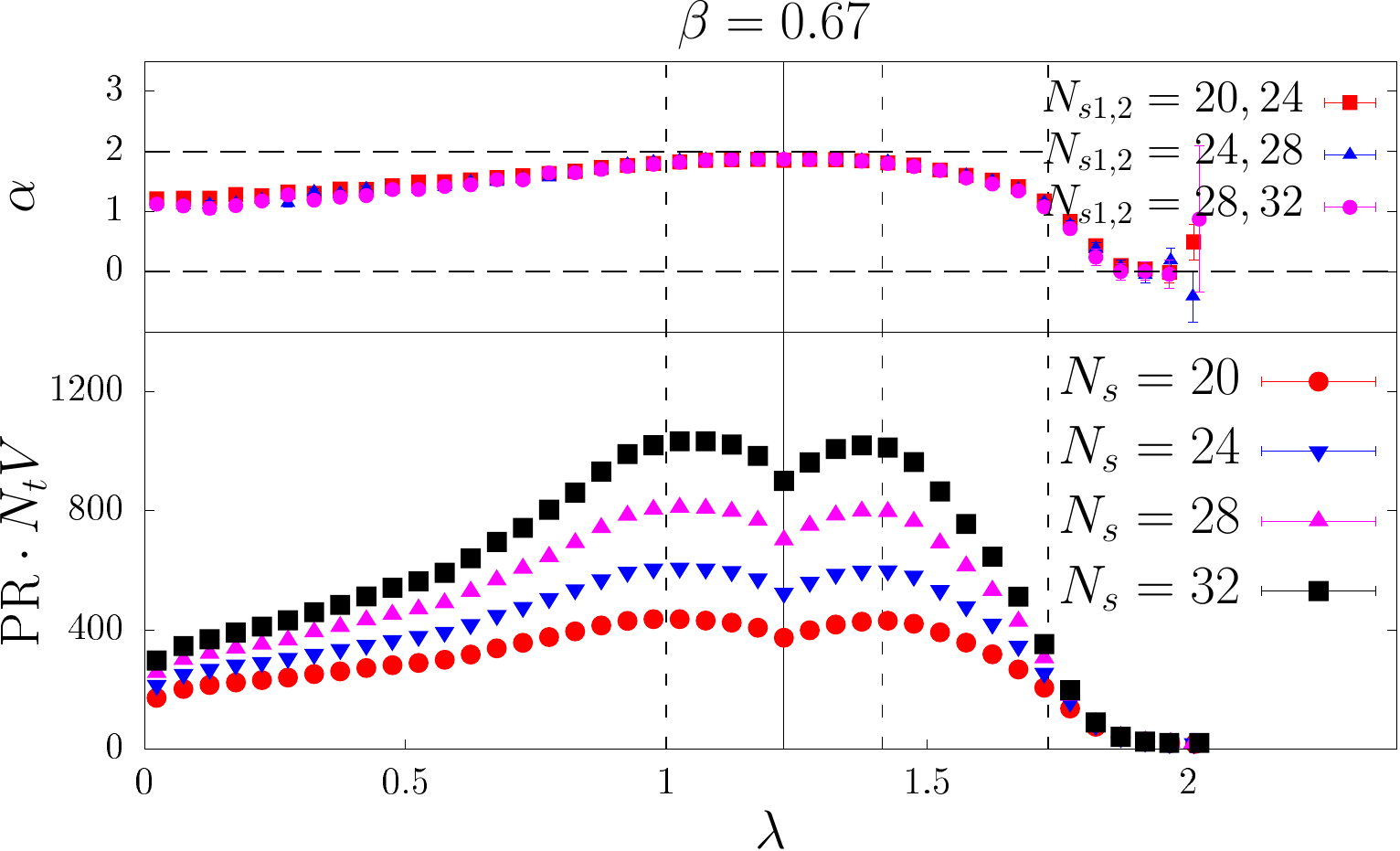}
   \hfil         
   \includegraphics[width=0.48\textwidth]{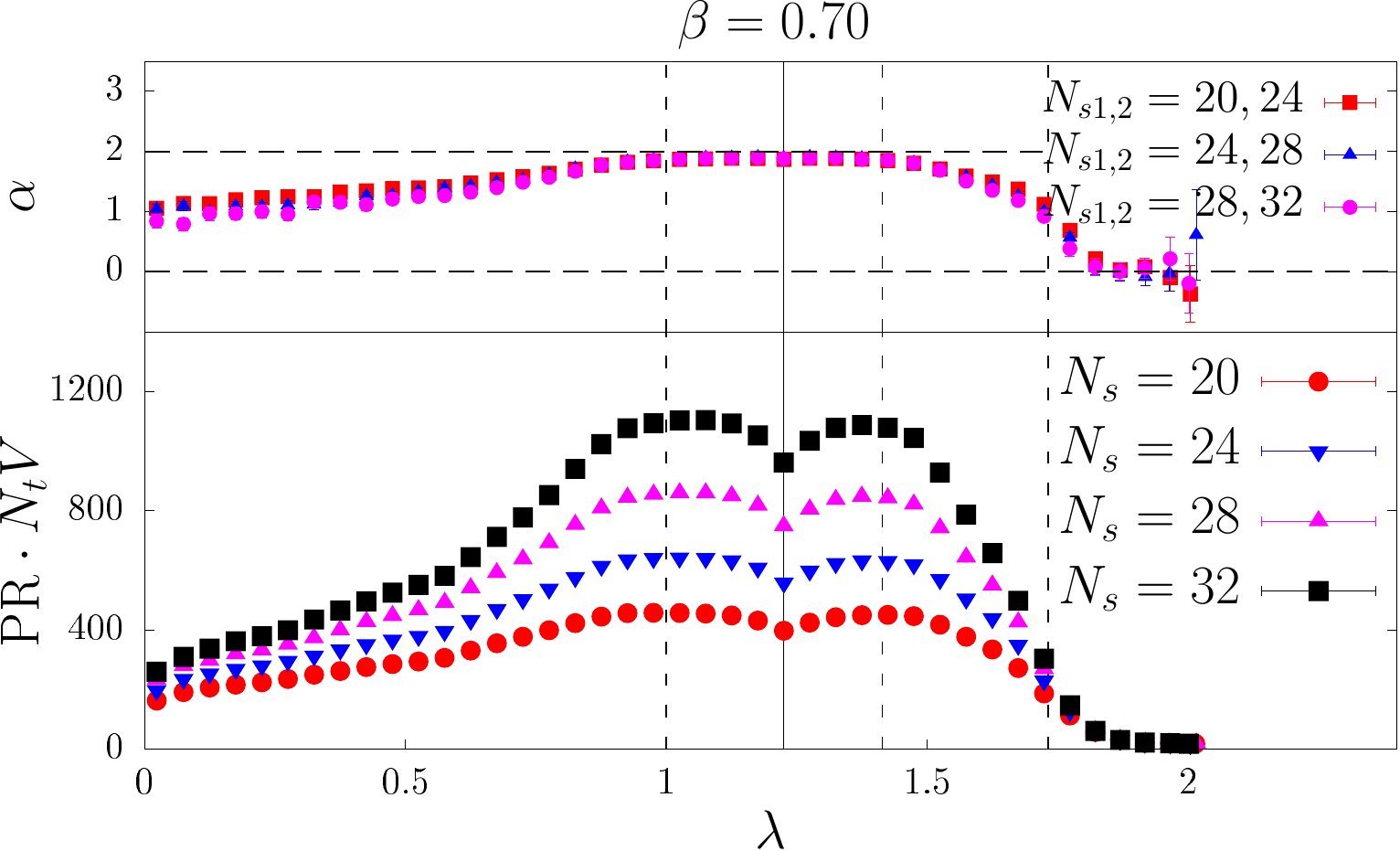}

   \vspace{0.1cm}
   \includegraphics[width=0.48\textwidth]{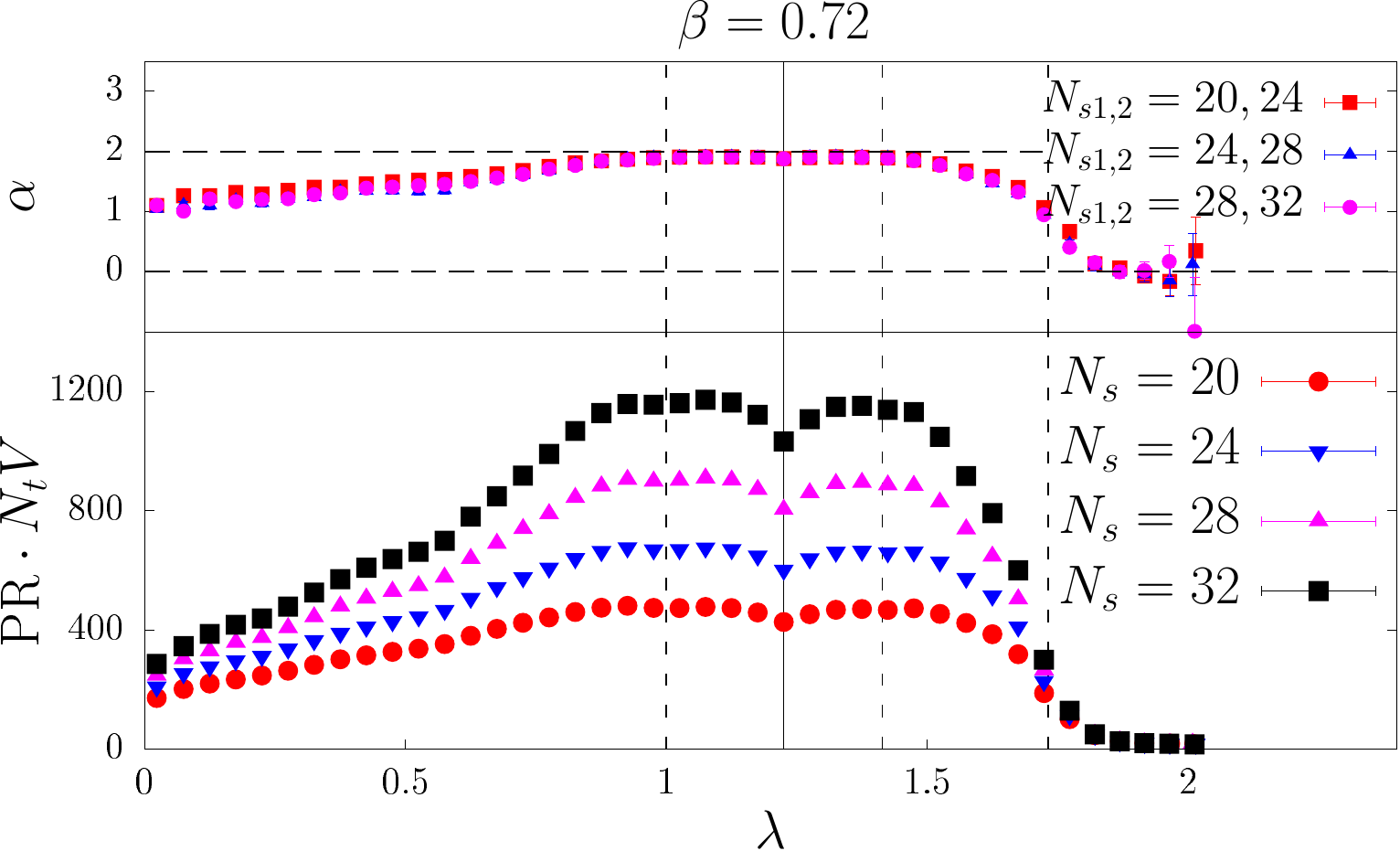}
   \hfil      
   \includegraphics[width=0.48\textwidth]{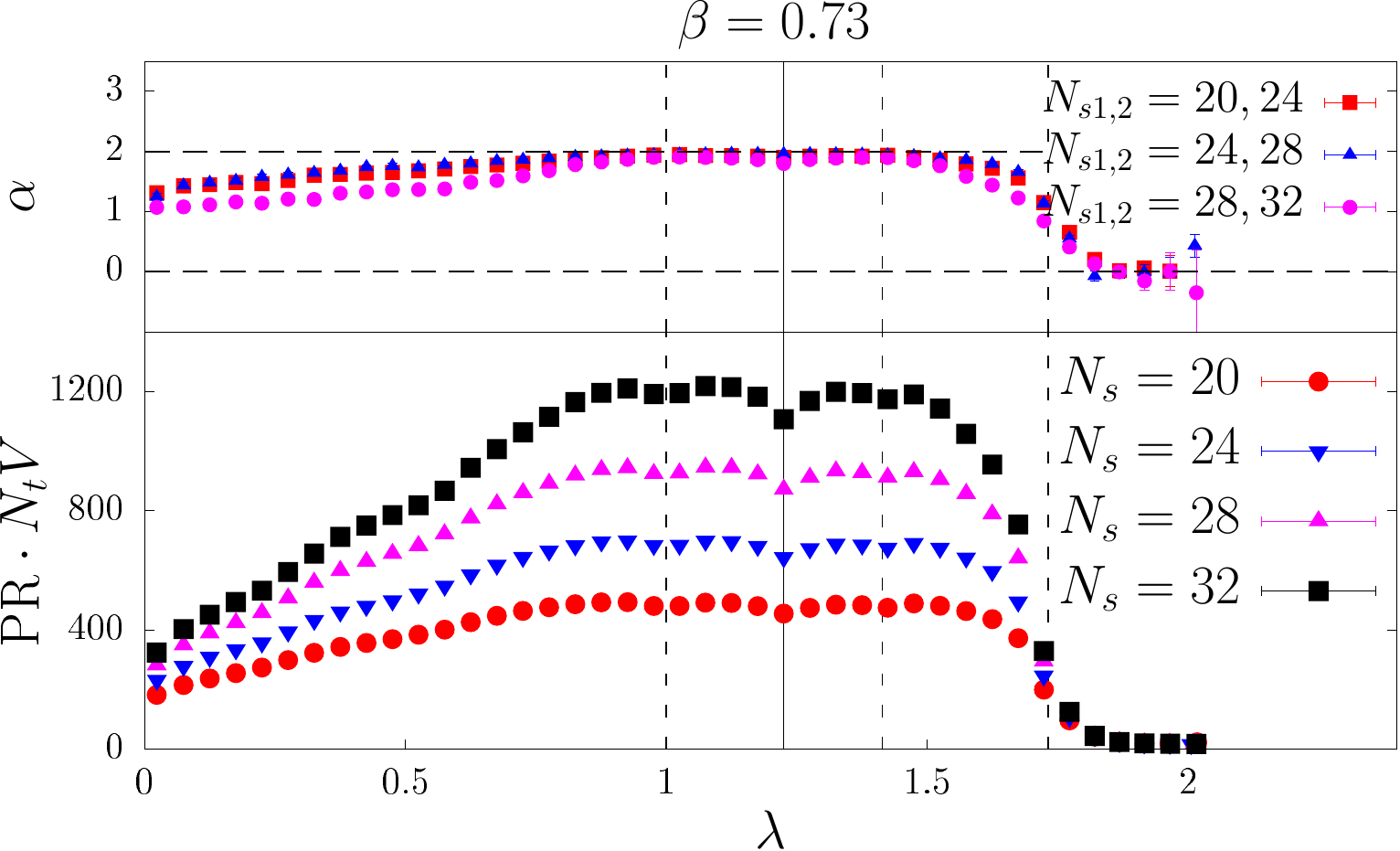}
   
   \caption{Confined phase, unphysical sector: mode size
     ${\rm PR} \cdot N_t V = {\rm IPR}^{-1}$ and fractal dimension
     $\alpha$. The points $\lambda^{\rm PBC}_{(1)}$, $\lambda_*$,
     $\lambda_{(2)}^{\rm PBC}$ and $\lambda_{(3)}^{\rm PBC}$ (see
     Eq.~\eqref{eq:ws_matsu}) are marked with vertical lines.}

   \label{fig:prVfdim_ws}
\end{figure}

\begin{figure}[t]
   \centering
   \includegraphics[width=0.48\textwidth]{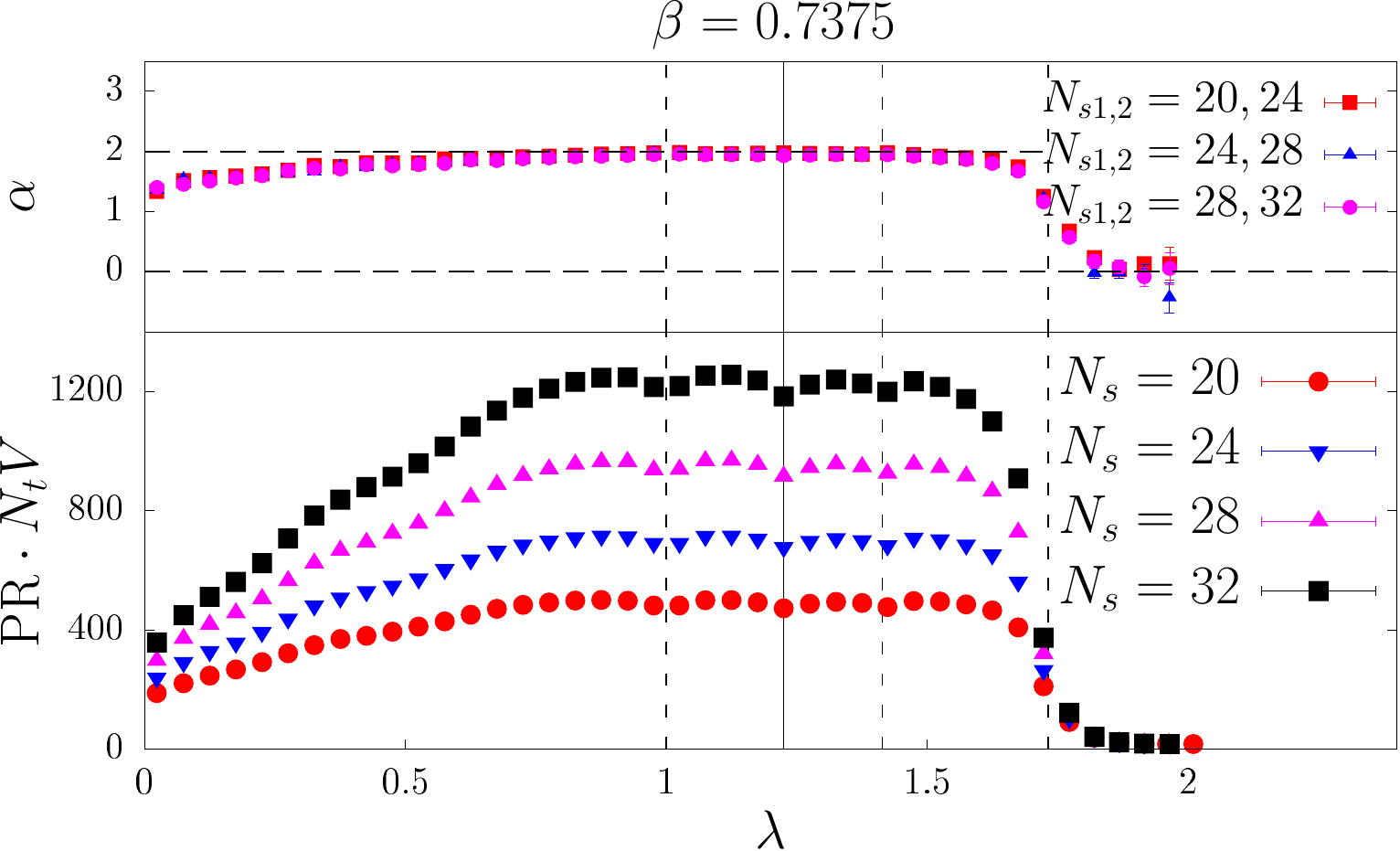}
   \hfil         
   \includegraphics[width=0.48\textwidth]{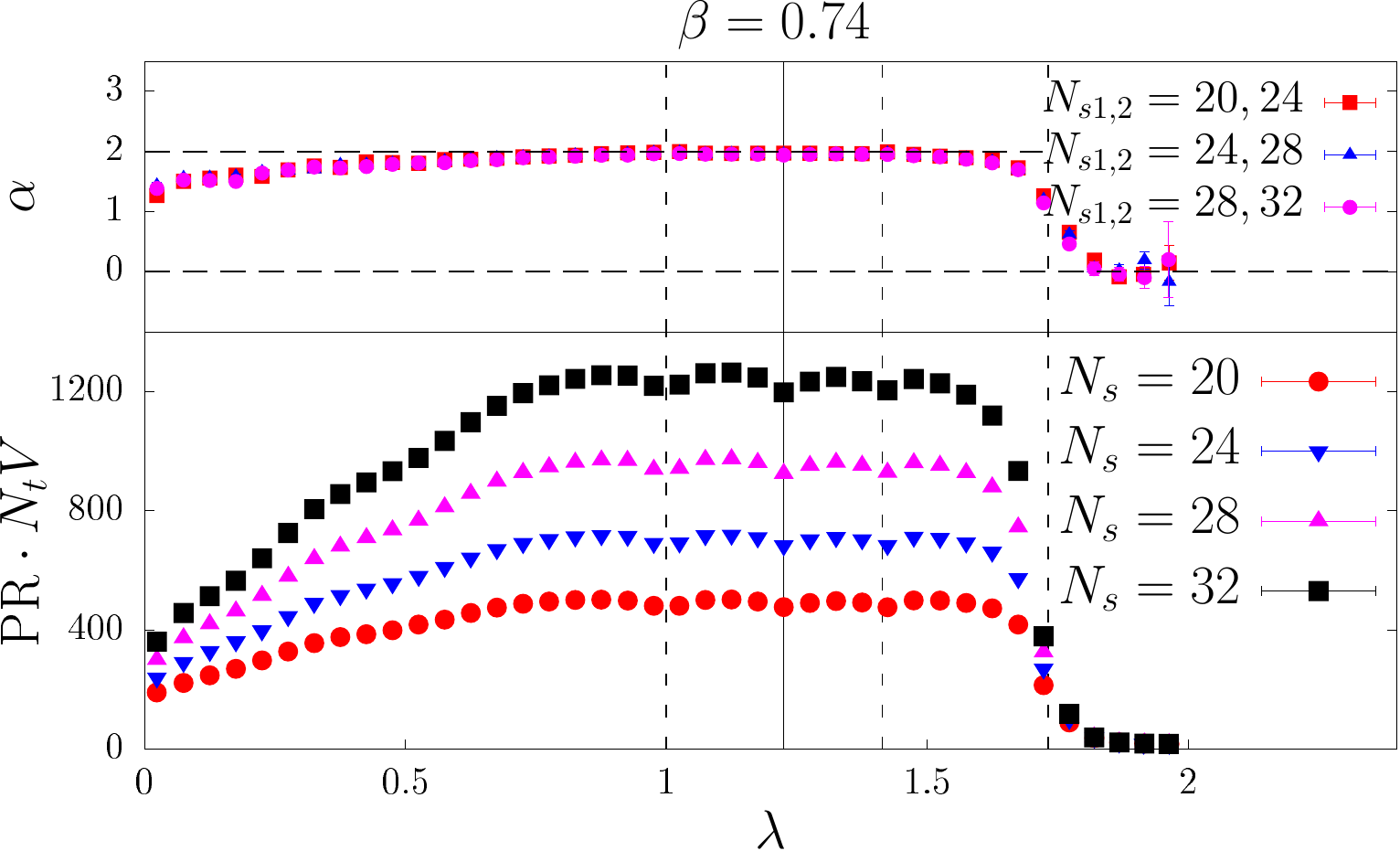}
   
   \vspace{0.1cm}
   \includegraphics[width=0.48\textwidth]{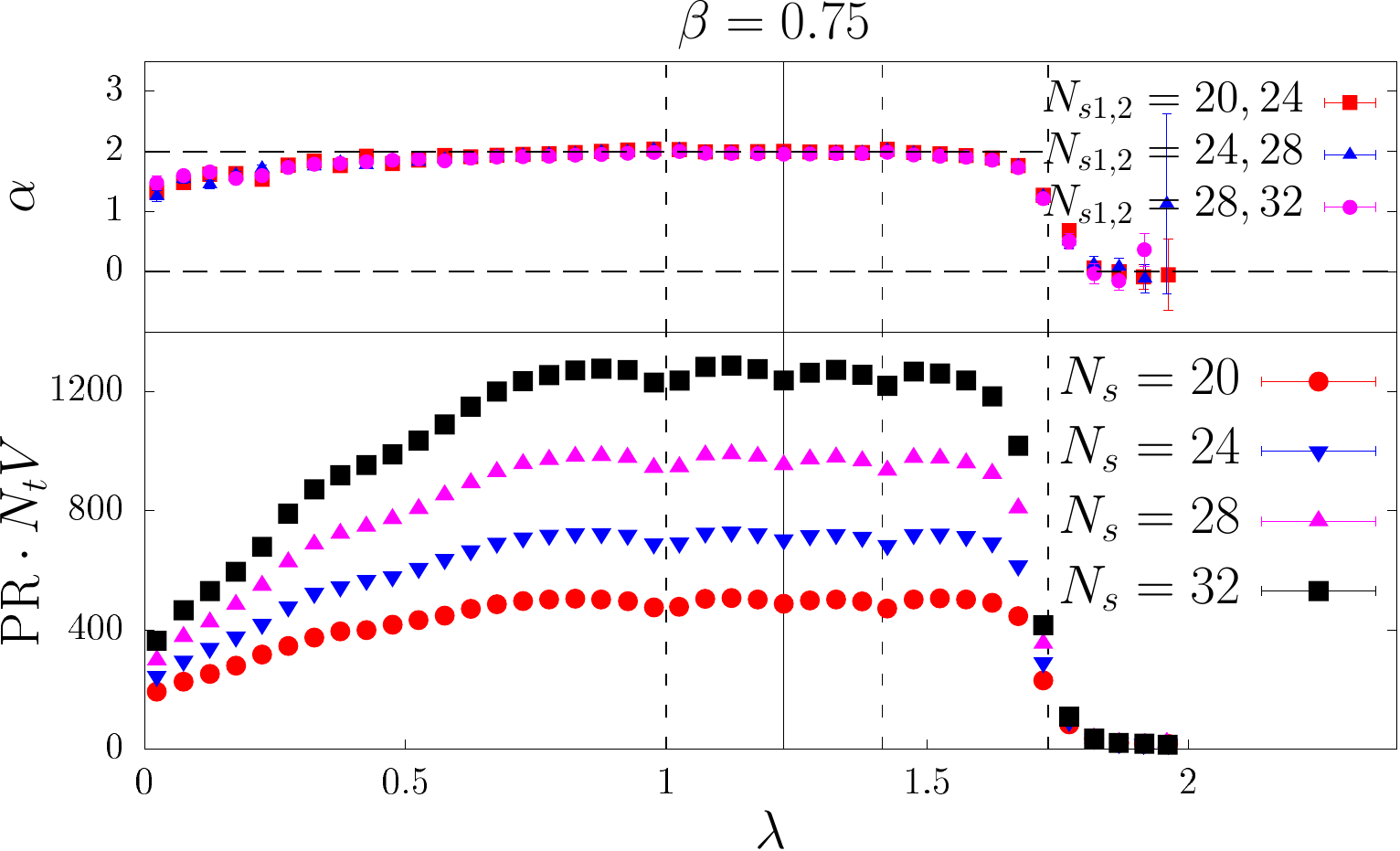}
   \hfil      
   \includegraphics[width=0.48\textwidth]{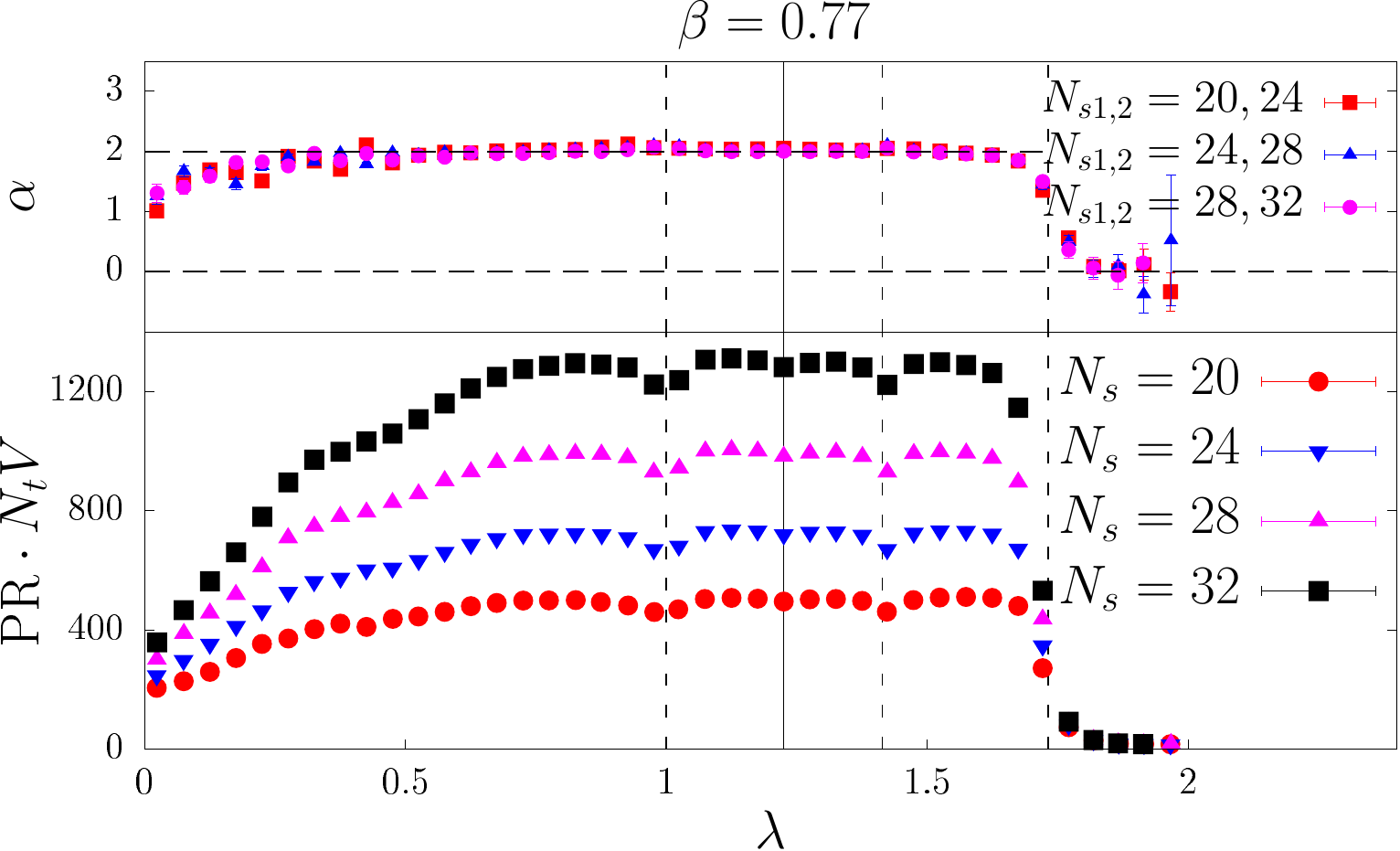}
   
   \caption{Deconfined phase, unphysical sector: mode size
     ${\rm PR} \cdot N_t V = {\rm IPR}^{-1}$ and fractal dimension
     $\alpha$.}
   
   \label{fig:prVfdim_ws2}
\end{figure}

Our numerical results for the ``size'' of the modes,
${\rm PR}\cdot N_t V = {\rm IPR}^{-1}$, are shown in
Figs.~\ref{fig:prVfdim}, \ref{fig:prVfdim2}, and \ref{fig:prVfdim2bis}
for the physical sector, and in Figs.~\ref{fig:prVfdim_ws} and
\ref{fig:prVfdim_ws2} for the unphysical sector, for several values of
$\beta$ both below and above $\beta_c$, and for all the available
lattice sizes. In the same figures we also show the fractal dimension
$\alpha$, computed locally in the spectrum using
Eq.~\eqref{eq:loc5}. Since the average eigenvalue in the various
spectral bins fluctuates slightly for different $N_s$, it is
understood that we compare the average PR in the same spectral bin.
Statistical errors were estimated by linear propagation of the
jackknife errors on ${\rm PR}(\lambda,N_s)$.

\begin{figure}[t]
   \centering
   \includegraphics[width=0.48\textwidth]{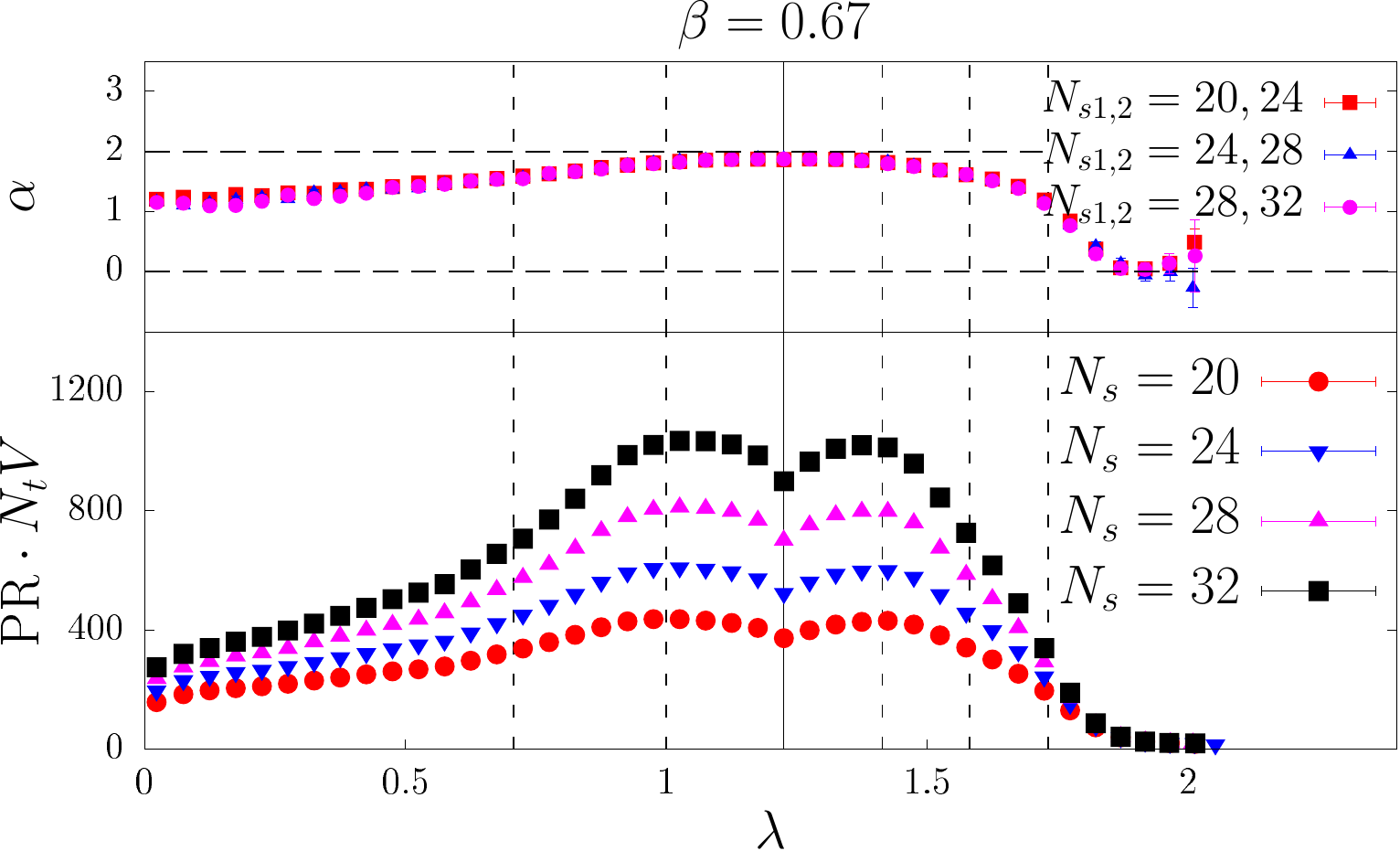}
   \hfil         
   \includegraphics[width=0.48\textwidth]{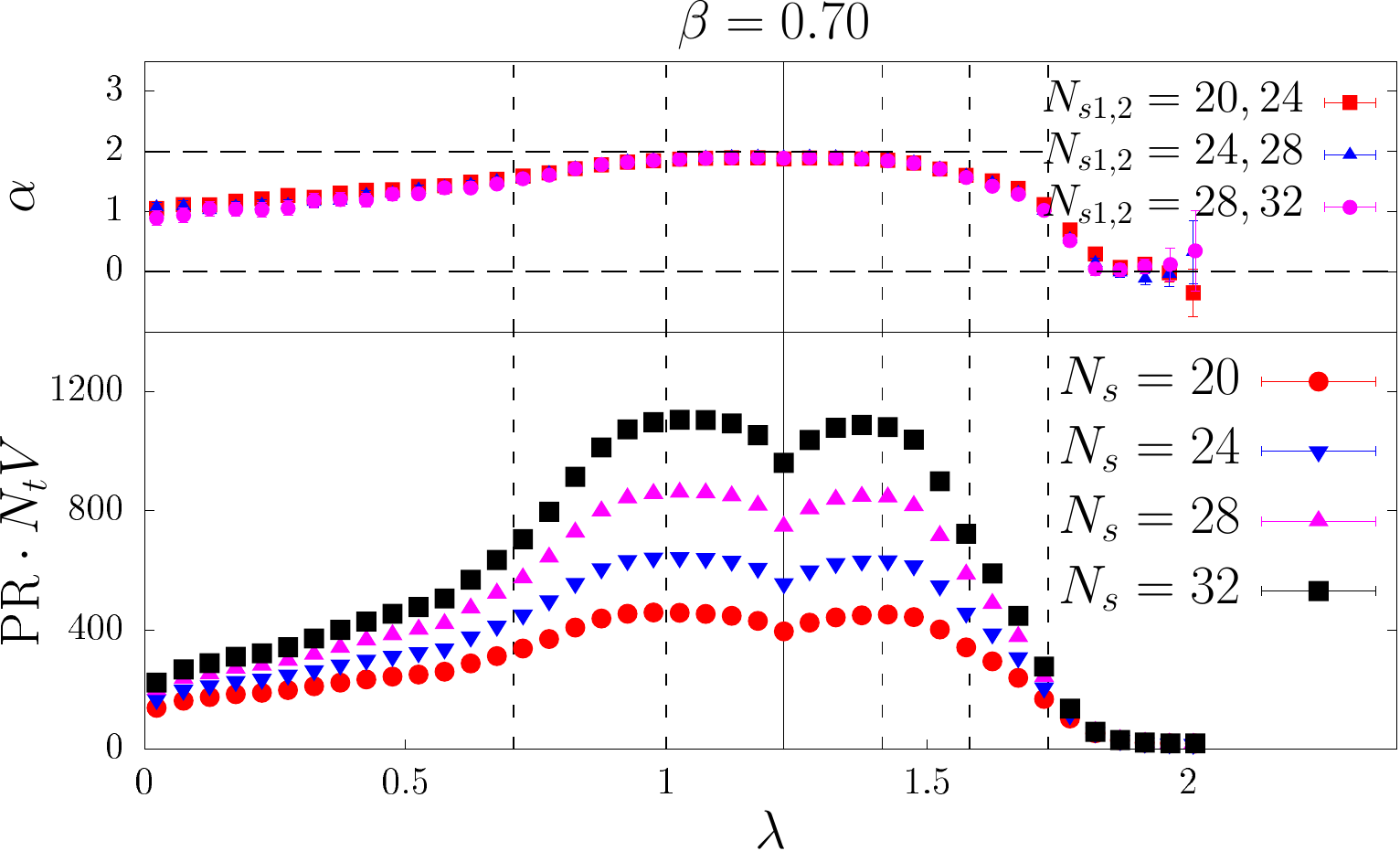}

   \vspace{0.1cm}
   \includegraphics[width=0.48\textwidth]{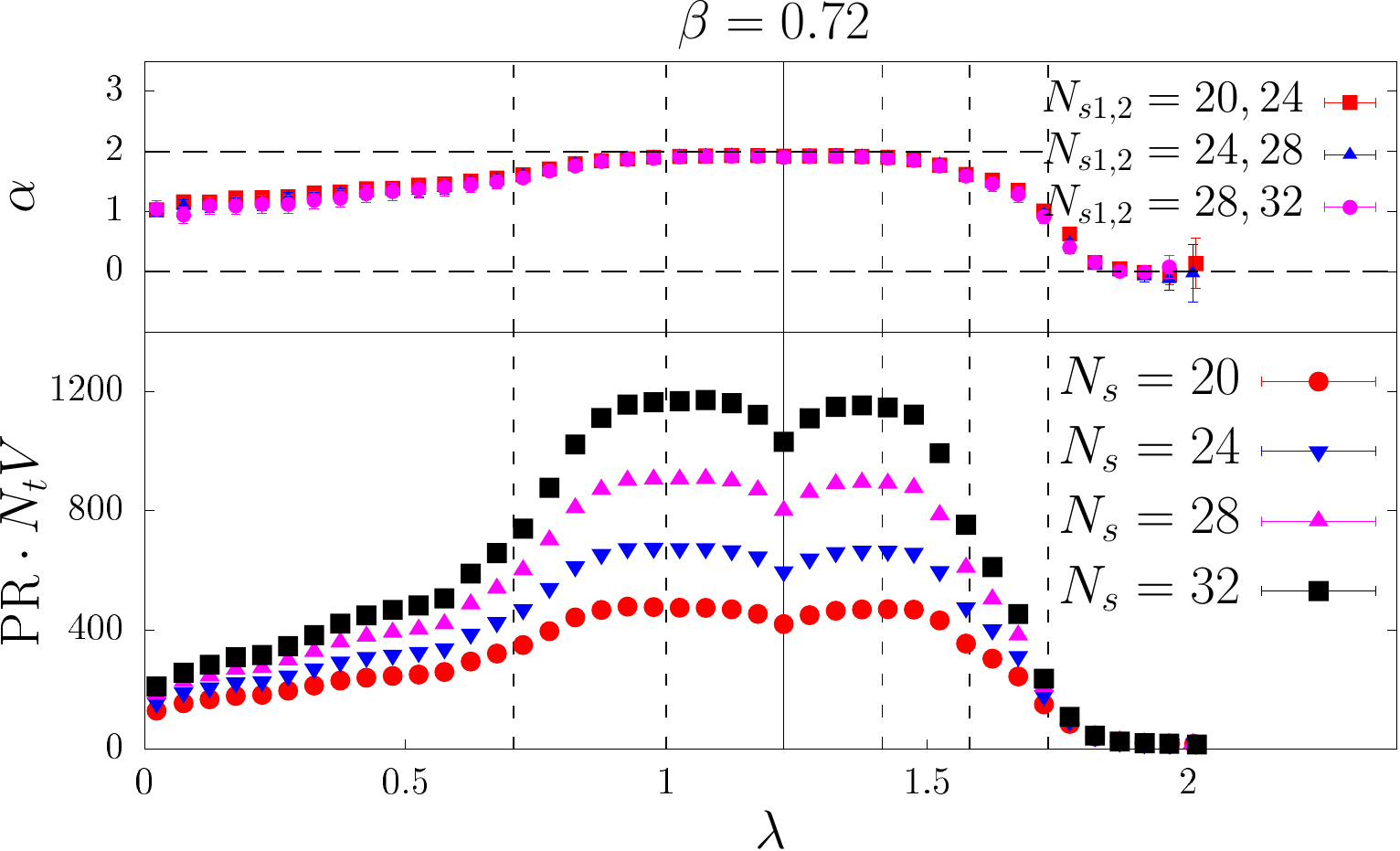}
   \hfil      
   \includegraphics[width=0.48\textwidth]{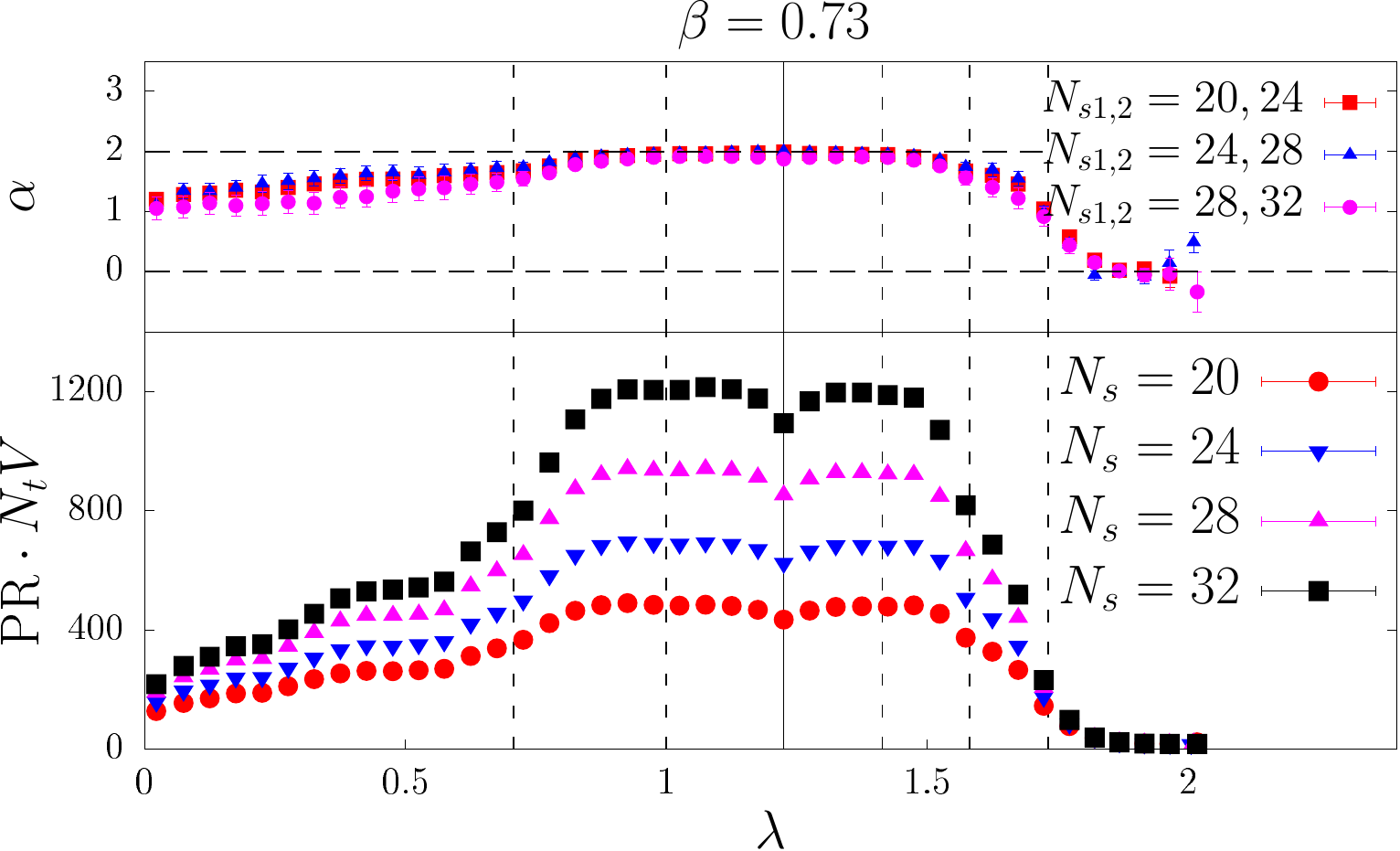}
   
   \caption{Confined phase, both sectors combined: mode size
     ${\rm PR} \cdot N_t V = {\rm IPR}^{-1}$ and fractal dimension
     $\alpha$.  The points $\lambda_{(0)}$, $\lambda^{\rm PBC}_{(1)}$,
     $\lambda_*$, $\lambda_{(2)}^{\rm PBC}$, $\lambda_{(1)}$ and
     $\lambda_{(3)}^{\rm PBC}$ (see Eqs.~\eqref{eq:ns_matsu} and
     \eqref{eq:ws_matsu}) are marked with vertical lines.}
   
   \label{fig:prVfdim_bs}
\end{figure}

In the confined phase the contributions of the two center sectors
should be combined to obtain the correct result. This is shown in
Fig.~\ref{fig:prVfdim_bs}. We first discuss the results in the two
sectors separately, and briefly comment on the combined result in the
confined phase afterwards.

\subsubsection{Physical sector}
\label{sec:num_pr_ns}

The eigenvalues of the free staggered operator (i.e., on the trivial
configuration $U_\mu(n)=1,\forall n,\mu$) in 2+1 dimensions read
\begin{equation}
  \label{eq:free_stag_spec}
 \lambda_{\rm free}=\pm\sqrt{(\sin\omega_k)^2 + (\sin p_1)^2 + (\sin p_2)^2}\,, 
\end{equation}
where $\omega_k= \f{(2k+1)\pi}{N_t}$, $k=0,\ldots,N_t-1$ are the
Matsubara frequencies and $p_j = \f{2j\pi}{N_s}$, $j=0,\ldots,N_s-1$
are the lattice momenta.  At fixed $k$, the positive eigenvalues lie
in the ``Matsubara sector'' $[\sin\omega_k,\sqrt{\sin\omega_k^2 +2}]$,
$k=0,\ldots,\f{N_t}{2}-1$ ($k=\f{N_t}{2},\ldots, N_t-1$ are reserved
instead for the negative spectrum), and so the whole positive spectrum
lies in the interval
$[\sin\omega_0,\sqrt{\sin\omega_{\rm max}^2 +2}]$, with
$\sin\omega_{\rm max} = \max_k|\sin\omega_k|$. For $N_t=4$ one has
$(\sin\omega_k)^2=\f{1}{\sqrt{2}}\forall k$. This suggests that one
should distinguish three spectral regions in the physical sector: the
low modes $\lambda<\lambda_{(0)}$, the bulk modes
$\lambda_{(0)}\le \lambda \le \lambda_{(1)}$, and the high modes
$\lambda>\lambda_{(1)}$, where
\begin{equation}
  \label{eq:ns_matsu}
  \lambda_{(0)}\equiv \sin\tf{\pi}{4}=\tf{1}{\sqrt{2}}\,, \qquad
  \lambda_{(1)}\equiv \sqrt{\left(\sin\tf{\pi}{4}\right)^2 + 2} =
  \sqrt{\tf{5}{2}}\,. 
\end{equation}
In both phases, the size of the mode, ${\rm IPR}^{-1}$, is larger in
the bulk than at the edges of the spectrum, as
Figs.~\ref{fig:prVfdim}--\ref{fig:prVfdim2bis} (bottom panels) show.
The volume scaling indicates that bulk modes are delocalized in both
phases, see Figs.~\ref{fig:prVfdim}--\ref{fig:prVfdim2bis} (top
panels).  The low modes, instead, while delocalized in the
low-temperature phase, become localized in the high-temperature phase,
up to some critical point in the spectrum. More precisely, in the
confined phase low modes are delocalized but with a nontrivial fractal
dimension, which starting from around $\alpha\approx 1$ for the lowest
modes increases toward 2 as one approaches the bulk. In the
deconfined phase, instead, for low modes $\alpha$ is close to 0, and
starts increasing toward 2 at some point $\lambda_c$, which we
identify as the mobility edge. While difficult to locate with the
current precision, the results shown in Figs.~\ref{fig:prVfdim2} and
\ref{fig:prVfdim2bis} suggest that $\lambda_c<\lambda_{(0)}$, up to
the largest available value of $\beta$, and that it increases with
$\beta$. Finally, the high modes are localized in both phases, above a
second mobility edge $\lambda_c'$, identified as the point where
$\alpha$, after decreasing from its bulk value, reaches again 0.  In
this case the results in
Figs.~\ref{fig:prVfdim}--\ref{fig:prVfdim2bis} suggest that
$\lambda_c'>\lambda_{(1)}$ up to the largest available value of
$\beta$. While in the confined phase there seems to be little or no
dependence on $\beta$, $\lambda_c'$ seems to decrease with $\beta$ in
the deconfined phase.

Very large fluctuations in the fractal dimension of the low modes, and
of the high modes near $\lambda_{(1)}$, are apparent near $\beta_c$.
These are a consequence of large finite-size effects, originating from
the tunnelling of the system between the two phases, where low modes
have different localization properties. A similar explanation holds
for the high modes right above $\lambda_{(1)}$: the second mobility
edge, $\lambda_c'$, seems in fact independent of $\beta$ below
$\beta_c$ while it decreases with $\beta$ above $\beta_c$, and
tunnelling between the two phases leads to contributions from the
``wrong'' type of modes.  The contribution of these tunnelling
configurations are however expected to become negligible as the system
size is increased.

Notice that the estimate of the fractal dimension of the bulk modes in
the deconfined phase, obtained from the available volumes, sometimes
exceeds 2. This is clearly a finite-size effect, indicating that the
volume scaling of the bulk modes has not yet reached its asymptotic
behavior (which must be such that $\alpha\le 2$). This effect is
present, and actually stronger, far away from $\beta_c$, and is most
likely of a different origin than tunneling configurations. A possible
explanation is that, as the system grows, not only the effective
support of the bulk modes keeps increasing at the same rate, but at
the same time these modes become also more uniformly distributed on
their support. In any case, our results indicate that $\alpha$ keeps
approaching 2 (from above) as the lattice volume grows.

As we mentioned above in Section \ref{sec:locD}, in the deconfined
phase of the present model we do not expect an abrupt change from
localized ($\alpha=0$) to fully delocalized ($\alpha=2$) modes even in
the thermodynamic limit. A spectral range with volume-independent
non-trivial fractal dimension seems to be present around
$\lambda_{(0)}$. This is consistent with the BKT nature of the
Anderson transition expected for this model. However, a detailed study
of this issue, as well as the precise location of the mobility edge,
is beyond our reach with the available statistics and system sizes.
It seems however beyond doubt that the localization properties of the
low modes change across the transition, and the presence of a mobility
edge seems very likely.

The clear separation of low, bulk and high modes at high temperature
is not surprising. In fact, as $\beta$ increases the Polyakov loop
becomes more and more ordered, inducing stronger and stronger
correlations among time slices. This brings the system to fluctuate
more and more closely around the trivial configuration, whose Dirac
spectrum lies exactly in the range $[\lambda_{(0)},\lambda_{(1)}]$. It
is then not unexpected that modes at both edges of the positive
spectrum tend to localize, as they are probably related to the local
fluctuations of the gauge configurations away from the trivial
one. This will be investigated in detail in Subsection
\ref{sec:num_corr}.  It is interesting to note that also the point
$\lambda_*$ is singled out, with a clear dip in the PR; however, we
have no explanation for this behavior.

\subsubsection{Unphysical sector}
\label{sec:num_pr_ws}

The analogue of the trivial configuration in the unphysical sector is
the gauge configuration with $U_{2,3}(n)=1~\forall n$, $U_{1}(n)=1$ if
$n_1<N_t-1$, $U_{1}(n)=-1$ if $n_1=N_t-1$. The spectrum of the
staggered operator on this configuration is equal to that obtained
with the trivial configuration but with temporal boundary conditions
switched to periodic instead of antiperiodic,
\begin{equation}
  \label{eq:free_stag_spec_pbc}
  \lambda_{\rm free}^{\rm PBC}=\pm\sqrt{(\sin\omega_k^{\rm PBC})^2 +
    (\sin p_1)^2 + (\sin p_2)^2}\,,  
\end{equation}
where $p_j$ have been defined under Eq.~\eqref{eq:free_stag_spec},
and
$\omega_k^{\rm PBC}= \f{2k\pi}{N_t}$, $k=0,\ldots,N_t-1$. For $N_t=4$
one has $(\sin\omega_{0,2}^{\rm PBC})^2=0$,
$(\sin\omega_{1,3}^{\rm PBC})^2=1$. In analogy with the physical
sector, one then expects that, at least at high temperature, the
points
\begin{equation}
  \label{eq:ws_matsu}
  \begin{aligned}
    & \lambda^{\rm PBC}_{(0)}\equiv 0\,,\qquad
    \lambda^{\rm PBC}_{(1)}\equiv   \sin\tf{\pi}{2}=1\,,\\
    & \lambda^{\rm PBC}_{(2)}\equiv \sqrt{2} \,, \qquad \lambda^{\rm
      PBC}_{(3)}\equiv \sqrt{\left(\sin\tf{\pi}{2}\right)^2 + 2} =
    \sqrt{3}\,,
  \end{aligned}
\end{equation}
that correspond to the boundaries of the Matsubara sectors, are
singled out in the spectrum. Indeed, as $\beta$ increases and the
system becomes more ordered, one expects the spectrum to resemble more
and more the ``free'' spectrum of Eq.~\eqref{eq:free_stag_spec_pbc}.

As a matter of fact, in the low-temperature phase the unphysical
sector does not differ much from the physical one: as shown in
Fig.~\ref{fig:prVfdim_ws} (bottom panels), all modes are extended,
except at the very high end ($\lambda>\lambda^{\rm PBC}_{(3)}$) where
they are localized. In contrast with the physical sector, though, this
does not change qualitatively as one crosses over into the
high-temperature phase. As it can be seen in
Fig.~\ref{fig:prVfdim_ws2} (bottom panels), the tendency is for the
mode size to increase everywhere below
$\lambda<\lambda^{\rm PBC}_{(3)}$ (especially below
$\lambda^{\rm PBC}_{(1)}$ and above $\lambda^{\rm PBC}_{(2)}$), and to
change very little at the high end $\lambda>\lambda^{\rm PBC}_{(3)}$.
Figs.~\ref{fig:prVfdim_ws} and \ref{fig:prVfdim_ws2} (bottom panels)
suggest that a mobility edge is present at the high end of the
spectrum, and that it depends on $\beta$ very little, or not at all,
in both phases of the theory. More detailed information is provided by
the fractal dimension, shown in Figs.~\ref{fig:prVfdim_ws},
\ref{fig:prVfdim_ws2} (top panels). In the confined phase one finds
the same situation as in the physical sector, with delocalized low
modes with nontrivial fractal dimension $\alpha\approx 1$, delocalized
bulk modes with $\alpha\approx 2$, and localized high modes
($\alpha=0$) above some $\lambda_c'$. While no qualitative change is
seen when crossing over into the deconfined phase, quantitatively one
observes that both the low and bulk modes tend to become more
delocalized, with $\alpha$ getting closer to 2 as $\beta$
increases. Not much seems to change for the high modes above
$\lambda^{\rm PBC}_{(3)}$, with $\lambda_c'$ barely moving, if at
all. As expected, the points $\lambda^{\rm PBC}_{(i)}$ correspond to
features in the PR at high $\beta$, as does again the point
$\lambda_*$: specifically, dips are present at
$\lambda^{\rm PBC}_{(1)}$, $\lambda^{\rm PBC}_{(2)}$, and $\lambda_*$,
and the mode size decreases dramatically above
$\lambda^{\rm PBC}_{(3)}$. In contrast with the physical sector, there
is no overshooting of the fractal dimension above 2 for the bulk
modes, and only near $\lambda^{\rm PBC}_{(1)}$ and
$\lambda^{\rm PBC}_{(2)}$ one can observe a small effect for the
largest $\beta$ values.

\subsubsection{Confined phase: both sectors combined}
\label{sec:num_pr_bs}

As already pointed out above, in the confined phase both sectors
contribute equally to physical observables. We show the resulting
mode size and fractal dimension in Fig.~\ref{fig:prVfdim_bs}. Given
the similar behavior in the two sectors when $\beta<\beta_c$, these
plots show little difference from those obtained for the two sectors
separately. These plots confirm the presence of localized modes only
at the high end of the spectrum, with a mobility edge that shows
little to no dependence on $\beta$; and that low modes have a
nontrivial fractal dimension.

\begin{figure}[t]
  \centering
  \includegraphics[width=0.6\textwidth]{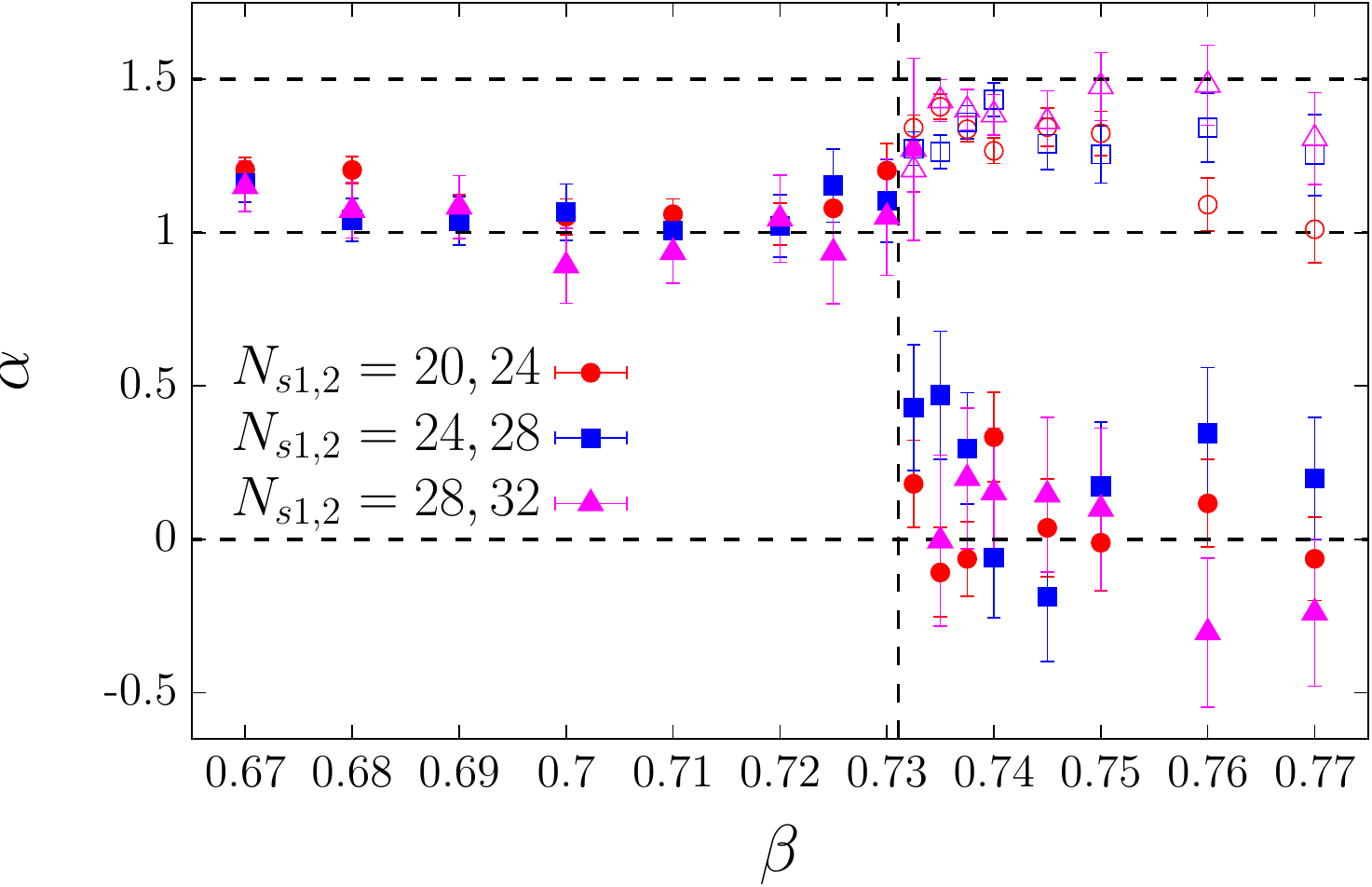}
  
  \caption{Fractal dimension in the lowest spectral bin, with physical
    and unphysical sectors combined in the confined phase, and shown
    separately in the deconfined phase, estimated with various pairs
    of system sizes. The critical coupling $\beta_c$ is marked by the
    dashed vertical line.}
  \label{fig:fdim_ens_lb}
\end{figure}
\begin{figure}[t]
  \centering
  \includegraphics[width=0.6\textwidth]{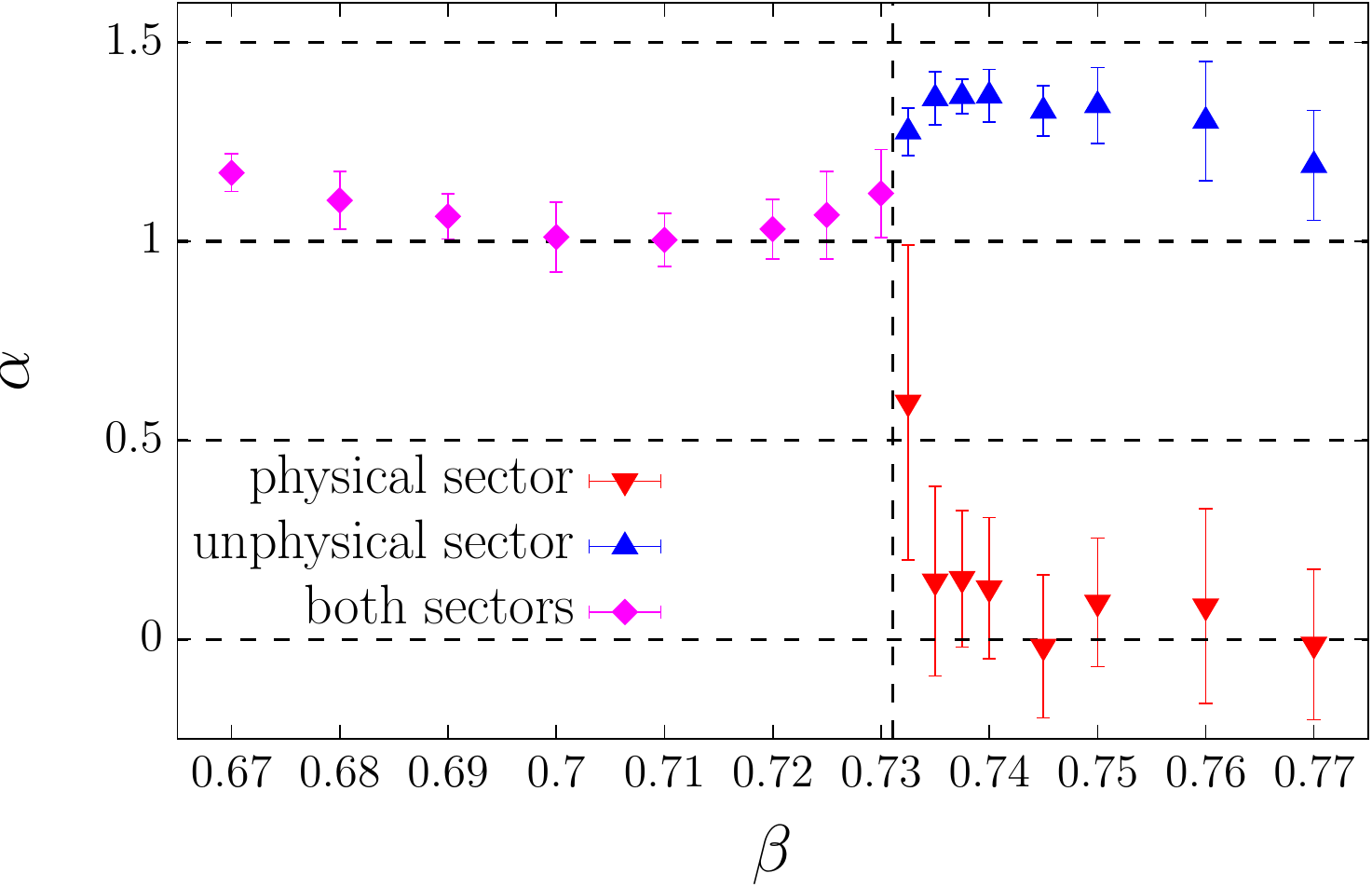}
  
  \caption{Fractal dimension in the lowest spectral bin, estimated by
    averaging over all pairs of system sizes with $20\le N_s \le
    32$. The two center sectors are combined for $\beta<\beta_c$, and
    shown separately for $\beta>\beta_c$, with $\beta_c$ marked by the
    dashed vertical line.}
  \label{fig:fdim_ens_lb_av}
\end{figure}

\subsubsection{Near-zero modes}
\label{sec:num_pr_nzm}

In order to summarize the dependence on temperature of the
localization properties of the low modes, in
Figs.~\ref{fig:fdim_ens_lb} and \ref{fig:fdim_ens_lb_av} we show the
fractal dimension of modes in the lowest spectral bin, as a function
of $\beta$. More precisely, in Fig.~\ref{fig:fdim_ens_lb} we show
estimates of the fractal dimension obtained from several pairs of
volumes, treating the two center sectors together in the confined
phase, and separately in the deconfined phase. For $\beta<\beta_c$,
$\alpha$ tends to slightly decrease as $\beta$ increases, from
slightly above 1 toward 1 up to around $\beta_c$. Above $\beta_c$ it
drops toward 0, in the physical sector, and increases to
approximately $1.2\div 1.4$, in the unphysical sector. Finite-size
effects are small at low $\beta$, expectedly large around $\beta_c$,
and again reasonably small above $\beta_c$, although larger than in
the low-temperature phase. In Fig.~\ref{fig:fdim_ens_lb_av} we show an
estimate of the fractal dimension $\alpha$ of the lowest modes
obtained after averaging over all pairs of available system sizes.
Error bars correspond to the sum in quadrature of the average
statistical error and of the finite-size systematic error, estimated
as the variance of $\alpha$ over the pairs of $N_s$ values. Our
results show clearly that the two center sectors respond very
differently to the phase transition, with low modes becoming more
delocalized in the unphysical sector, and localized in the physical
sector.

Although the use of only moderately large volumes, for which large
finite-size effects are still present near $\beta_c$, does not allow
us to make a conclusive statement, it seems likely that in the
thermodynamic limit an abrupt transition from $\alpha\approx 1$ to
$\alpha=0$ will take place at $\beta_c$ in the physical
sector. Conclusive evidence requires the extension of the present
study to larger system sizes.

\begin{figure}[t!]
  \centering
  \includegraphics[width=0.67\textwidth]{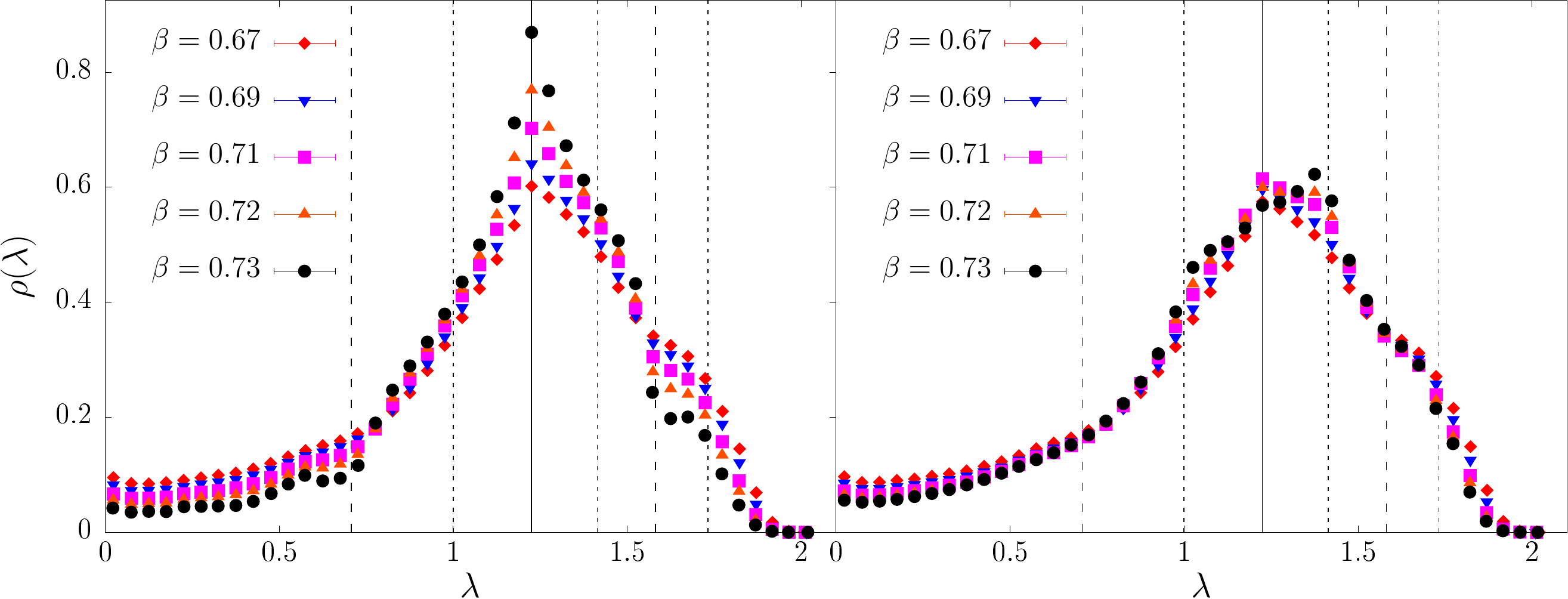}
    
  \vspace{0.15cm}
  \includegraphics[width=0.43\textwidth]{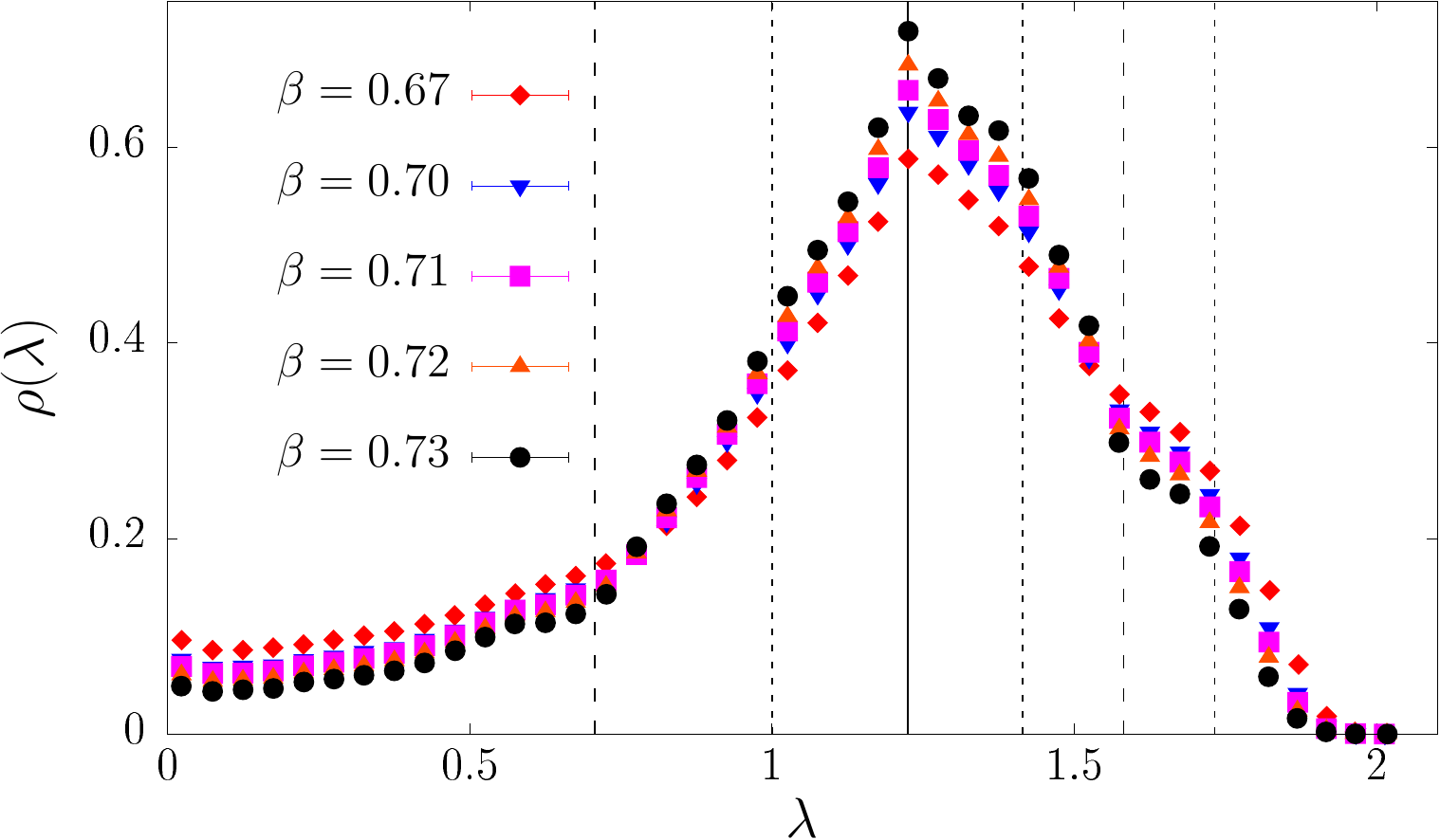}
  
  \vspace{0.15cm}
  \includegraphics[width=0.67\textwidth]{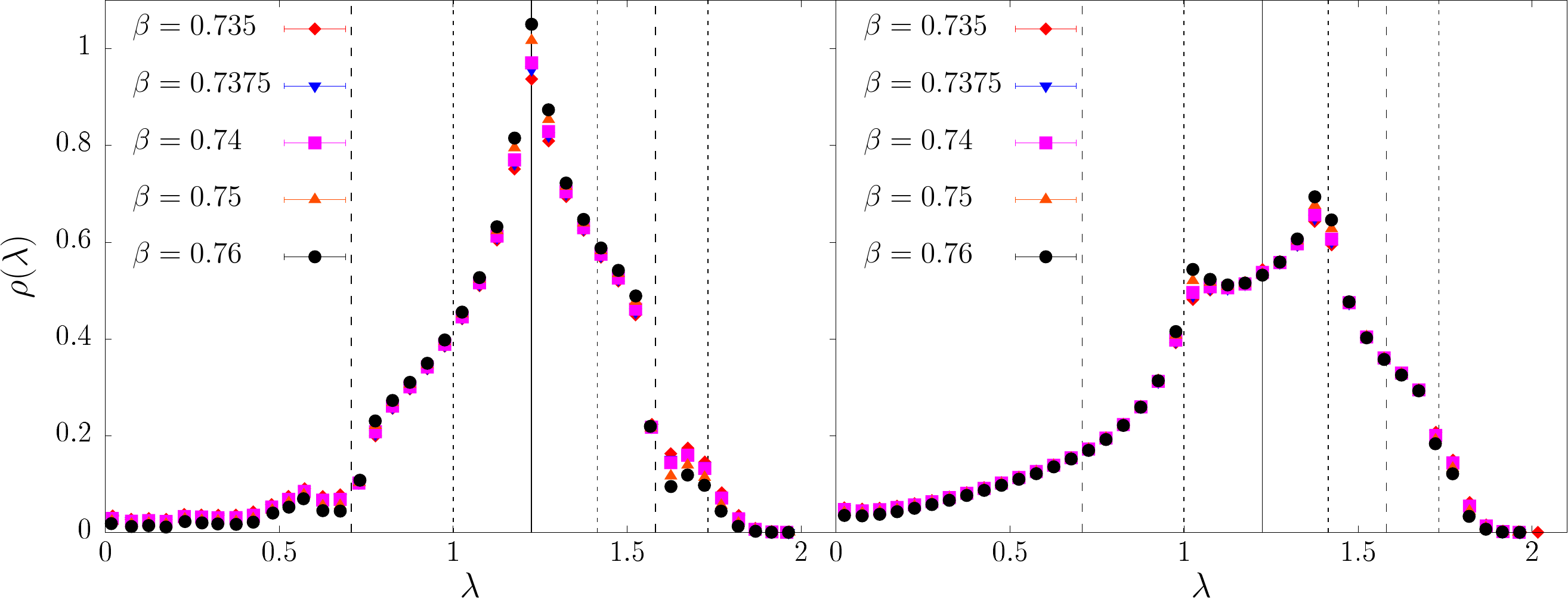}
  
  \caption{Spectral density -- confined phase, physical (top left) and
    unphysical (top right) sector, and both sectors together (center);
    deconfined phase, physical (bottom left) and unphysical (bottom
    right) sector.  The points $\lambda_{(0)}$,
    $\lambda^{\rm PBC}_{(1)}$, $\lambda_*$, $\lambda_{(2)}^{\rm PBC}$,
    $\lambda_{(1)}$ and $\lambda_{(3)}^{\rm PBC}$ (see
    Eqs.~\eqref{eq:ns_matsu} and \eqref{eq:ws_matsu}) are marked with
    vertical lines.  Here $N_s=32$.}
  \label{fig:spec}
\end{figure}

\subsection{Spectral density}
\label{sec:num_spd}

\begin{figure}[t!]
  \centering
  \includegraphics[width=0.55\textwidth]{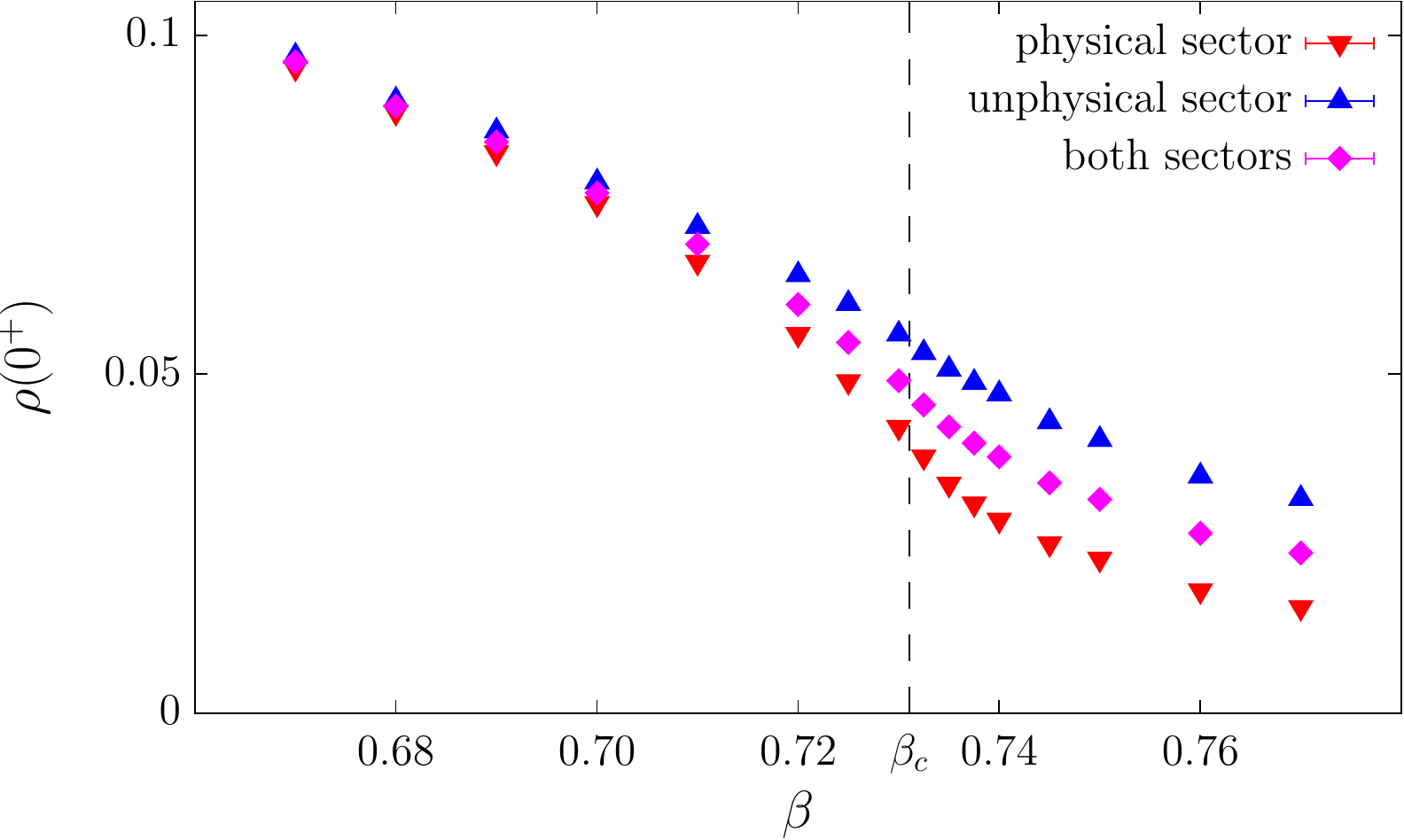}
  
  \caption{Spectral density of near-zero modes. The case of the two
    center sectors combined in the deconfined phase is shown only for
    comparison. Here $N_s=32$.}
  \label{fig:sd_lb}
\end{figure}

The spectral density per unit lattice volume, $\rho$,
\begin{equation}
  \label{eq:specdens_def}
    \rho(\lambda) \equiv (N_t V)^{-1} \left\la{\textstyle\sum_l}
      \delta(\lambda-\lambda_l)\right\ra\,, 
\end{equation}
is shown in Fig.~\ref{fig:spec} for several values of $\beta$, for the
physical and unphysical sectors separately, and for both sectors
together in the confined phase. The dependence on the system size is
negligible, so only the largest size $N_s=32$ is shown. The points
$\lambda_{(i)}$, $i=0,1,2$ and $\lambda_{(i)}^{\rm PBC}$, $i=1,2,3$,
as well as $\lambda_*$, are marked. The spectral density is large in
the bulk and small at the low and high ends; remarkably, it does not
vanish at the low end $\lambda\simeq 0$ in either phase. The spectral
density $\rho(0^+)$ in the lowest bin $[0,\Delta\lambda]$ is shown in
Fig.~\ref{fig:sd_lb}, again for $N_s=32$, for the two center sectors
separately and for both sectors combined (also in the deconfined phase
for comparison). While $\rho(0^+)$ keeps decreasing with $\beta$, it
remains nonzero in both sectors up to the largest $\beta$ studied
here. Loosely speaking, this indicates chiral symmetry breaking by a
(valence) quark-antiquark condensate both in the confined and in the
deconfined phase of the theory.  Furthermore, comparing the physically
meaningful results, i.e., both sectors added up in the confined phase
and treated separately in the deconfined phase, one sees that
$\rho(0^+)$ jumps at $\beta_c$, downwards (resp.\ upwards) if the
physical (resp.\ unphysical) sector is chosen in the deconfined
phase. Notice that since the spectral density near zero is smaller
(resp.\ larger) in the physical (resp.\ unphysical) sector, this
sector will be favored in the presence of dynamical fermions (resp.\
pseudofermions), as anticipated (see Ref.~\ref{foot:polloop}).  This
means, again in a loose sense, that chiral symmetry remains broken in
the high-$\beta$ phase but a first-order phase transition is displayed
in the chiral properties.  Moreover, in the physical sector in the
deconfined phase the low modes are localized.

A similar situation is found in SU(3) pure-gauge theory in 3+1
dimensions, where however the near-zero localized modes stand out as a
prominent peak~\cite{Kovacs:2017uiz,Vig:2020pgq}, rather than being
part of a plateau as in the present case.\footnote{Localization is
  understood here in the sense of the usual fractal dimension $\alpha$
  (see Eq.~\eqref{eq:loc4}) being equal to zero. For a 3+1 dimensional
  theory this is obtained as follows from the scaling of the PR,
  ${\rm PR}(\lambda,N_s) \simeq c(\lambda)N_s^{\alpha(\lambda)-3}$.
  However, using a different definition $d_{\rm IR}$ of fractal
  dimension, the authors of Ref.~\cite{Alexandru:2021pap} find for 3+1
  dimensional SU(3) pure gauge theory that $d_{\rm IR}=2$ for
  near-zero modes, differing both from $d_{\rm IR}=3$ found for exact
  zero modes and from $d_{\rm IR}=1$ found for low modes above the
  peak. The relation between the two definitions of fractal dimensions
  is not clear yet.}  The inclusion of dynamical fermions is expected
to lower the density of near-zero modes, but it is possible that even
in this case the $\mathbb{Z}_2$ theory displays a first-order
transition to a phase with a non-zero density of localized near-zero
modes.  If this scenario survived also in the chiral limit (taken
after the continuum limit, also assumed to exist), it would lead to
the disappearance of the massless Goldstone excitations associated
with the spontaneous breaking of the ${\rm SU}(N_f = 4)_A$ symmetry of
the theory at finite temperature~\cite{giordano_GT_lett}.  The
connection between localization and disappearance of Goldstone modes
was originally made in Ref.~\cite{McKane:1980fs}, in the context of
Anderson models, and later rediscovered in
Ref.~\cite{Golterman:2003qe}, in the context of lattice field theories
at zero temperature.

A curious feature is the presence of a peak at $\lambda_*$ at all
$\beta$ in the physical sector, and at low $\beta<\beta_c$ in the
unphysical sector. At high $\beta>\beta_c$ in the unphysical sector
this peak disappears, and is replaced by two peaks near
$\lambda^{\rm (PBC)}_{(1)}$ and $\lambda^{\rm (PBC)}_{(2)}$. We do not
know the reason behind these peaks and the analogous presence of
features at $\lambda_*$ (as well as at $\lambda_{(i)}$ or
$\lambda^{\rm (PBC)}_{(i)}$) in most observables considered in this
work.

\subsection{Correlation of localized modes with gauge observables}
\label{sec:num_corr}

\subsubsection{Polyakov loop}
\label{sec:num_corr_pol}

\begin{figure}[t]
  \centering
  \includegraphics[width=0.48\textwidth]{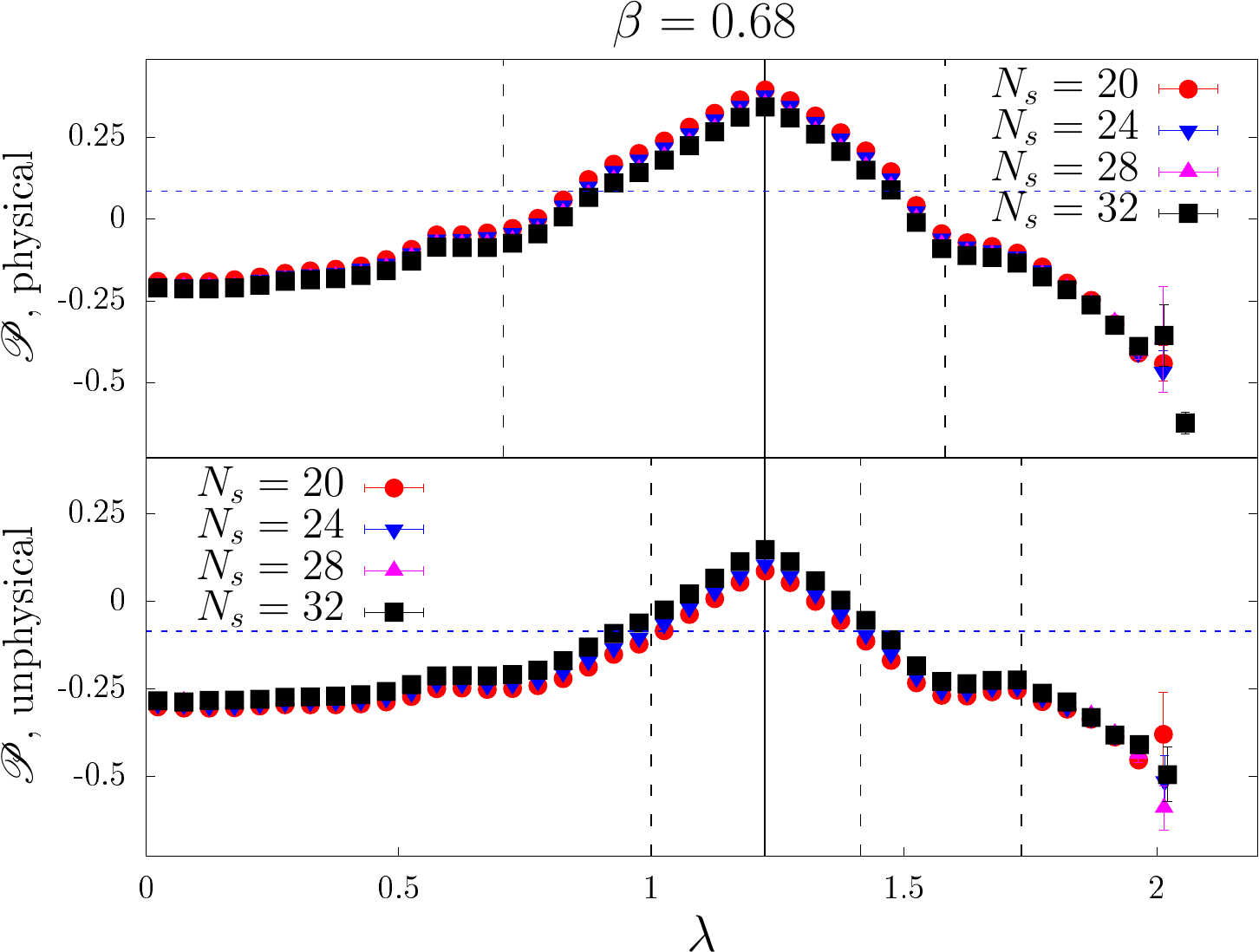}
  \hfil
  \includegraphics[width=0.48\textwidth]{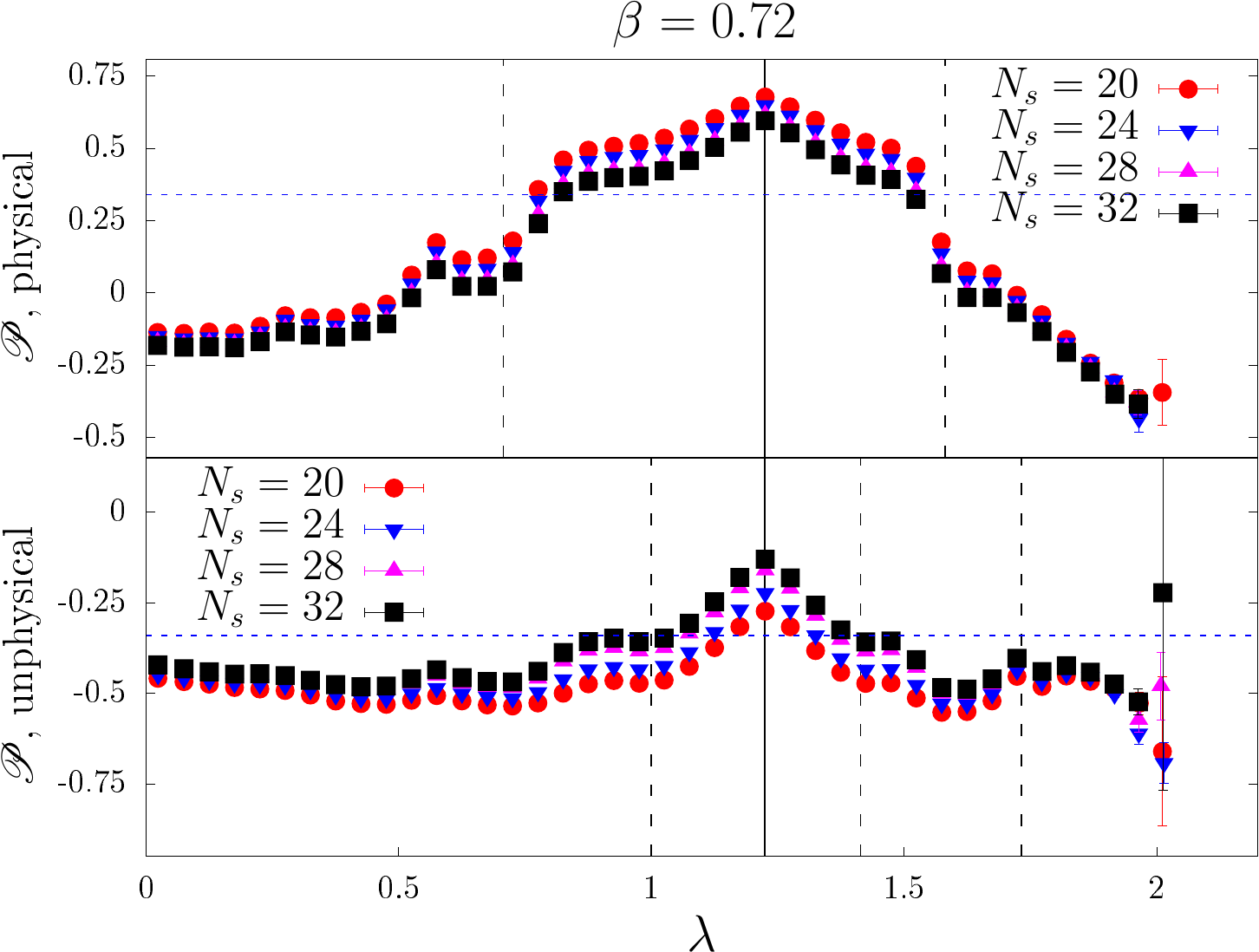}

  \caption{Confined phase: Polyakov loop weighted by the modes.  Here
    and in the following plots, the physical (resp.\ unphysical)
    sector is shown in the top (resp.\ bottom) panel.  Horizontal
    lines correspond to $\pm\la |\bar{P}|\ra$ on the largest available
    volume ($N_s=32$).}
  
  \label{fig:polbymode}
\end{figure}
\begin{figure}[th!]
  \centering
  \includegraphics[width=0.48\textwidth]{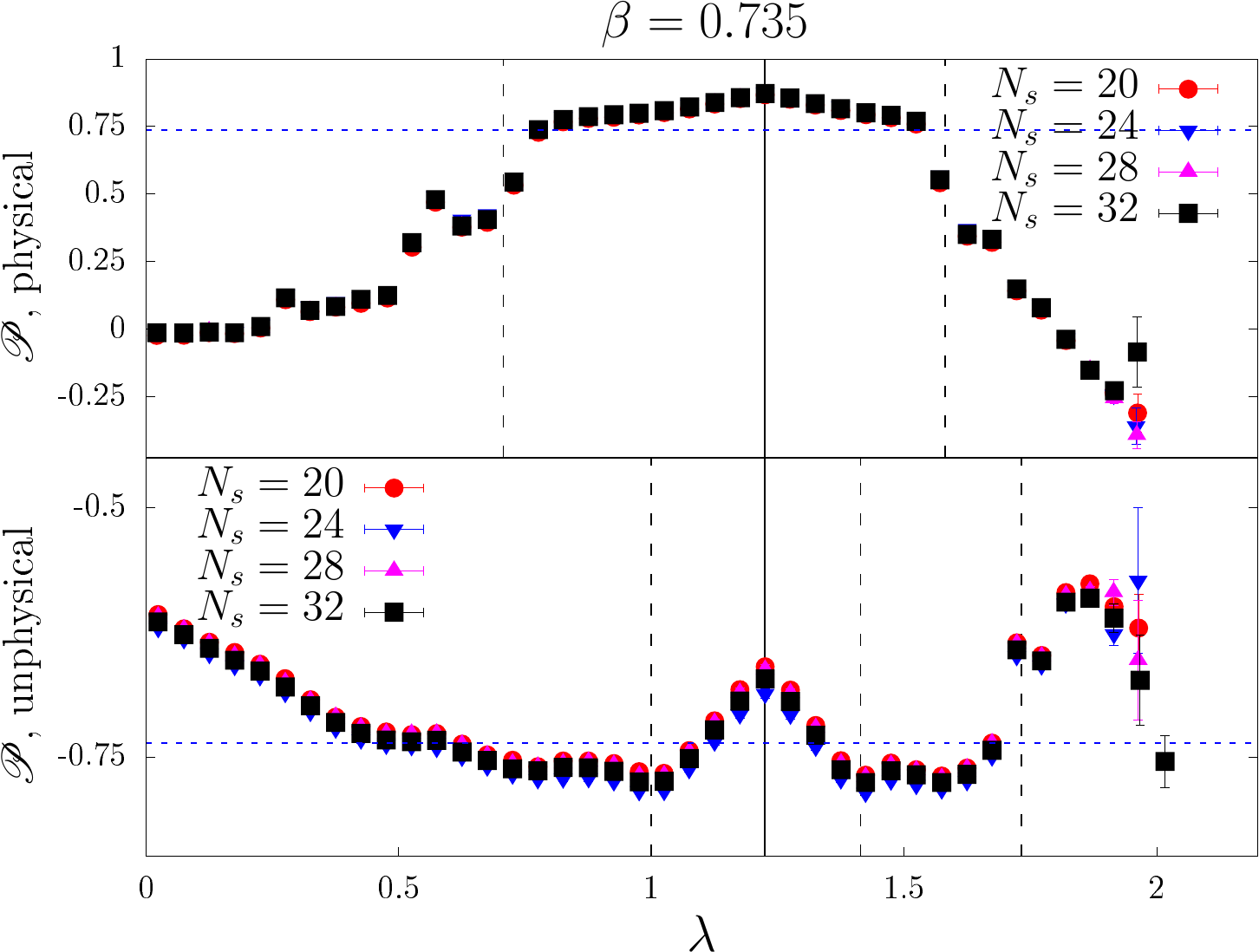}
  \hfil
  \includegraphics[width=0.48\textwidth]{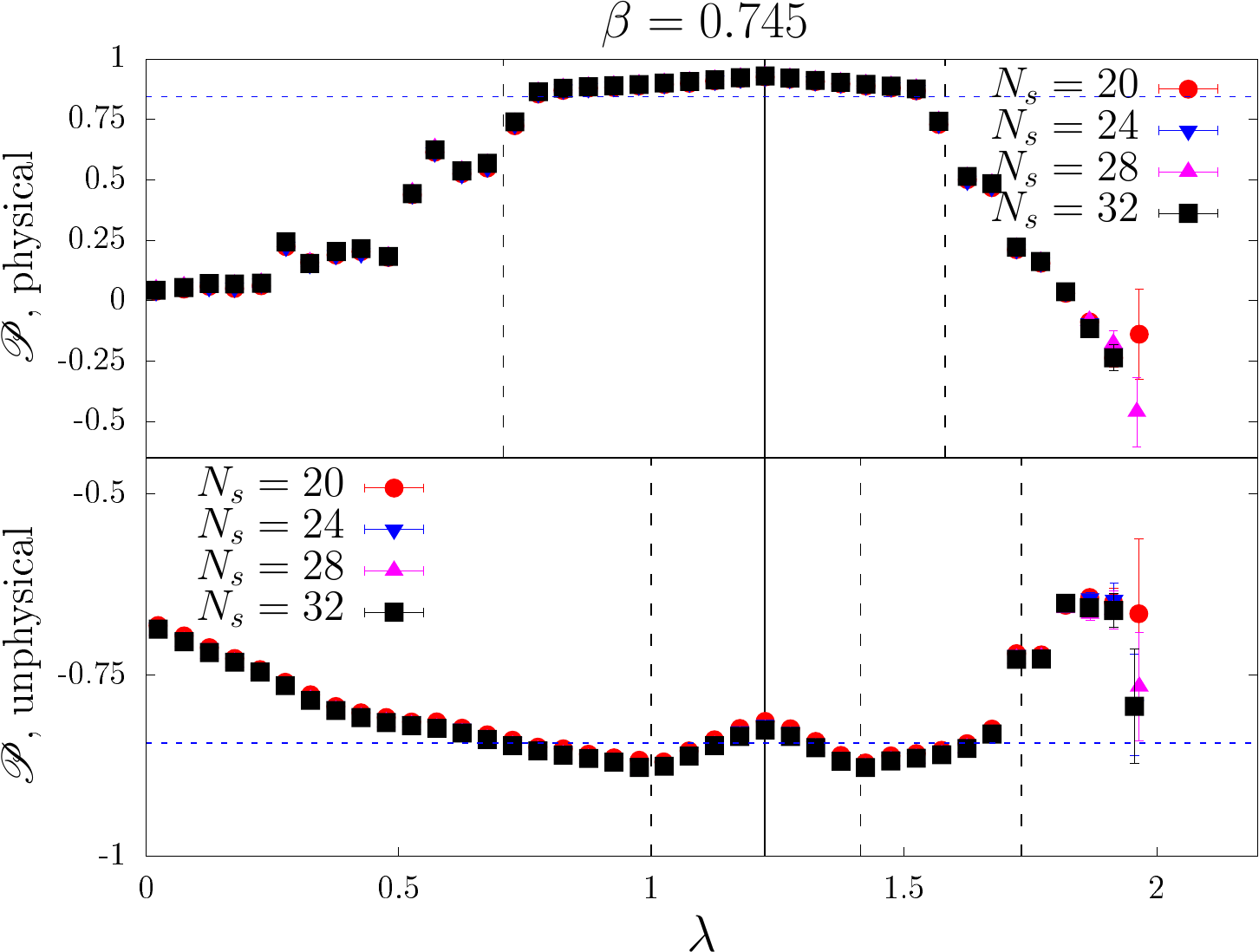}
  
  \caption{Deconfined phase: Polyakov loop weighted by the modes.}
  
  \label{fig:polbymode2}
\end{figure}

According to the sea/islands picture of
localization~\cite{Bruckmann:2011cc,Giordano:2015vla,Giordano:2016cjs},
localized modes in the physical sector of the deconfined phase are
expected to live near fluctuations of the Polyakov loops away from the
ordered value. For $\mathbb{Z}_2$ the Polyakov loop can take only two
values, and we expect the localized modes to prefer sites where
$P(\vec{x})=-1$. In Figs.~\ref{fig:polbymode} and \ref{fig:polbymode2}
we show the Polyakov loop weighted by the eigenmodes $\Po(\lambda)$,
Eq.~\eqref{eq:loc6}, for the physical (top panels) and unphysical
(bottom panels) sectors.  This is compared to the average of $\bar{P}$
in the respective sectors. By construction, these sector averages are
simply $\pm \protect\la |\bar{P}| \protect\ra$, where the average is
over the ensemble of configurations generated in the Monte-Carlo
simulation.

Let us discuss first the physical sector. In the confined phase, while
delocalized, the low modes show a moderate but clear preference for
negative Polyakov loops; the bulk modes show a moderate but clear
preference for positive Polyakov loops, with $\Po$ above the average
of $\bar{P}$ in the physical sector; and the high modes, which are
localized, show a clear preference for negative Polyakov loops.
Throughout the spectrum, $\Po(\lambda)$ decreases slightly with the
volume, probably reflecting the fact that $\la |\bar{P}|\ra\to 0$ in
the thermodynamic limit; it seems likely though that a nontrivial
$\Po(\lambda)$ will be found also in this limit. In the deconfined
phase the low modes, which are now localized, have roughly half of
their weight on sites with positive Polyakov loop and half on the now
much rarer sites with negative Polyakov loop. This is in agreement
with expectations.  More precisely, the ratio between the average
fraction of sites with negative Polyakov loop among those ``occupied''
by a mode and the average fraction of sites with negative Polyakov
loop on the whole lattice, i.e.,
$(1-\Po(\lambda))/(1-\la |\bar{P}|\ra)$, increases steadily with
$\beta$ in the deconfined phase.  Bulk modes tend instead to avoid
negative Polyakov loops, as shown by the fact that
$1-\Po(\lambda) < 1-\la |\bar{P}|\ra$.  High modes are again clearly
preferring negative Polyakov loops, with more than half of the mode's
weight concentrated on the corresponding sites. In the deconfined
phase the dependence of $\Po$ on the volume seems very mild.

We now discuss the unphysical sector. In the confined phase, negative
Polyakov loops seem again to be preferred everywhere in the spectrum
except in the very middle, where positive Polyakov loops are slightly
preferred. In fact, also in this case $\Po(\lambda)$ for the bulk
modes is above the average of $\bar{P}$ in the given center sector,
although in this case it means that they favor the slightly less
frequent positive loops to the negative ones (in the physical sector
they prefer instead the more frequent positive loops). Low and high
modes instead have $\Po(\lambda)< -\la |\bar{P}|\ra$. The tendency is
this time for $\Po(\lambda)$ to increase as the volume increases. In
the deconfined phase $\Po(\lambda)$ becomes flatter in the bulk, where
$\Po(\lambda)$ gets closer to $-\la |\bar{P}|\ra$, while low and high
modes show now clearly $0>\Po(\lambda)>-\la |\bar{P}|\ra$, thus
favoring positive Polyakov loops; the volume dependence is again very
mild. The correlation of low modes with positive Polyakov loops in the
deconfined phase cannot be simply explained in terms of
``energetically'' favorable islands ``attracting'' the mode. It is,
however, not in contrast with the sea/islands picture of
localization. We elaborate on this point in the conclusions.

\begin{figure}[t]
  \centering
  \includegraphics[width=0.48\textwidth]{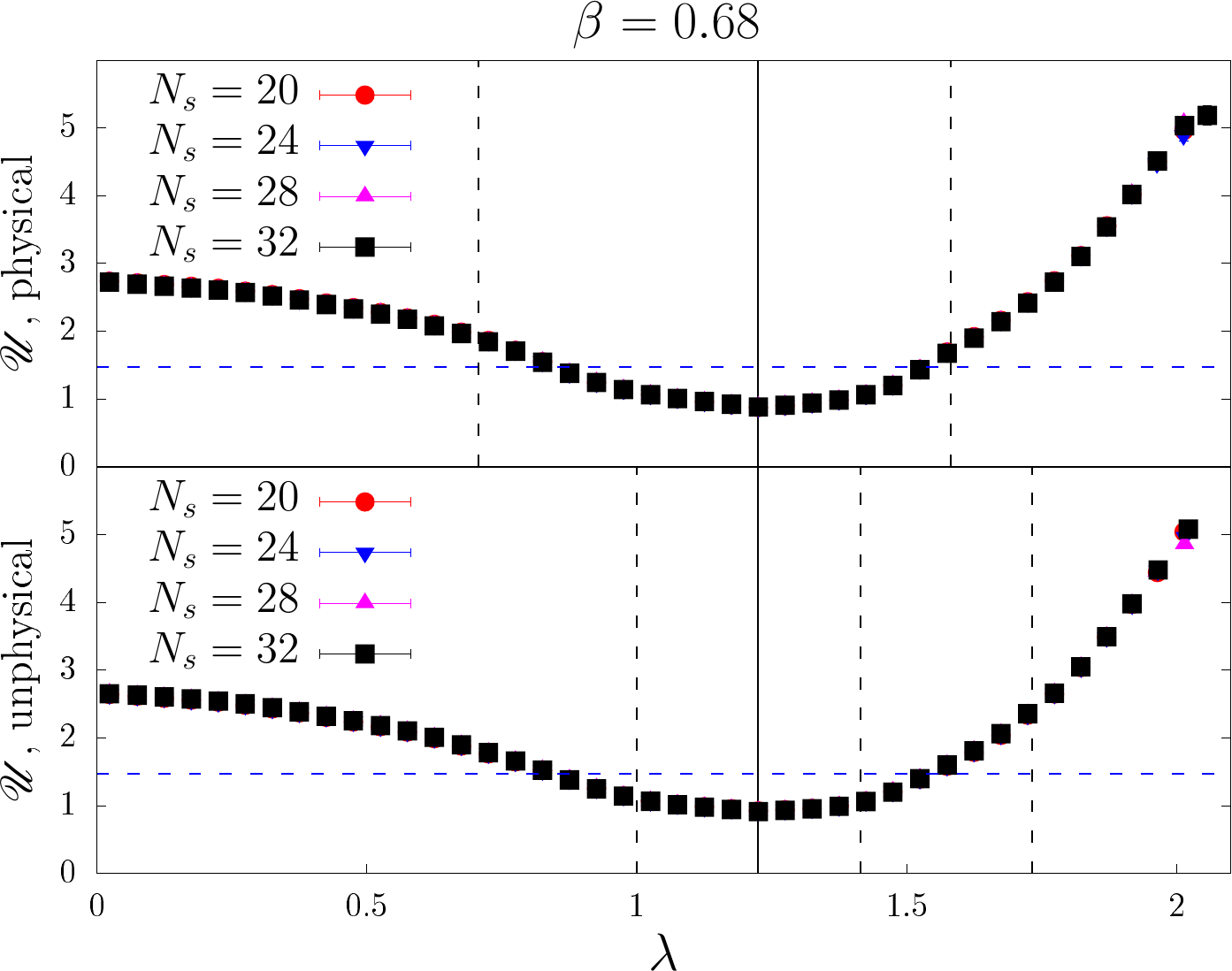}
  \hfil
  \includegraphics[width=0.48\textwidth]{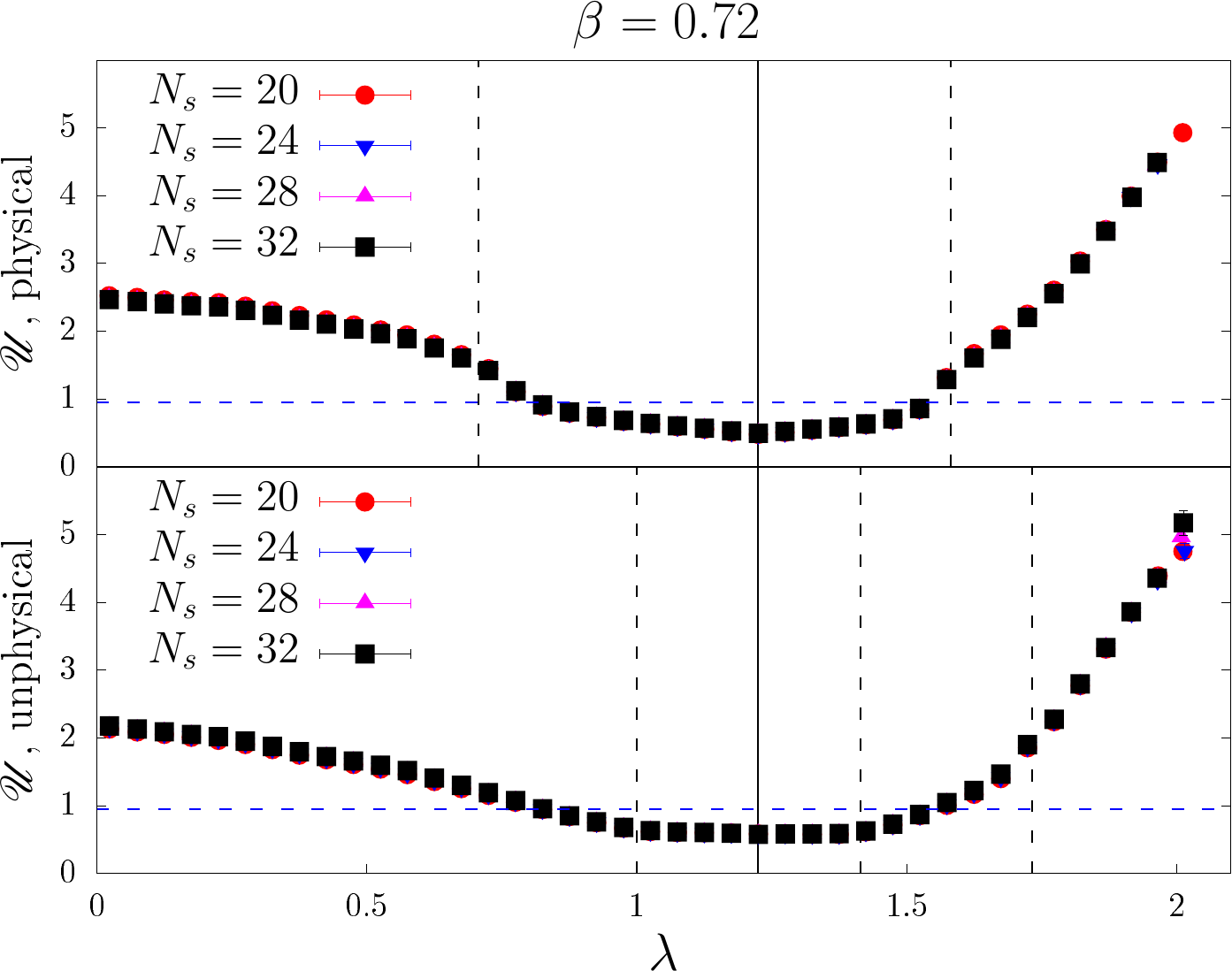}
  
  \caption{Confined phase: average number of negative plaquettes
    touched by a mode. Here and in the following plot, horizontal
    lines correspond to $6\la 1- U_{\mu\nu}\ra$ on the largest
    available volume ($N_s=32$).}

  \label{fig:weightnegplaq}
\end{figure}
\begin{figure}[t!]
  \centering
  \includegraphics[width=0.48\textwidth]{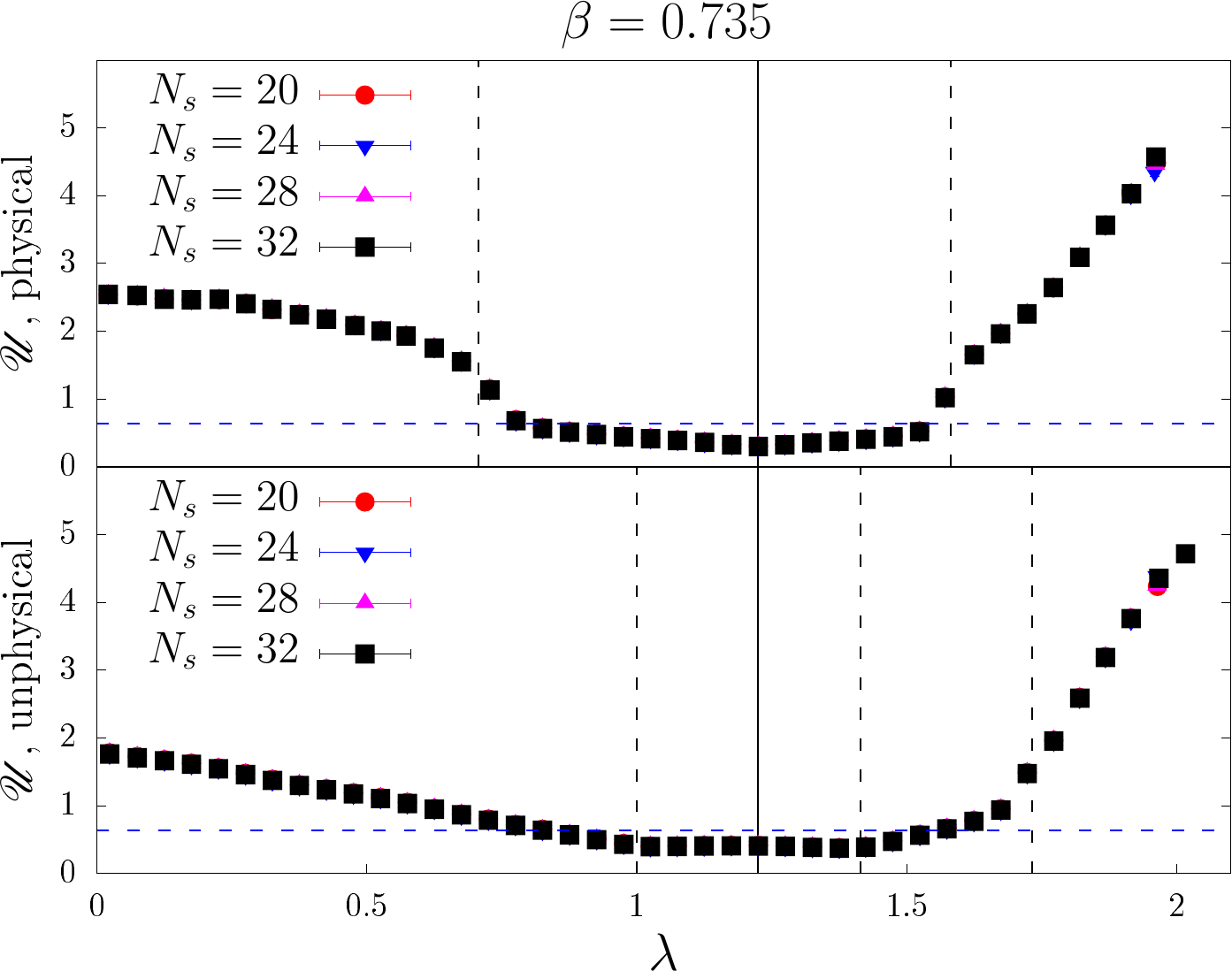}
  \hfil
  \includegraphics[width=0.48\textwidth]{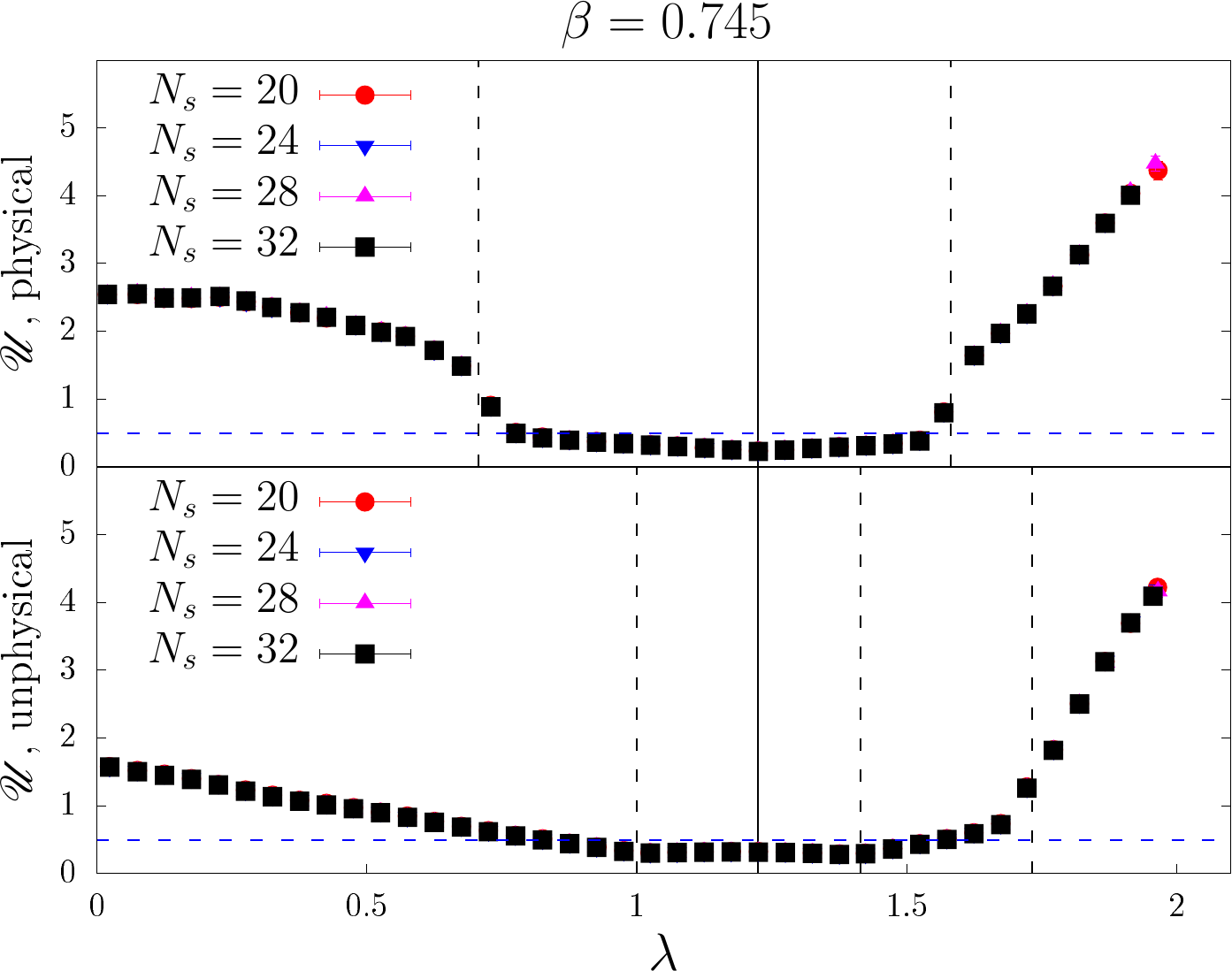}
  
  \caption{Deconfined phase: average number of negative plaquettes
    touched by a mode.}

  \label{fig:weightnegplaq2}
\end{figure}

\begin{figure}[t]
  \centering
  \includegraphics[width=0.48\textwidth]{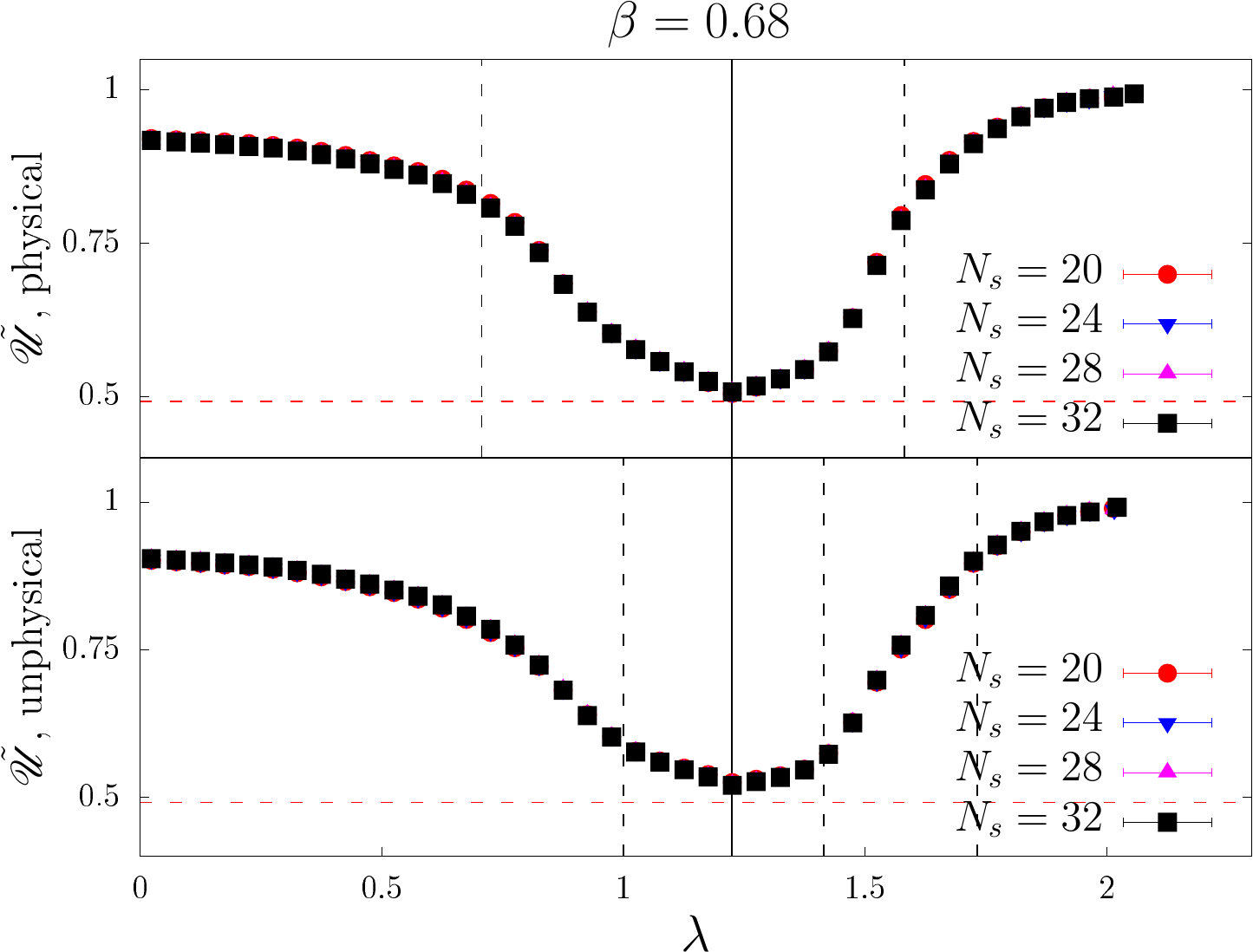}
  \hfil
  \includegraphics[width=0.48\textwidth]{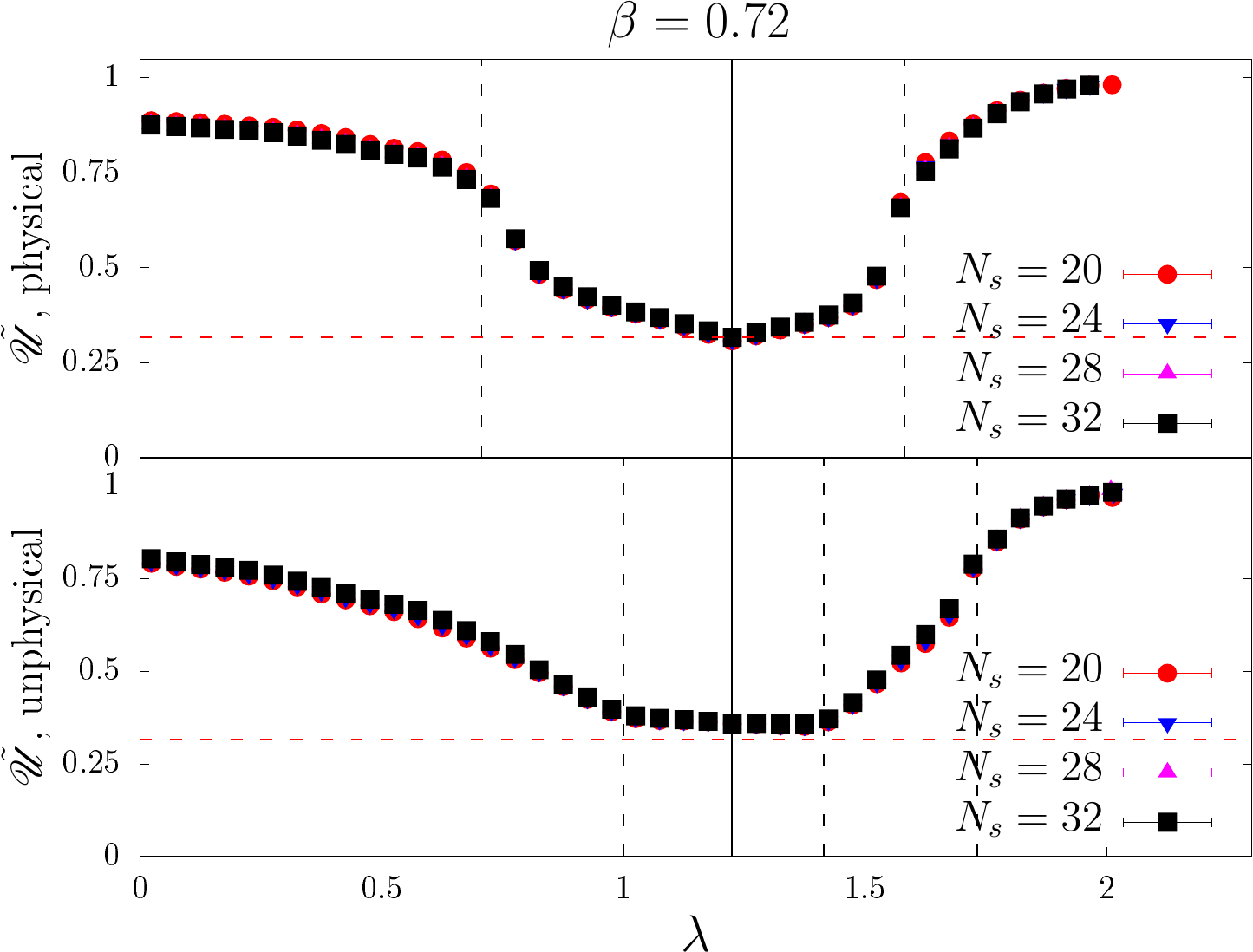}
  
  \caption{Confined phase: fraction of mode touched by negative
    plaquettes. Here and in the following plot, horizontal lines
    correspond to $2(1-\la U_{\mu\nu}\ra)$ on the largest available
    volume ($N_s=32$).}

  \label{fig:negplaq}
\end{figure}
\begin{figure}[t!]
  \centering
  \includegraphics[width=0.48\textwidth]{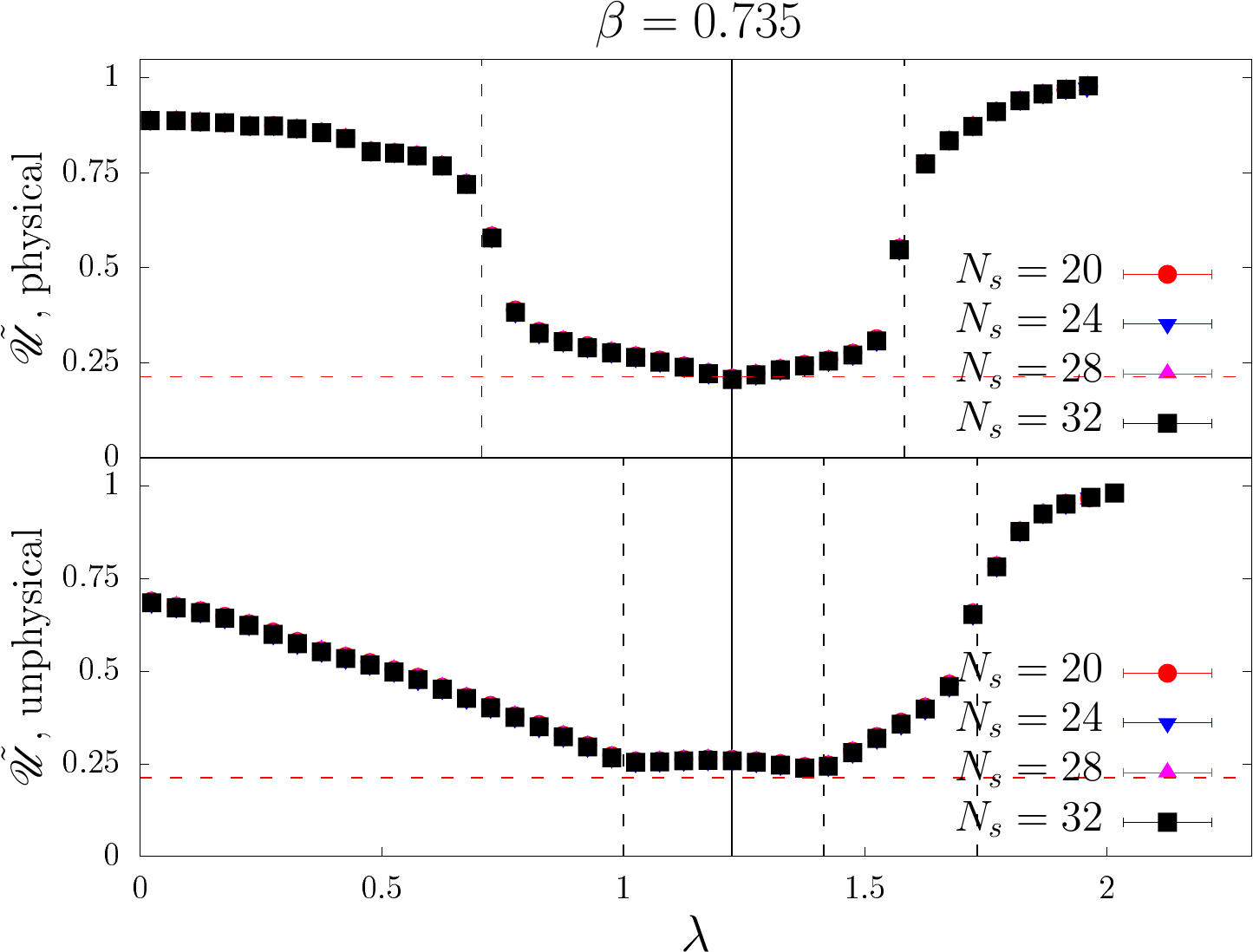}
  \hfil
  \includegraphics[width=0.48\textwidth]{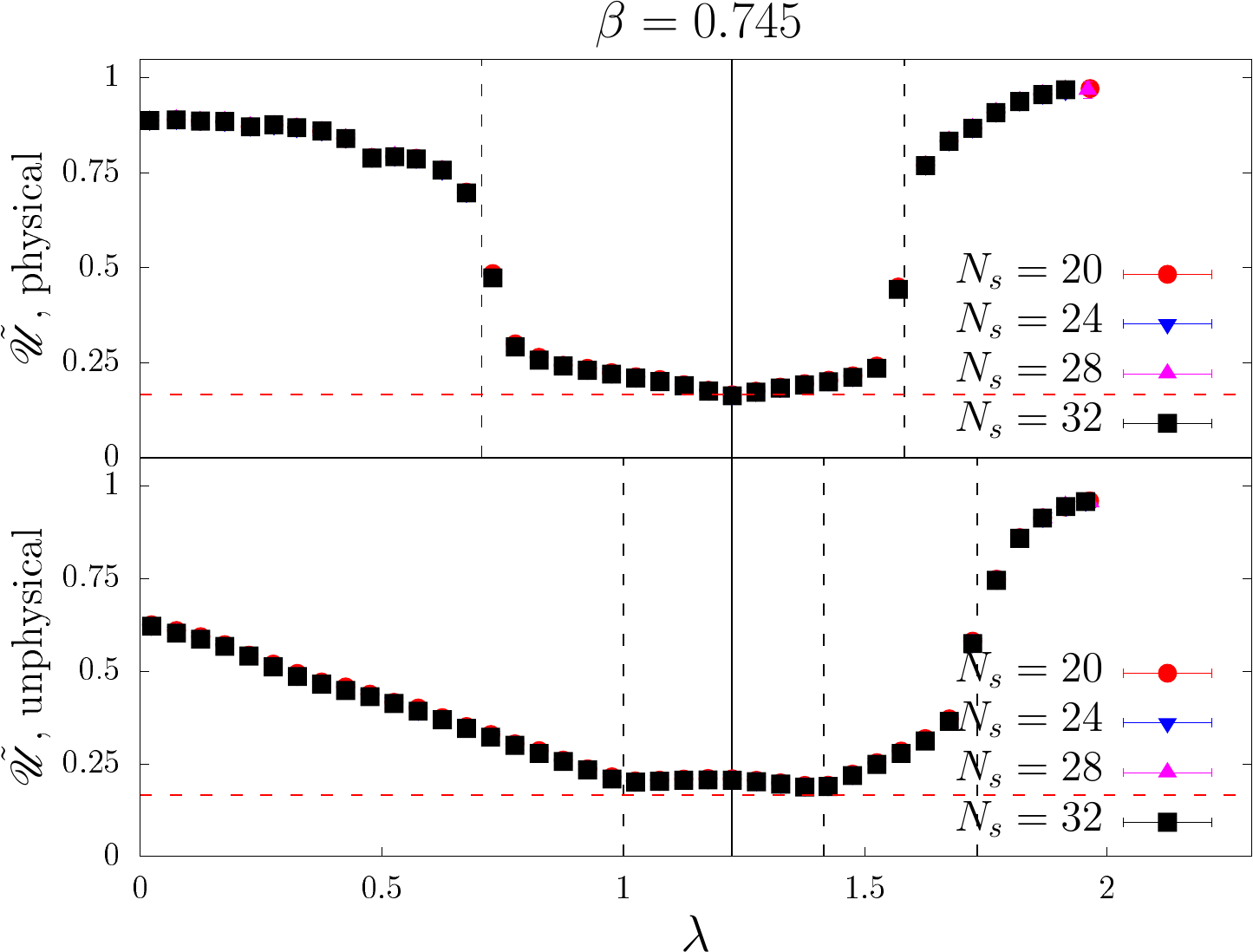}
  
  \caption{Deconfined phase: fraction of mode touched by negative
    plaquettes.}

  \label{fig:negplaq2}
\end{figure}

\subsubsection{Negative plaquettes}
\label{sec:num_corr_plaq}

We now turn to the correlation between localized modes and negative
plaquettes, for which we provide two different measures. In
Figs.~\ref{fig:weightnegplaq} and \ref{fig:weightnegplaq2} we show
$\U(\lambda)$, which measures the average number of negative
plaquettes touched by the modes (see Eq.~\eqref{eq:loc6_bis}) or,
equivalently, the ``negative plaquettes seen by the modes''. In
Figs.~\ref{fig:negplaq} and \ref{fig:negplaq2} we show instead
$\tilde\U(\lambda)$, which measures what fraction of the modes lives
on sites touched by at least one negative plaquette (see
Eq.~\eqref{eq:loc6_ter}).  The two observables provide slightly
different information: $\tilde\U(\lambda)$ tells us ``how much'' of
the modes is found near negative plaquettes, while $\U(\lambda)$ tells
us how much the modes like to live near {\it clusters} of negative
plaquettes.

We discuss $\U(\lambda)$ first. In the physical sector
(Figs.~\ref{fig:weightnegplaq} and \ref{fig:weightnegplaq2}, top
panels) the low and high modes prefer to be closer to negative
plaquettes ($\U(\lambda)>6\la 1- U_{\mu\nu}\ra$), while bulk modes
prefer to avoid them ($\U(\lambda)< 6\la 1- U_{\mu\nu}\ra$), in both
phases of the theory.  More precisely, the low and, especially, the
high modes have a larger weight on sites touching more than one
negative plaquette, and so prefer to live close to where negative
plaquettes tend to cluster. No clear dependence on the volume is
visible.  $\U(\lambda)$ shows little dependence on $\beta$ as well,
the only notable features being that for bulk modes the tendency to
avoid negative plaquettes is much less pronounced in the deconfined
phase, where $\U(\lambda)$ also becomes flatter. For the high modes
$\U(\lambda)$ changes very little with $\beta$; for the low modes it
slightly decreases as the deconfined phase is approached, remaining
essentially constant in the deconfined phase. Since the average number
of negative plaquettes decreases as $\beta$ is increased, this means
that the localized low and high modes are more and more localized near
clusters of negative plaquettes as one gets deeper in the deconfined
phase.

The situation is quite similar in the unphysical sector
(Figs.~\ref{fig:weightnegplaq} and \ref{fig:weightnegplaq2}, bottom
panels): low and high modes still prefer to be near clusters of
negative plaquettes, and bulk modes prefer to avoid them. Again, no
clear volume dependence is visible.  The dependence on $\beta$ is
slightly stronger than in the physical sector: here not only
$\U(\lambda)$ for bulk modes becomes flatter and closer to
$6\la 1- U_{\mu\nu}\ra$ as $\beta$ increases, but also the upward
deviation of $\U(\lambda)$ from $6\la 1- U_{\mu\nu}\ra$ for low modes
becomes less pronounced. As low and bulk modes are delocalized, this
is in line with the fact that negative plaquettes become less frequent
in the deconfined phase. For the localized high modes the change of
$\U(\lambda)$ with $\beta$ is small, indicating again that they prefer
to localize more near negative plaquettes as $\beta$ increases.

We now turn to $\tilde\U(\lambda)$.  In the physical sector
(Figs.~\ref{fig:negplaq} and \ref{fig:negplaq2}, top panels), and for
all $\beta$, 80\% or more of the weight of low and high modes is on
sites touching at least one negative plaquette, while for bulk modes
this fraction is 80\% or less, and decreasing rapidly as one
approaches $\lambda_*$ from either side.  For low and high modes there
seems to be little variation as $\beta$ increases, despite the fact
that negative plaquettes become rarer, while for bulk modes
$\tilde\U(\lambda)$ clearly decreases with $\beta$. This again
indicates that localized modes tend to localize near negative
plaquettes. In the unphysical sector (Figs.~\ref{fig:negplaq} and
\ref{fig:negplaq2}, bottom panels), low and high modes have again a
much larger value of $\tilde\U(\lambda)$ than bulk modes at all
$\beta$s.  Differently from the physical sector, $\tilde\U(\lambda)$
clearly decreases as $\beta$ increases both for low and bulk modes,
while for high modes it shows very little dependence on $\beta$. This
again indicates that the localized high modes prefer to localize near
negative plaquettes. As already mentioned in Section \ref{sec:locD},
it is not straightforward to estimate what one should expect for
$\tilde\U(\lambda)$ in the case of fully delocalized modes
($|\psi|^2\sim 1/(N_tV)$). It turns out that $2(1-\la U_{\mu\nu}\ra)$
provides an accurate lower bound on $\tilde\U(\lambda)$ in the bulk of
the spectrum.

The stronger correlation between localized modes and negative
plaquettes than between localized modes and Polyakov loop fluctuations
does not necessarily mean that the former are better candidates as the
relevant source of disorder than the latter. On the one hand, negative
plaquettes are strongly correlated with Polyakov loop fluctuations.
On the other hand, localized modes in the deconfined phase, physical
sector, are not strictly confined to Polyakov loop fluctuations away
from the ordered value, as that would be ``energetically''
expensive. One can then reconcile the observed stronger correlation
with negative plaquettes and the Polyakov loop fluctuations being the
actual source of disorder, if localization takes place on clusters of
negative plaquettes near the Polyakov loop fluctuations. This is
particularly interesting in the light of the
observation~\cite{Gliozzi:2002ht} that the largest negative-plaquette
cluster scales as the system size in the confined phase, while it
remains finite in the deconfined phase. This may help explain the
relation between deconfinement and localization (and between
confinement and delocalization) of the low modes in the physical
sector. This point deserves a more detailed study, which is however
outside of the scope of the present paper.

\subsection{Inertia tensor of the eigenmodes}
\label{sec:intens}

\begin{figure}[t]
  \centering
  \includegraphics[width=0.48\textwidth]{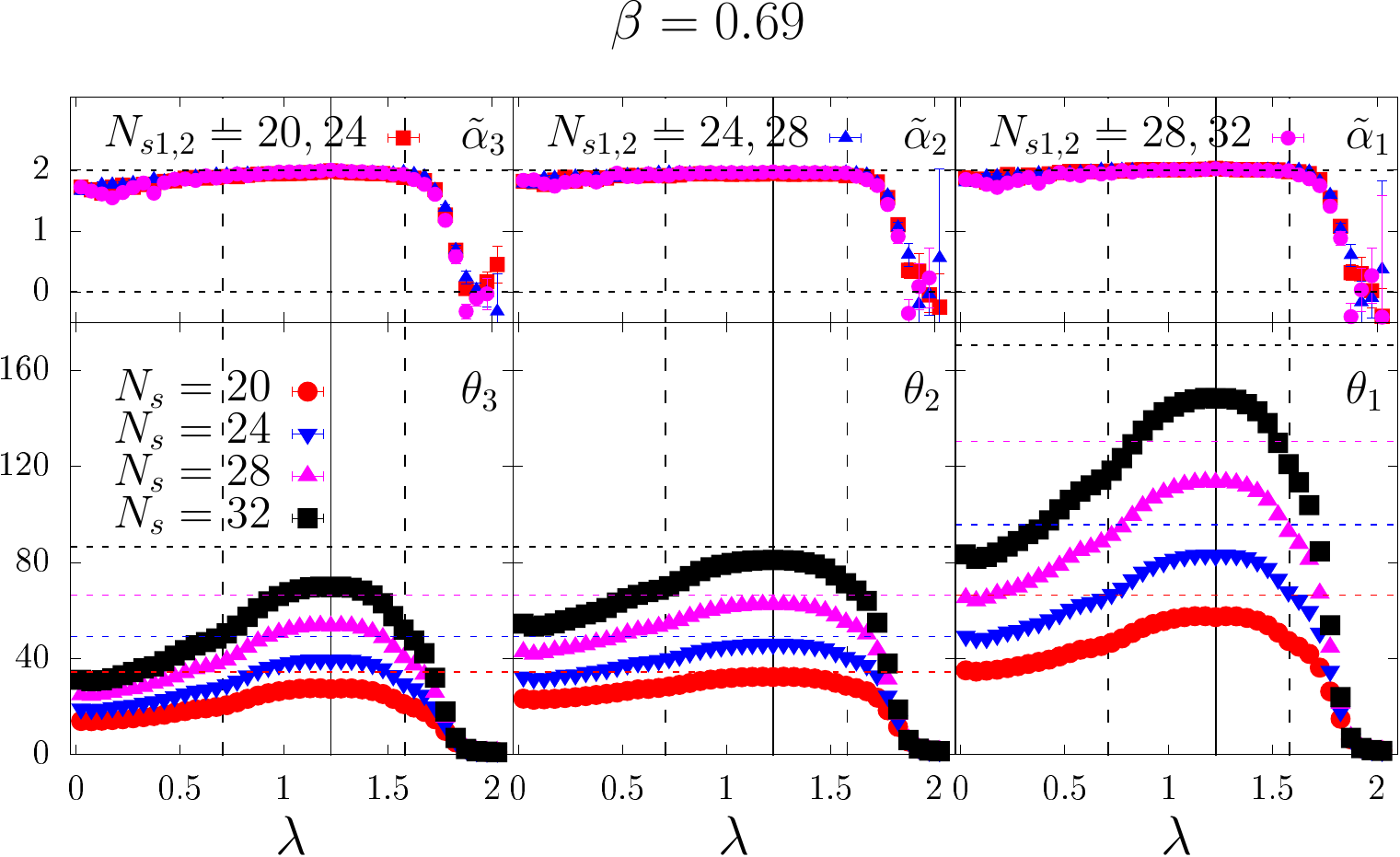}
  \hfil
  \includegraphics[width=0.48\textwidth]{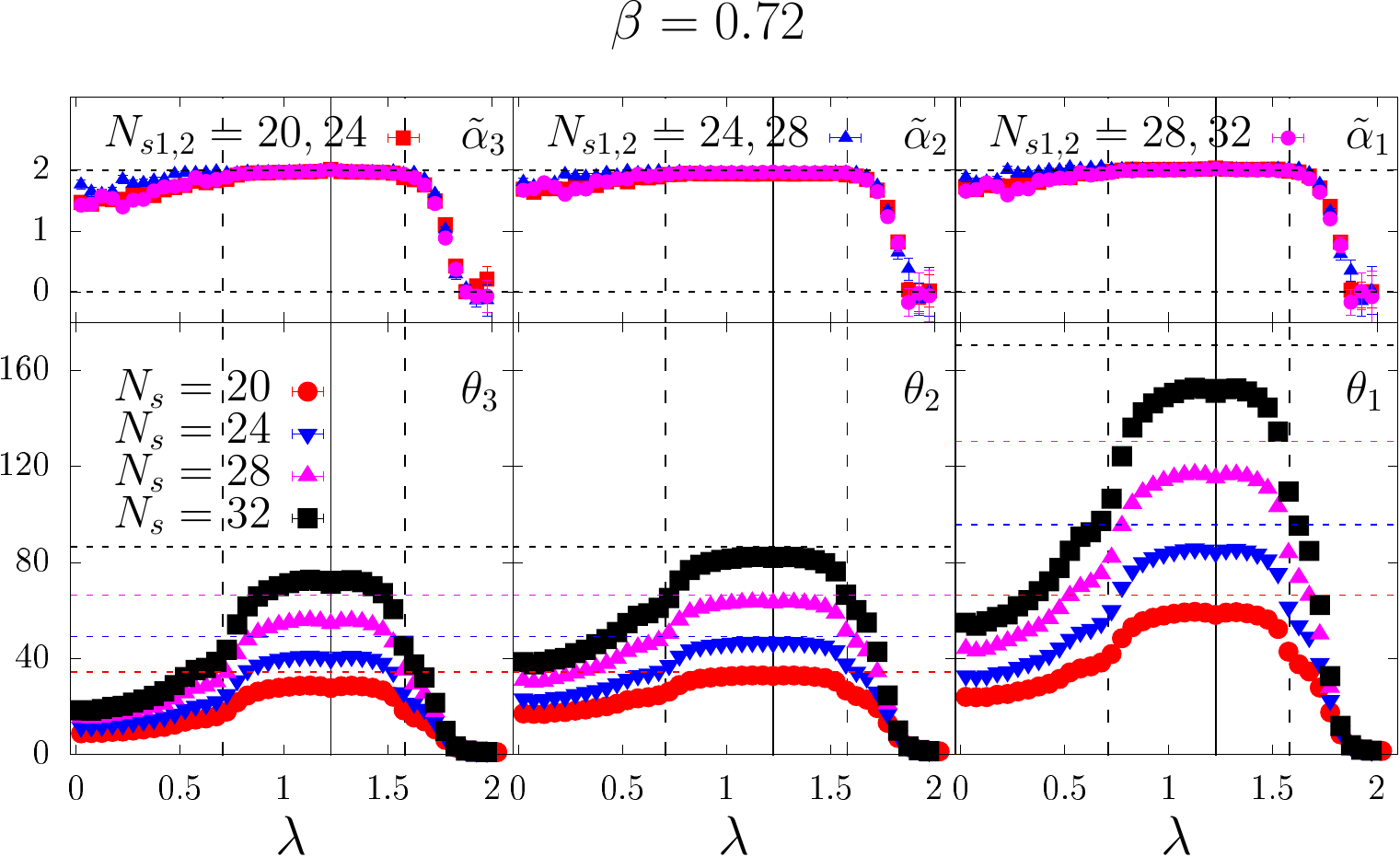}

  \caption{Confined phase, physical sector. Bottom panels: average
    principal moments of inertia $\theta_i$. Here and in the following
    plots, the expectation for a cuboid of size
    $N_t\times N_s\times N_s$ is also shown (color code matching that
    of the numerical data for the various volumes). Top panels:
    scaling dimension $\tilde{\alpha}_i$ of moment $\theta_i$ for
    various pairs of volumes.}

  \label{fig:inmom}
\end{figure}

\begin{figure}[t]
  \centering
    \includegraphics[width=0.48\textwidth]{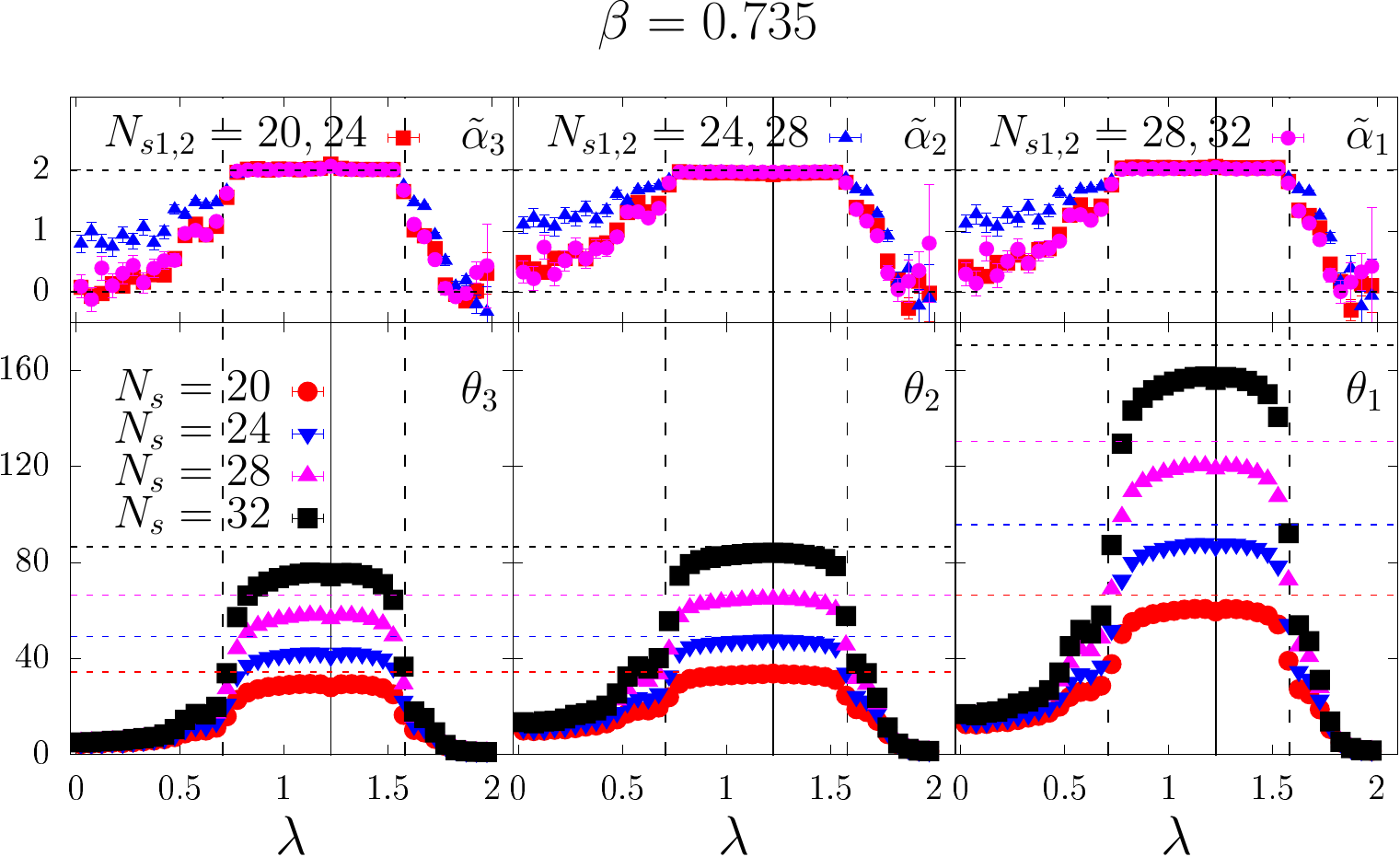}
    \hfil
  \includegraphics[width=0.48\textwidth]{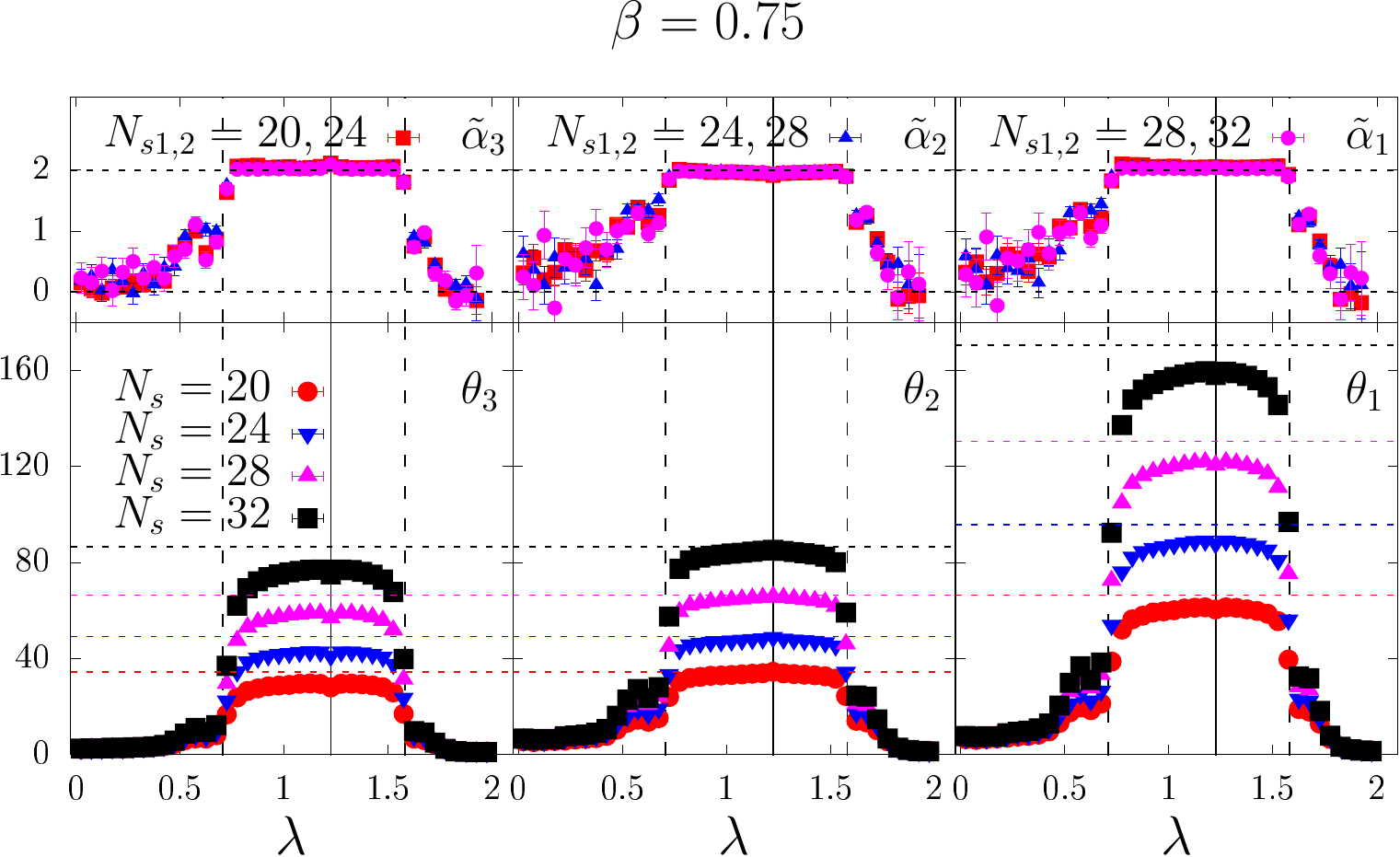}

  \caption{Deconfined phase, physical sector. Bottom panels: average
    principal moments of inertia $\theta_i$.  Top panels: scaling
    dimension $\tilde{\alpha}_i$ of moment $\theta_i$ for various
    pairs of volumes.}

  \label{fig:inmom2}
\end{figure}

\begin{figure}[t]
  \centering
  \includegraphics[width=0.48\textwidth]{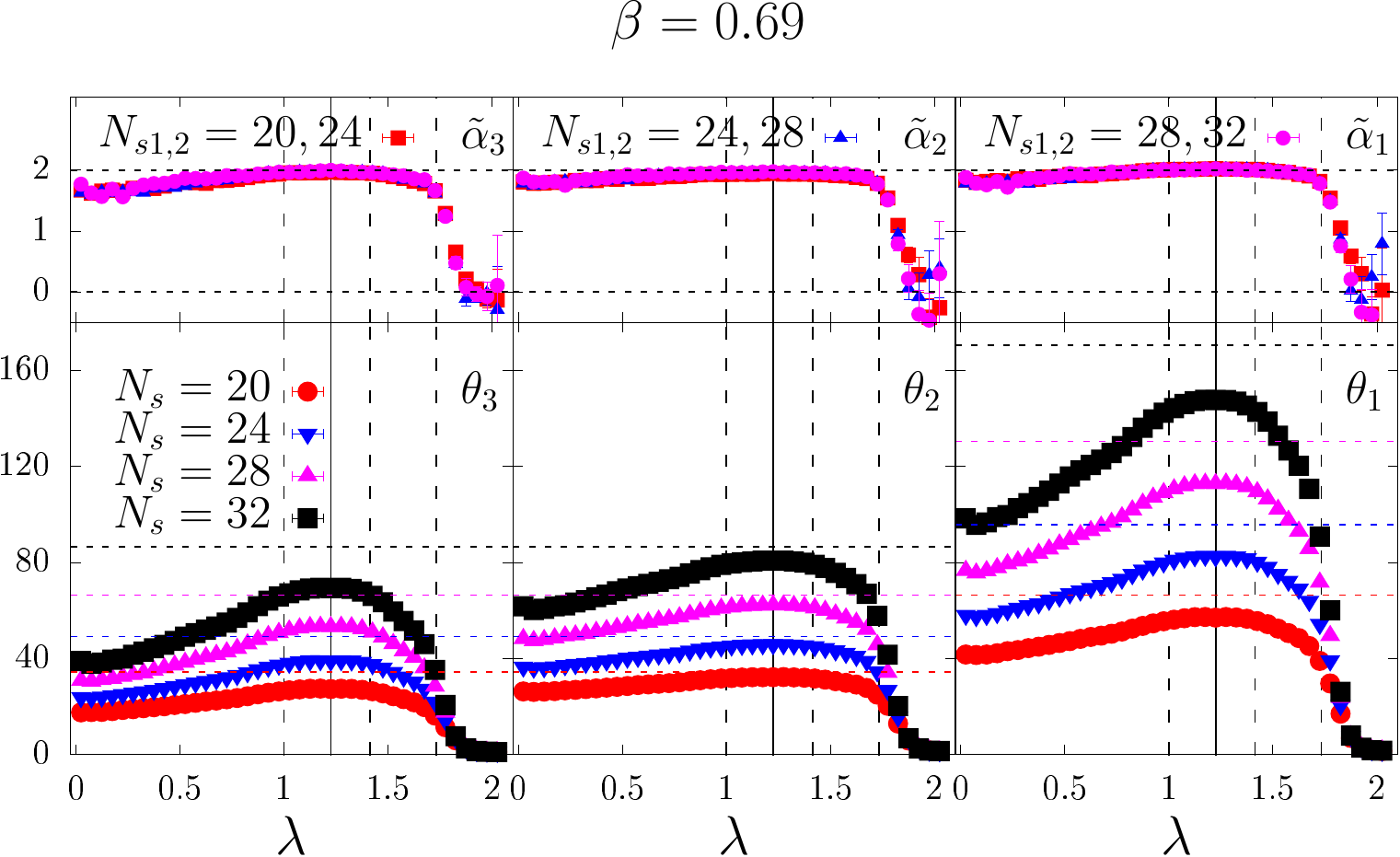}
  \hfil
  \includegraphics[width=0.48\textwidth]{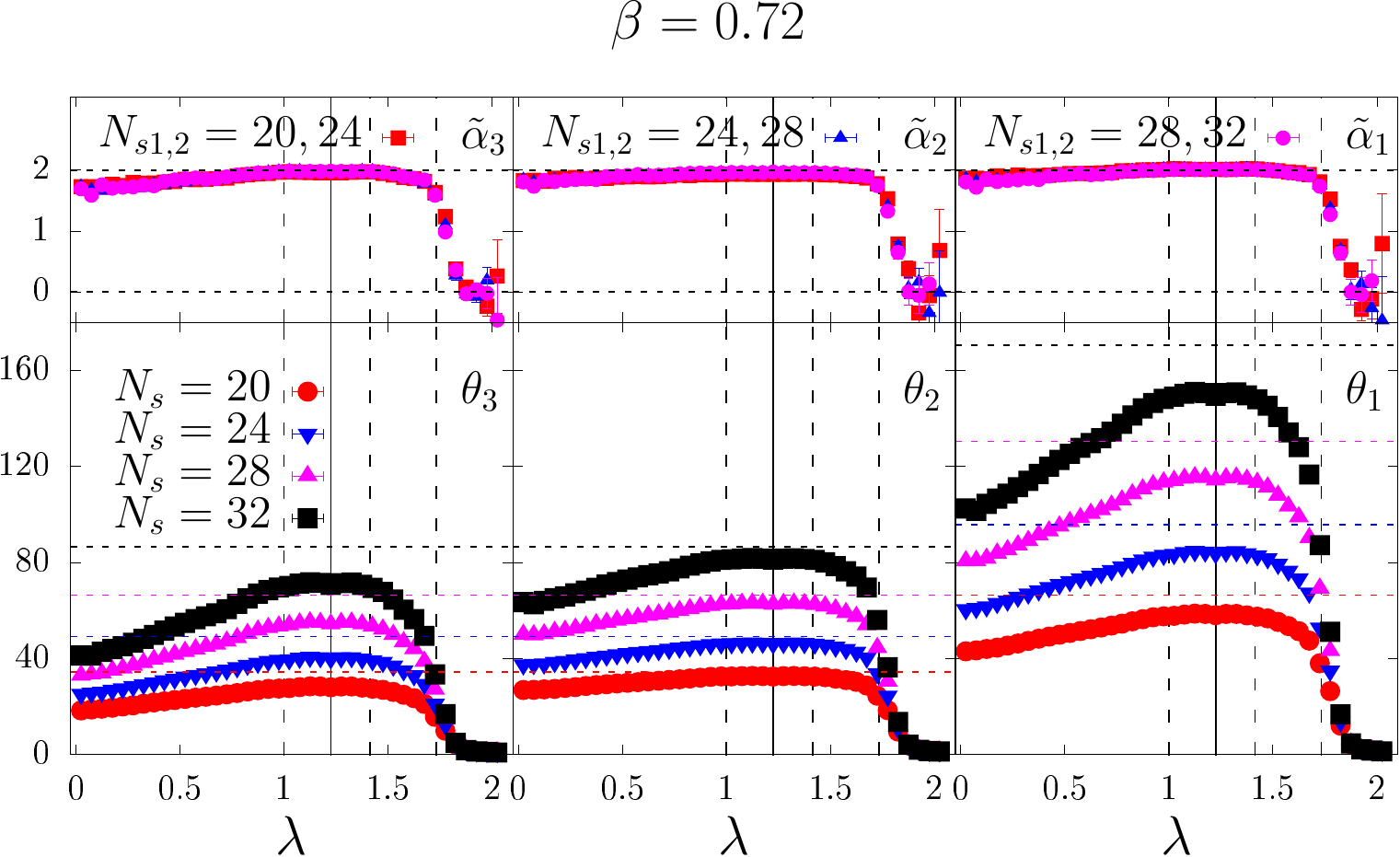}
  
  \caption{Confined phase, unphysical sector. Bottom panels: average
    principal moments of inertia $\theta_i$.  Top panels: scaling
    dimension $\tilde{\alpha}_i$ of moment $\theta_i$ for various
    pairs of volumes.}

  \label{fig:inmom_ws}
\end{figure}

\begin{figure}[t]
  \centering
  \includegraphics[width=0.48\textwidth]{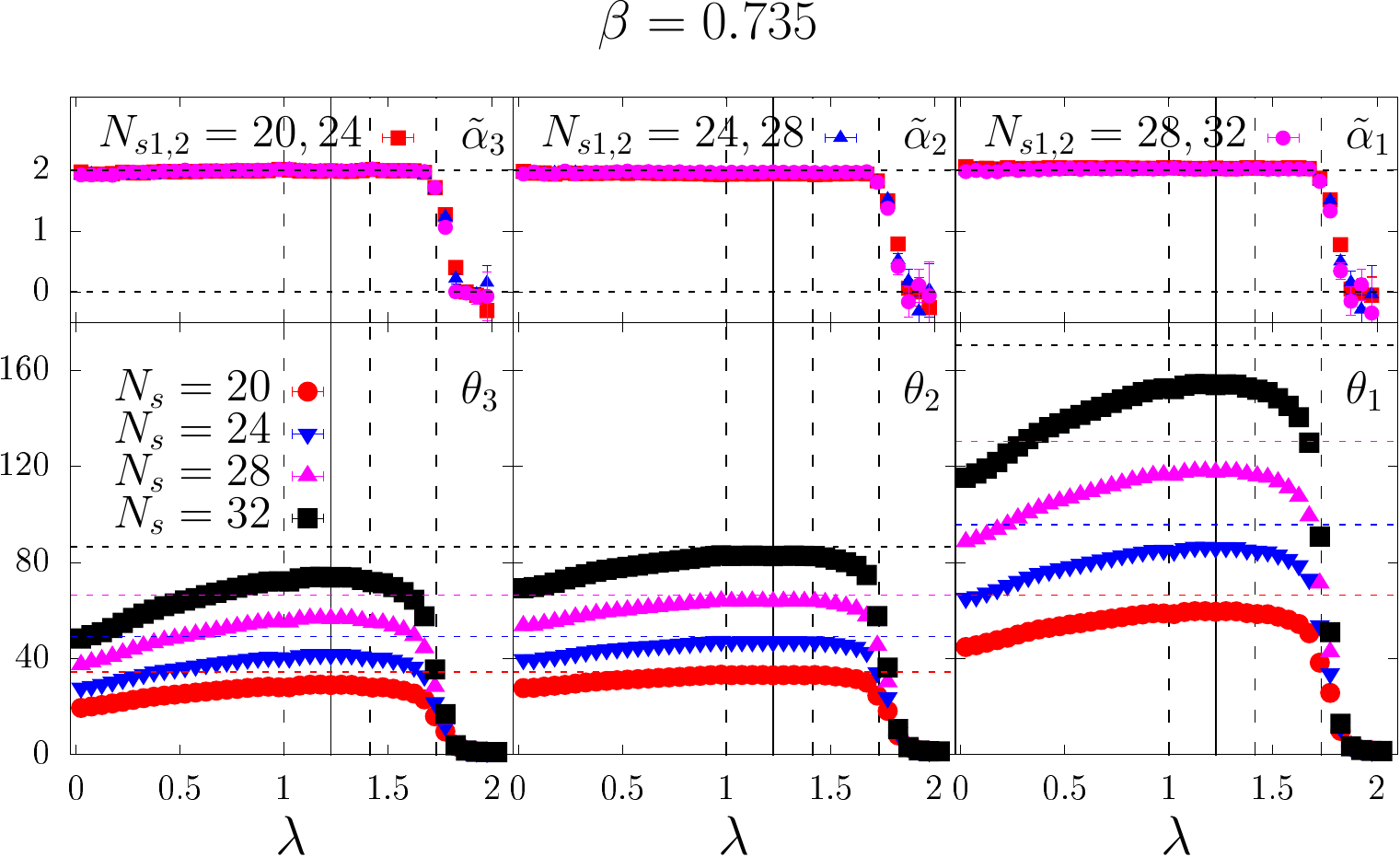}
  \hfil
  \includegraphics[width=0.48\textwidth]{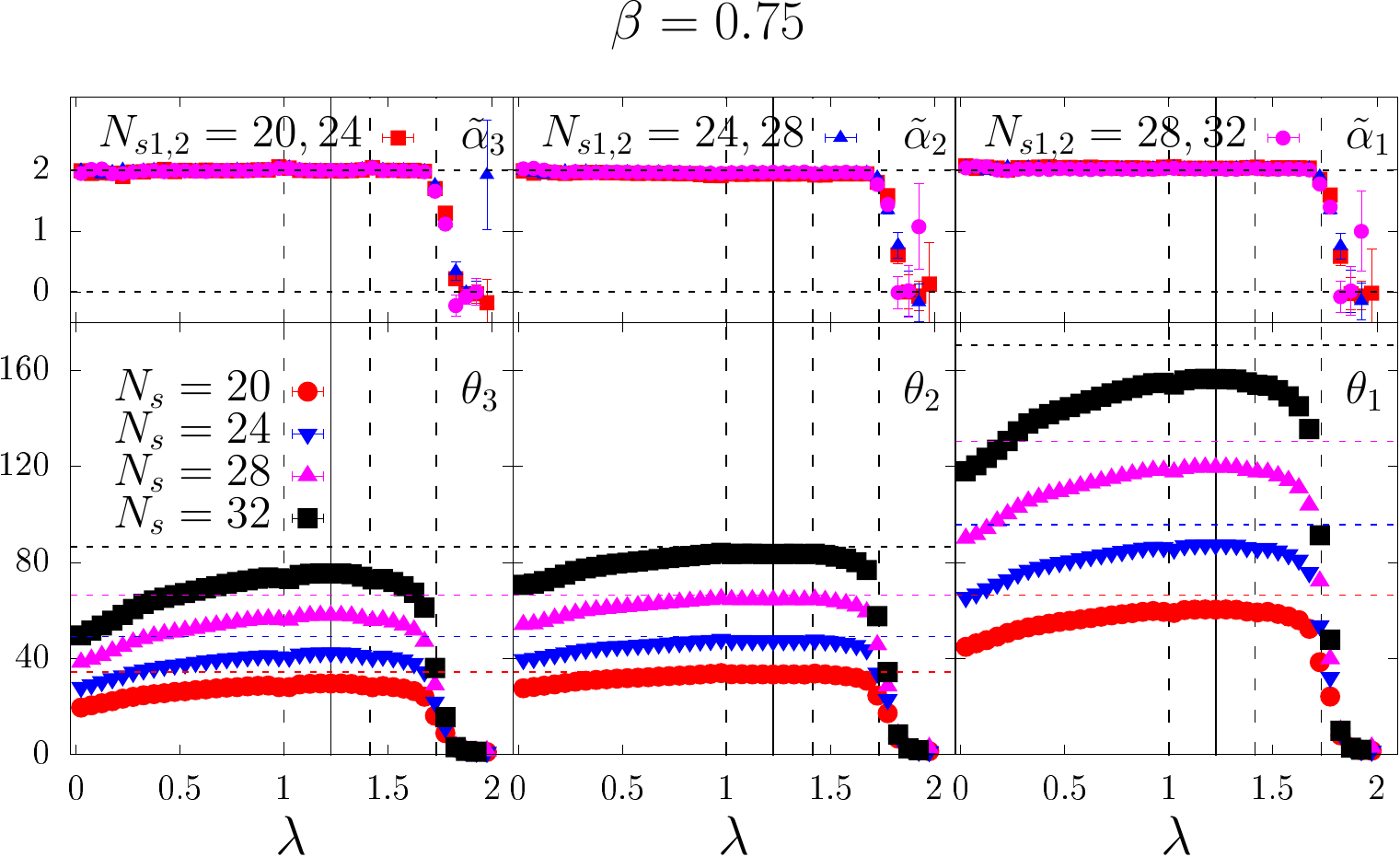}
  
  \caption{Deconfined phase, unphysical sector. Bottom panels: average
    principal moments of inertia $\theta_i$.  Top panels: scaling
    dimension $\tilde{\alpha}_i$ of moment $\theta_i$ for various
    pairs of volumes.}

  \label{fig:inmom_ws2}
\end{figure}

\begin{figure}[t]
  \centering
  \includegraphics[width=0.45\textwidth]{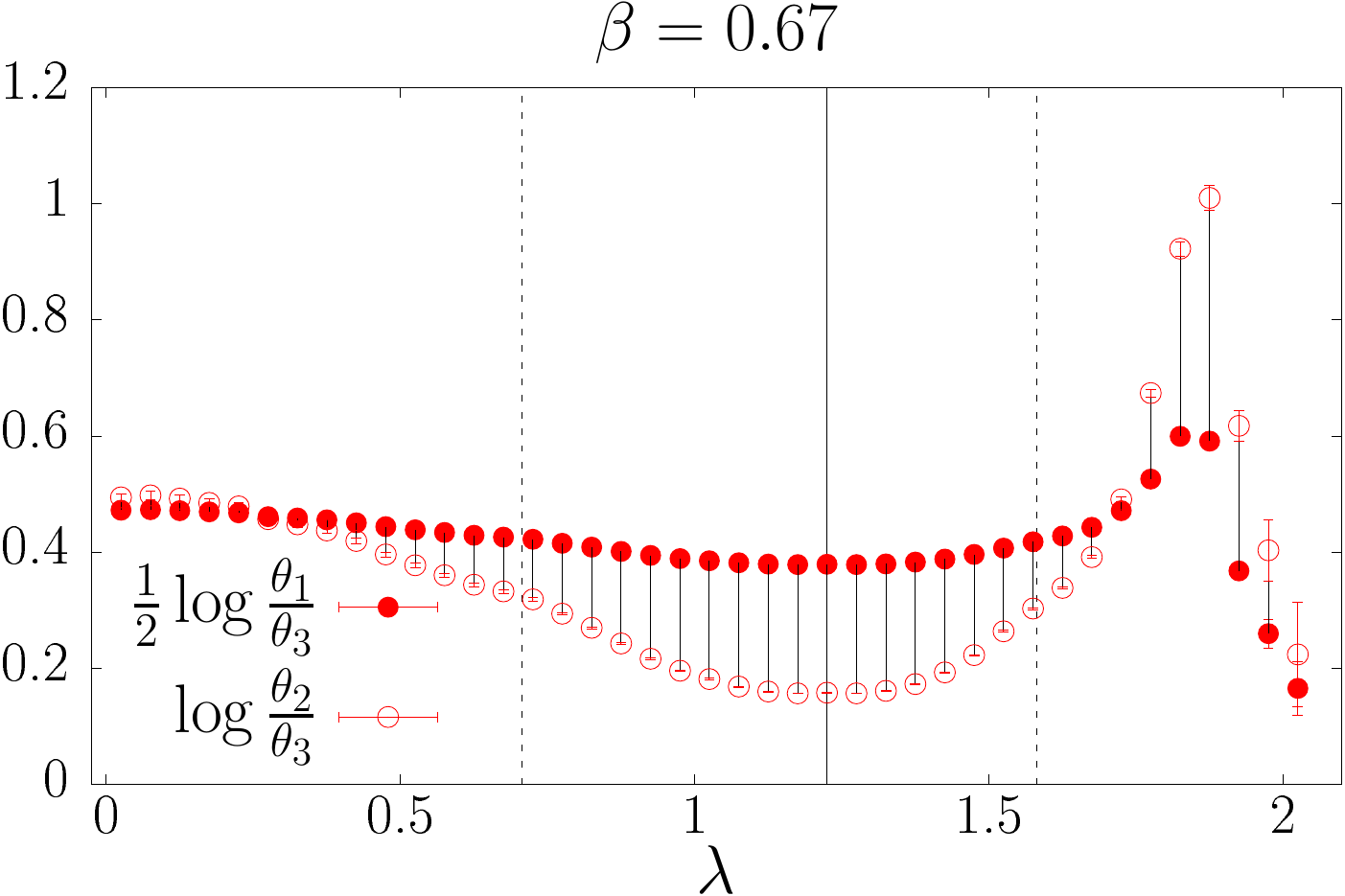}
  \hfil
  \includegraphics[width=0.45\textwidth]{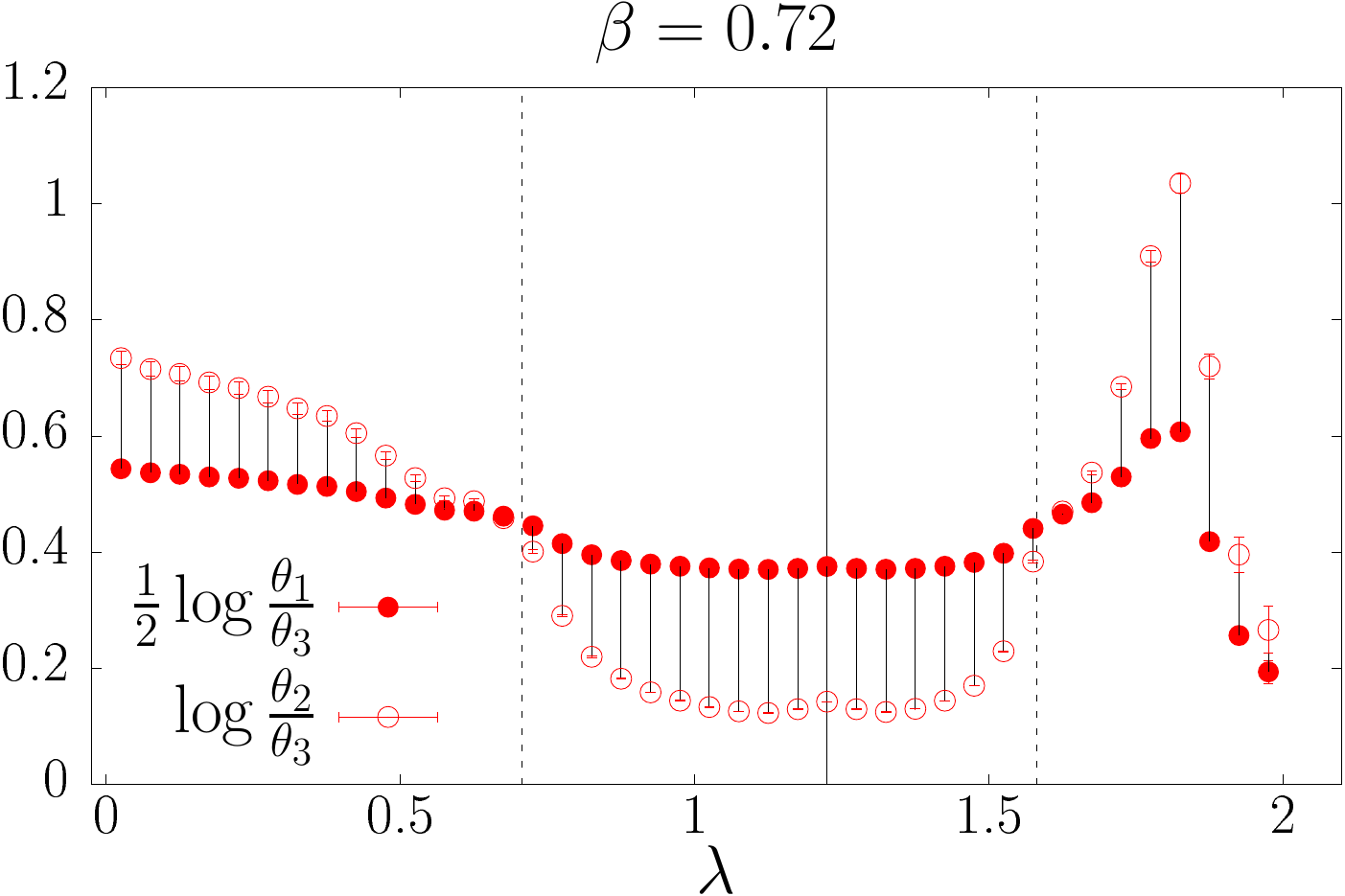}
  
  \vspace{0.2cm}
  \includegraphics[width=0.45\textwidth]{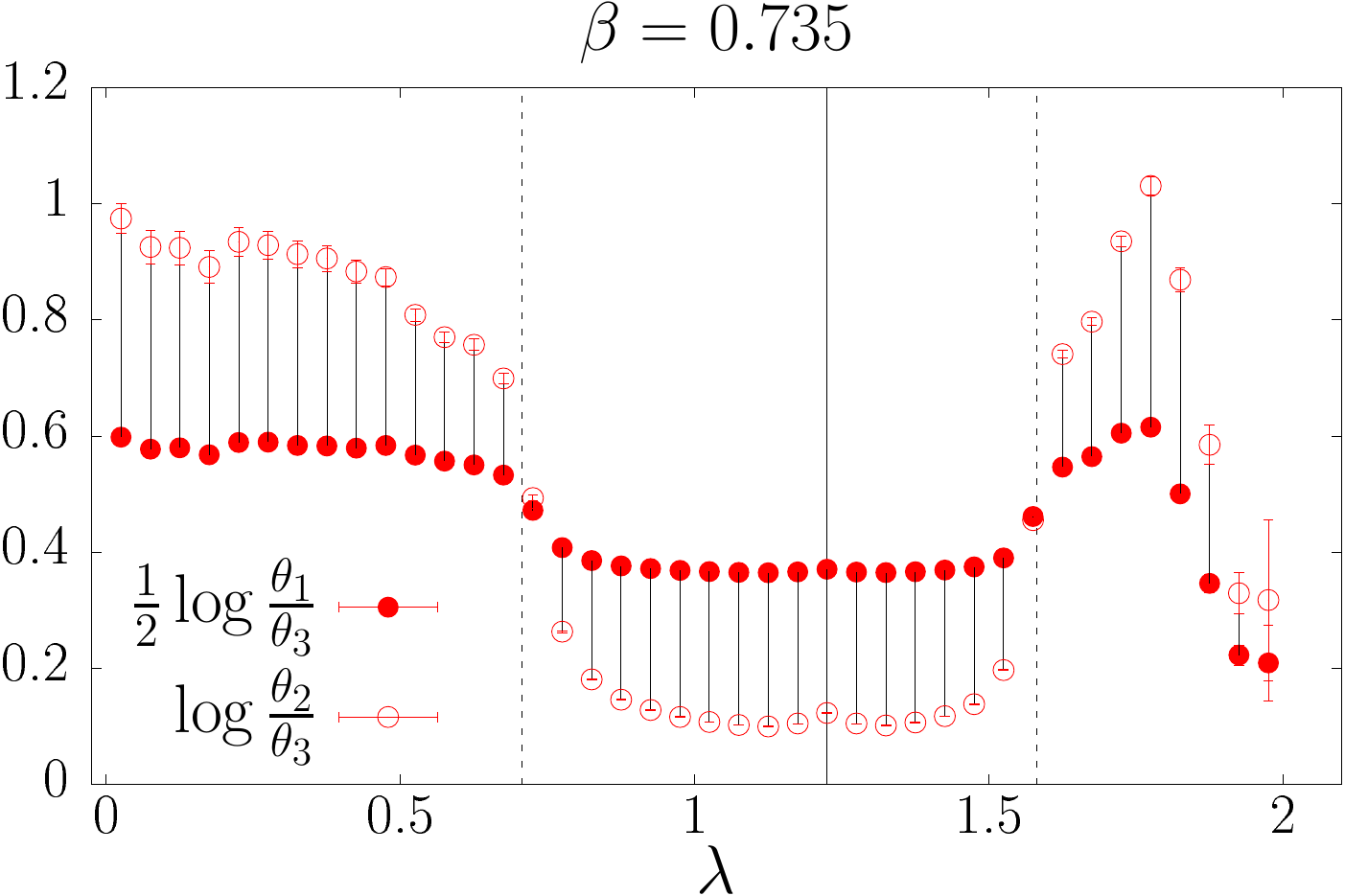}
  \hfil
  \includegraphics[width=0.45\textwidth]{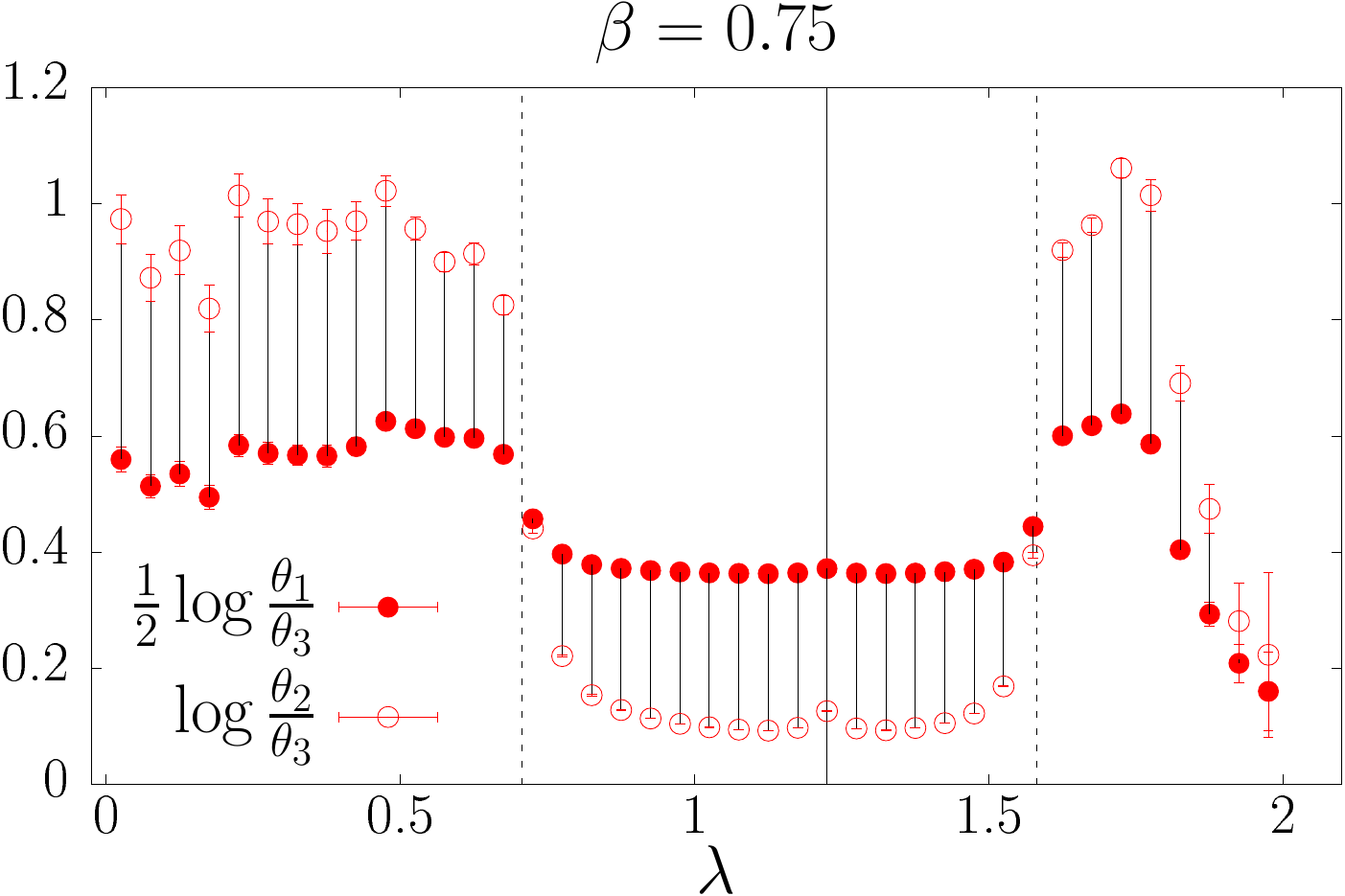}

  \caption{Confined (top row) and deconfined (bottom row) phase,
    physical sector: shape of the eigenmodes. Here and in the next
    plot, the quantity $\f{1}{2}\log\f{\theta_1}{\theta_3}$ measures
    the sphericity, and the difference
    $\log\f{\theta_2}{\theta_3}-\f{1}{2}\log\f{\theta_1}{\theta_3}=
    \f{1}{2}\log\f{\theta_2^2}{\theta_1\theta_3}$ measures the
    prolateness (if $>0$) or oblateness (if $<0$) of the modes. The
    system size is fixed to $N_s=32$.}

  \label{fig:shape}
\end{figure}
\begin{figure}[t]
  \centering
  \includegraphics[width=0.45\textwidth]{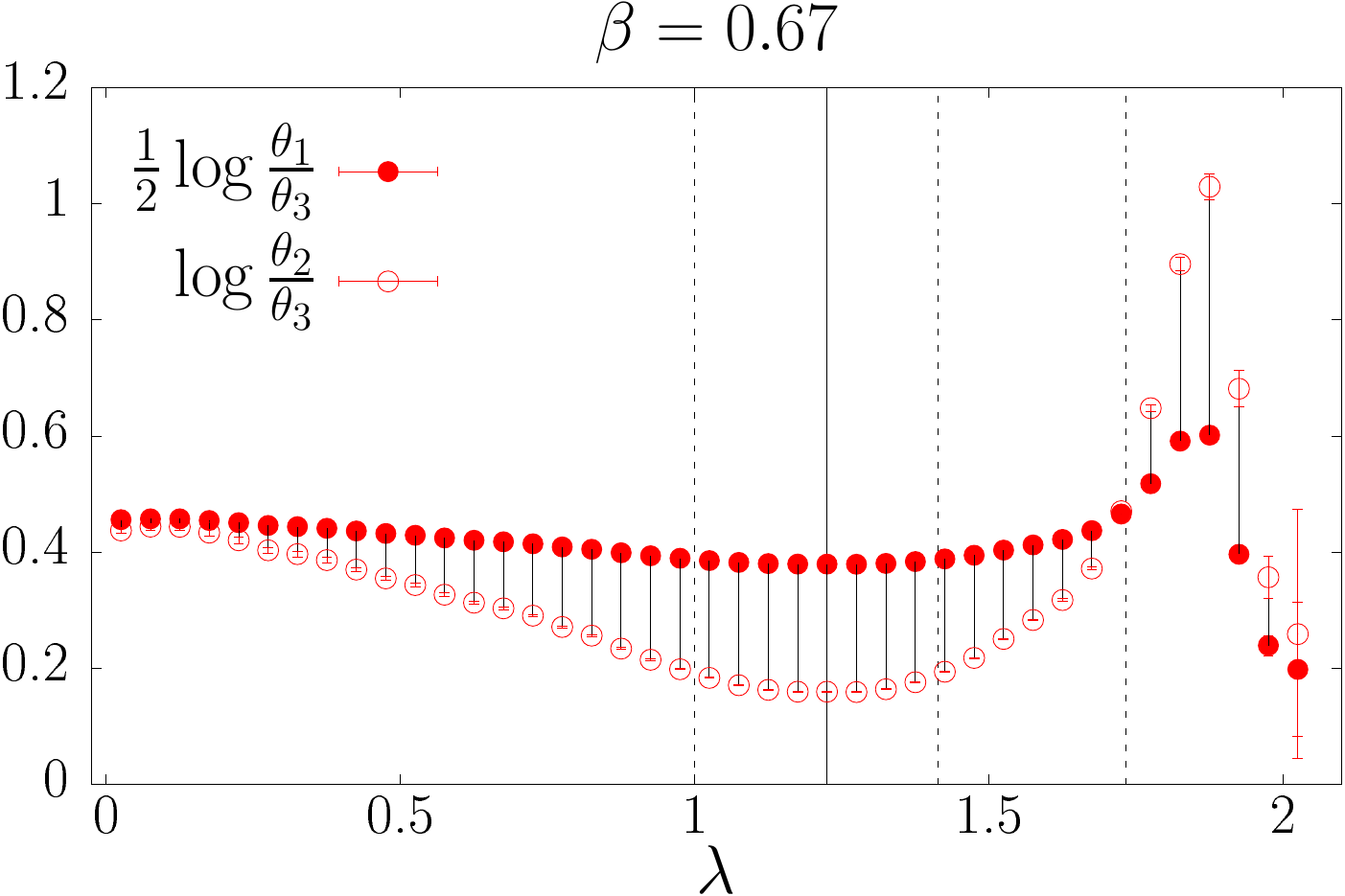}
  \hfil
  \includegraphics[width=0.45\textwidth]{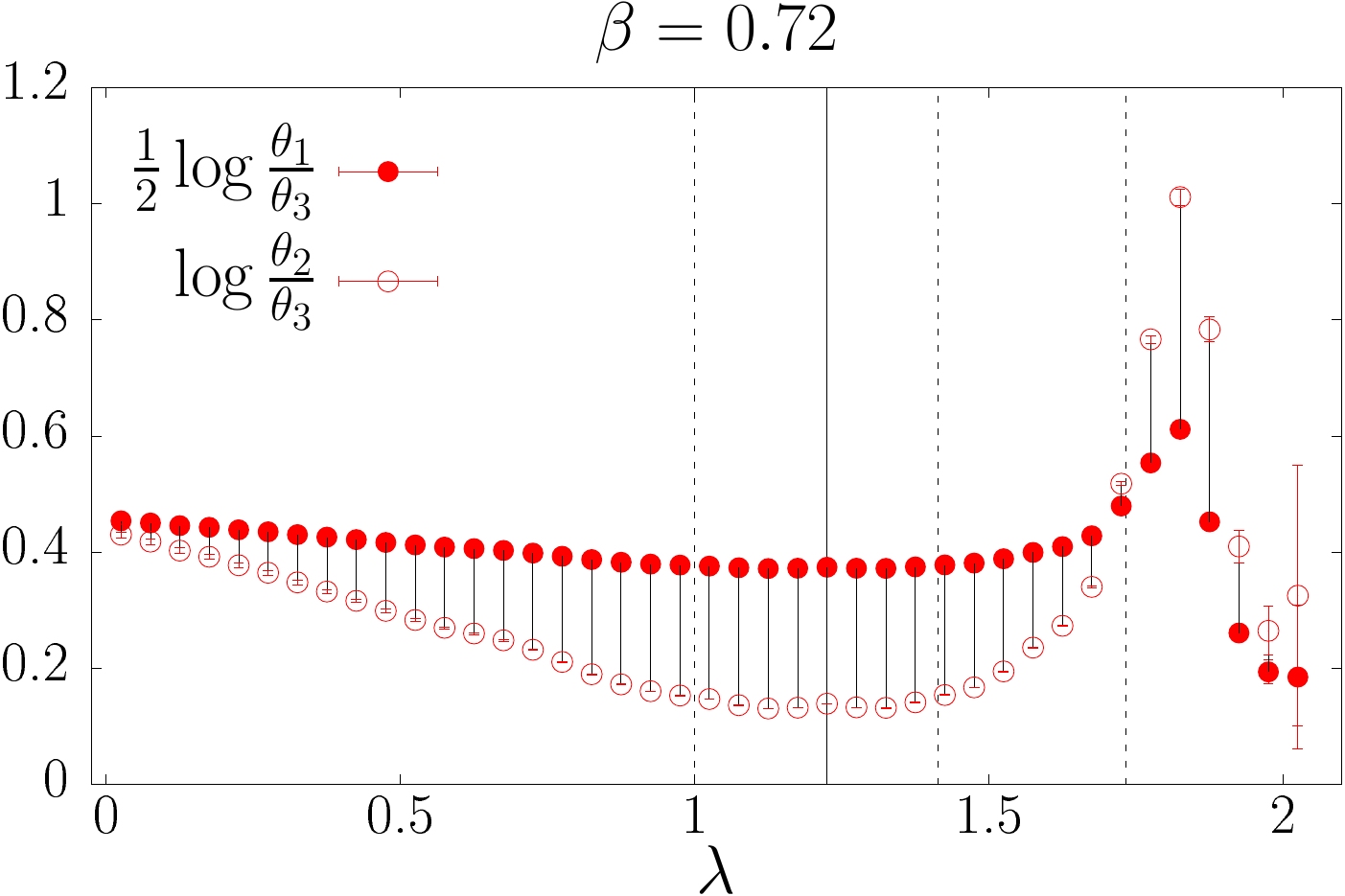}
  
  \vspace{0.2cm}
  \includegraphics[width=0.45\textwidth]{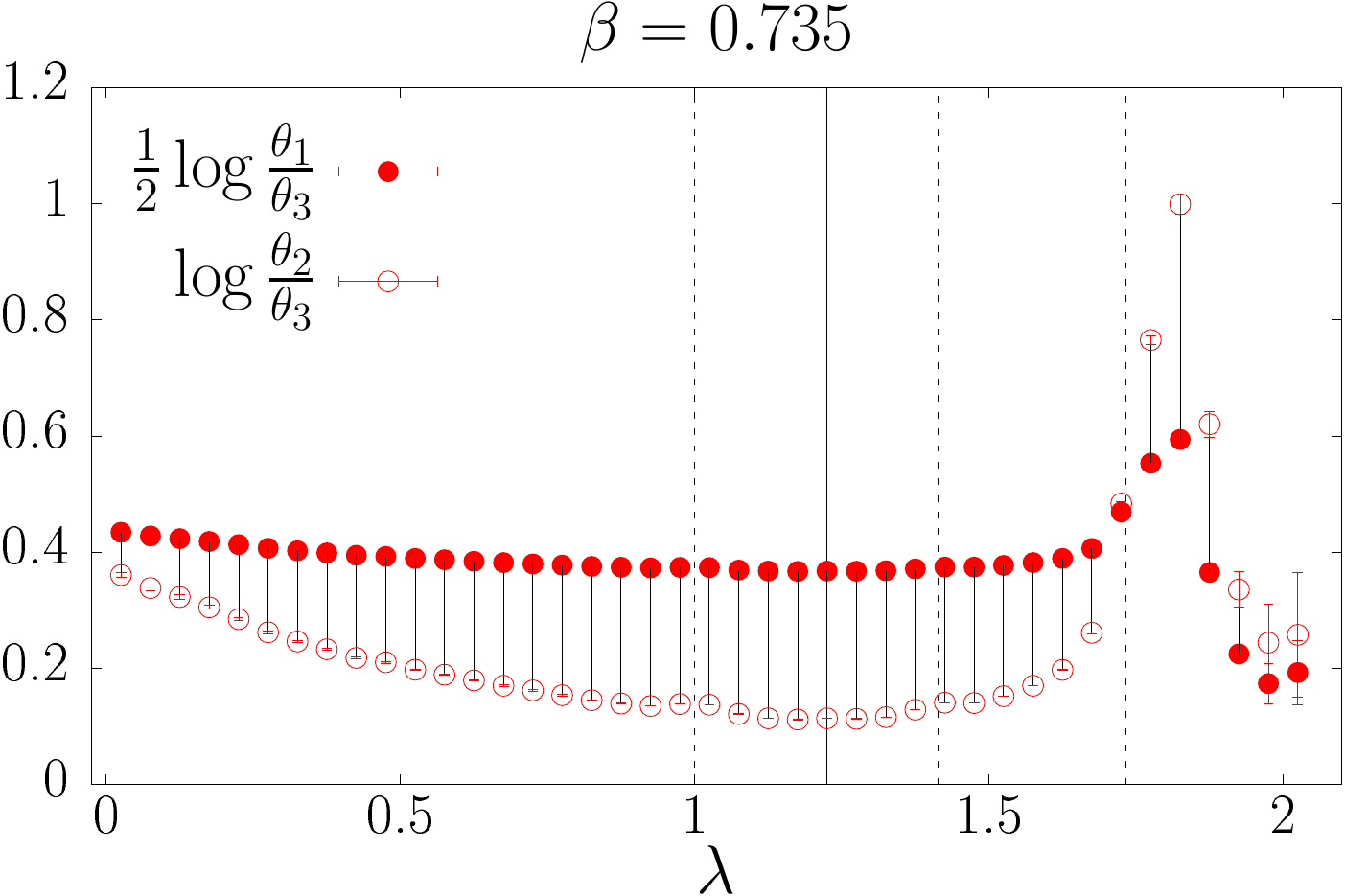}
  \hfil
  \includegraphics[width=0.45\textwidth]{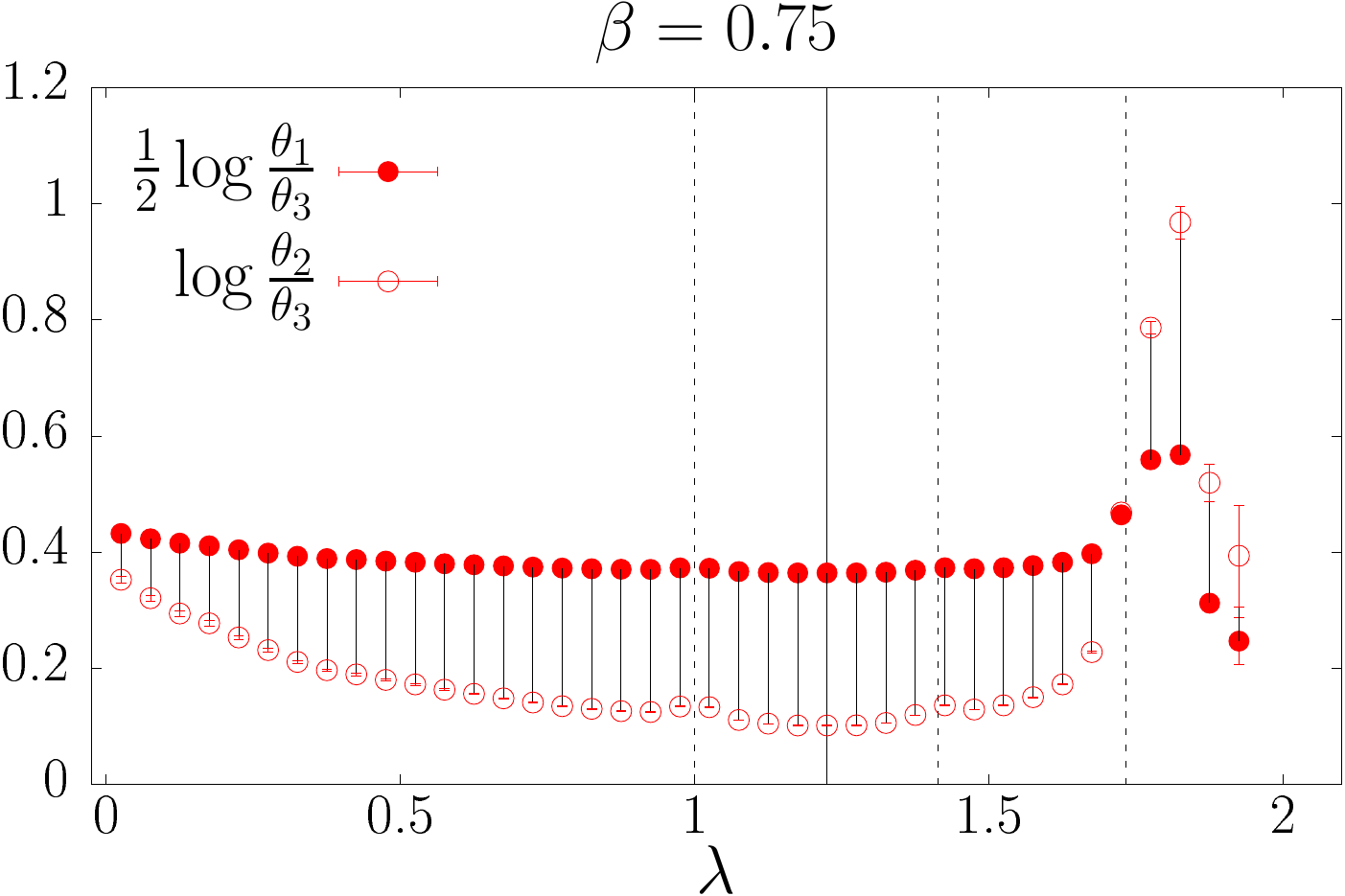}

  \caption{Confined (top row) and deconfined (bottom row) phase,
    unphysical sector: shape of the eigenmodes.}

  \label{fig:shape_ws}
\end{figure}

The inertia tensor of the eigenmodes is expected not only to
distinguish between localized and delocalized modes, but also to
provide more detailed information on their shape. This requires that a
general model for the shape of the eigenmodes is given first, so that
the principal axes of the mode, $v_{1,2,3}$, and the associated
principal moments of inertia, $\theta_1\ge\theta_2\ge \theta_3$, can
be directly related to geometric features.  What specific model is
used to describe the geometry of the mode is, however, not so
important, since we are mostly after generic features such as the
scaling of the mode with the lattice size, and, for localized modes,
whether they are isotropic or not.

For delocalized modes the natural model is a cuboid of unit mass and
uniform density extending throughout the whole lattice. For a cuboid
of linear size $\ell_i$ in direction $v_i$, one finds for the
principal moment of inertia $\theta_1$ corresponding to the principal
axis $v_1$ the value
$\theta_{1}^{\rm cont} = \f{1}{12}(\ell_2^2 + \ell_3^2)$ for a
continuous mass distribution, and
$\theta_{1}^{\rm lat} = \f{1}{12}(\ell_2^2 + \ell_3^2 -2)$ on a
periodic lattice; similar expressions hold for the other two principal
moments $\theta_{2,3}$ corresponding to the principal axes $v_{2,3}$.
Localized modes can also be modelled as cuboids, this time confined to
a limited spatial region, or perhaps more naturally as ellipsoids of
extension $\ell_i'$ in direction $v_i$. In the latter case one finds
$\theta_{1}^{\rm cont} = \f{1}{20}(\ell_2^{\prime 2} + \ell_3^{\prime
  2})$ (for a continuous mass distribution), and similar expressions
for the other moments.

For either model, the localized or delocalized nature of the modes
should be reflected in the scaling of the moments of inertia with the
spatial size $N_s$ of the lattice,
$\theta_i\sim N_s^{\tilde{\alpha}_i}$.  In fact, using either a cuboid
or an ellipsoid model, for a spatially delocalized mode at least one
of the two contributions to any of the principal moments should scale
like $N_s^2$, resulting in $\tilde{\alpha}_i=2$. For localized modes,
instead, all the contributions should remain finite as $N_s$ grows,
resulting in $\tilde{\alpha}_i=0$.

Information about how the modes are stretched or flattened in the
various directions can be obtained from $\theta_i$. The isotropy, or
sphericity, of the mode, can be measured by
$\f{1}{2}\log\f{\theta_1}{\theta_3}$: the closer to zero, the closer
the mode is to isotropic/spherical
($\theta_1\approx \theta_2 \approx \theta_3$). The prolateness or
oblateness of a mode can be measured instead by
$\log\f{\theta_2}{\theta_3}-\f{1}{2}\log\f{\theta_1}{\theta_3}=
\f{1}{2}\log\f{\theta_2^2}{\theta_1\theta_3}$. Indeed, for a
``rod-like'' cuboid or a prolate spheroid
($\theta_1\approx \theta_2 > \theta_3$) this quantity is positive, and
the larger in magnitude the more elongated the mode is.  Instead, for
a ``slab-like'' cuboid or an oblate spheroid
($\theta_1 > \theta_2 \approx \theta_3$) this quantity is negative,
and the larger in magnitude the more flattened the mode is.

Finally, from the principal axes one can determine the typical
orientation of the eigenmodes with respect to the temporal direction.
For delocalized modes extended throughout the lattice, one expects
$v_1$ to lie along the temporal direction, and $v_{2,3}$ to lie in the
spatial plane.  For localized modes one can further distinguish
between modes extended in the temporal direction, and modes localized
in the temporal direction (possibly on a comparable scale). In the
first case one similarly expects one principal axis along the temporal
direction and two in the spatial plane, while in the second case the
orientation of the mode with respect to the temporal direction can
fluctuate.

We now discuss our numerical results, separating physical and
unphysical sector, and looking at how the modes change across the
transition.
\paragraph{Moments of inertia}
In Figs.~\ref{fig:inmom}--\ref{fig:inmom_ws2}, bottom panels, we show
the average principal moments of inertia $\theta_i$ of the eigenmodes,
computed locally in the spectrum, for various $\beta$ and spatial
sizes $N_s$. For comparison, we also show the theoretical expectation
for a uniform $N_t\times N_s \times N_s$ cuboid extended throughout
the lattice. In the top panels we show the corresponding scaling
dimension $\tilde{\alpha}_i$.

In the physical sector in the confined phase, Fig.~\ref{fig:inmom},
the low modes clearly differ from the theoretical prediction for a
fully delocalized cuboid. However, all $\tilde{\alpha}_i$ being close
to 2 clearly shows that they are delocalized. Modes in the bulk are
closer to the theoretical prediction, and correspondingly the
$\tilde{\alpha}_i$ are much closer to 2. The deviations from the
theoretical expectation clearly visible in $\theta_1$ and $\theta_3$
probably reflect the inhomogeneities in the modes due to the tendency
to prefer positive Polyakov loops (see Figs.~\ref{fig:polbymode} and
\ref{fig:polbymode2}) and avoid negative plaquettes (see
Figs.~\ref{fig:weightnegplaq} and \ref{fig:weightnegplaq2}). For high
modes, on the other hand, the $\theta_i$ are independent of $N_s$, and
correspondingly $\tilde{\alpha}_i\approx 0$, consistently with their
localized nature. Notice that as $\beta$ increases, the difference
between low and bulk modes becomes sharper, somehow indicating that a
phase transition is approaching.

Still in the physical sector but in the deconfined phase,
Fig.~\ref{fig:inmom2}, the moments of inertia of the low modes are now
$N_s$-independent, indicating that they have become localized. This is
confirmed by $\tilde{\alpha}_i$ becoming approximately 0 above the
critical temperature. Notice that sizeable finite-size effects are
visible in $\tilde{\alpha}_i$ close to $\beta_c$. For bulk and high
modes the moments of inertia and their volume scaling remain
approximately the same as in the confined phase.

In the unphysical sector, both in the confined
(Fig.~\ref{fig:inmom_ws}) and in the deconfined phase
(Fig.~\ref{fig:inmom_ws2}), the moments of inertia show that low and
bulk modes are delocalized, with the bulk modes very close to a fully
delocalized cuboid, and low modes clearly deviating from it. For high
modes the moments of inertia are instead $N_s$-independent,
consistently with localization, in both phases. Correspondingly,
$\tilde{\alpha}_i\approx 2$ for bulk modes, and
$\tilde{\alpha}_i\approx 0$ for high modes, in both phases. For low
modes $\tilde{\alpha}_i\approx 2$ in the deconfined phase, while they
are close to 2 but clearly deviate from it in the confined phase,
taking values similar to those found in the physical sector.

\begin{figure}[t]
  \centering
  \includegraphics[width=0.45\textwidth]{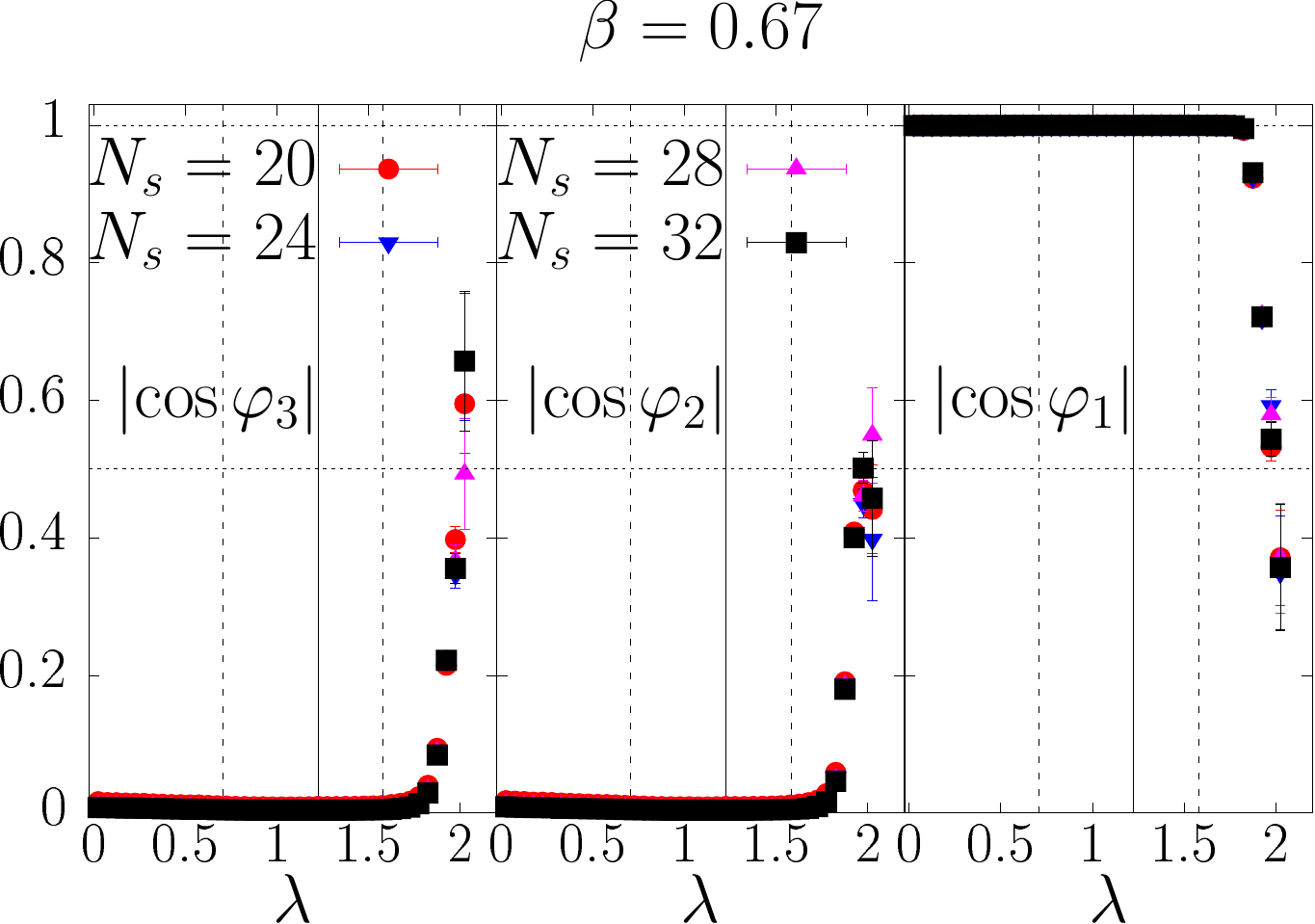}
  \hfil
  \includegraphics[width=0.45\textwidth]{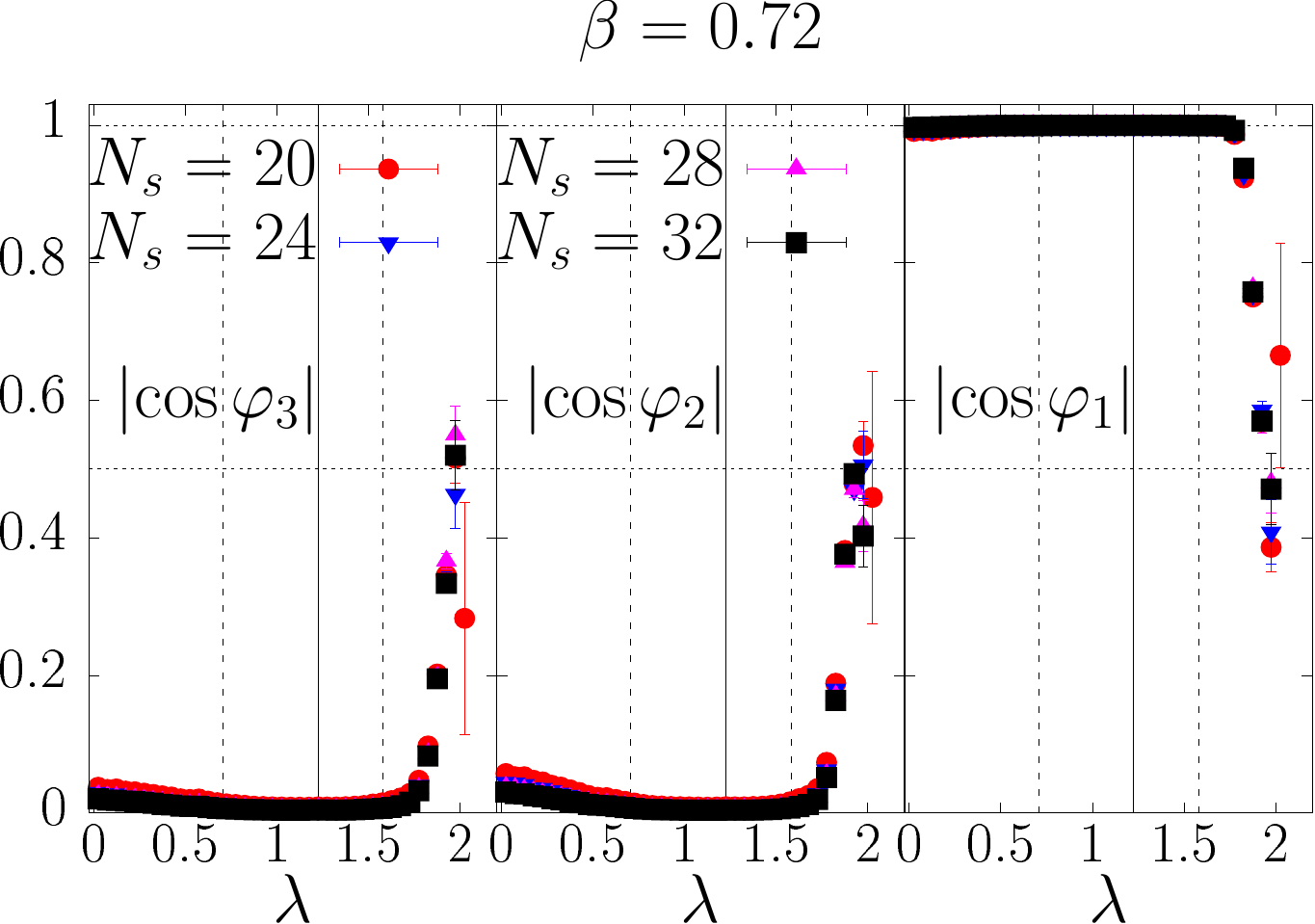}

  \vspace{0.2cm}
  \includegraphics[width=0.45\textwidth]{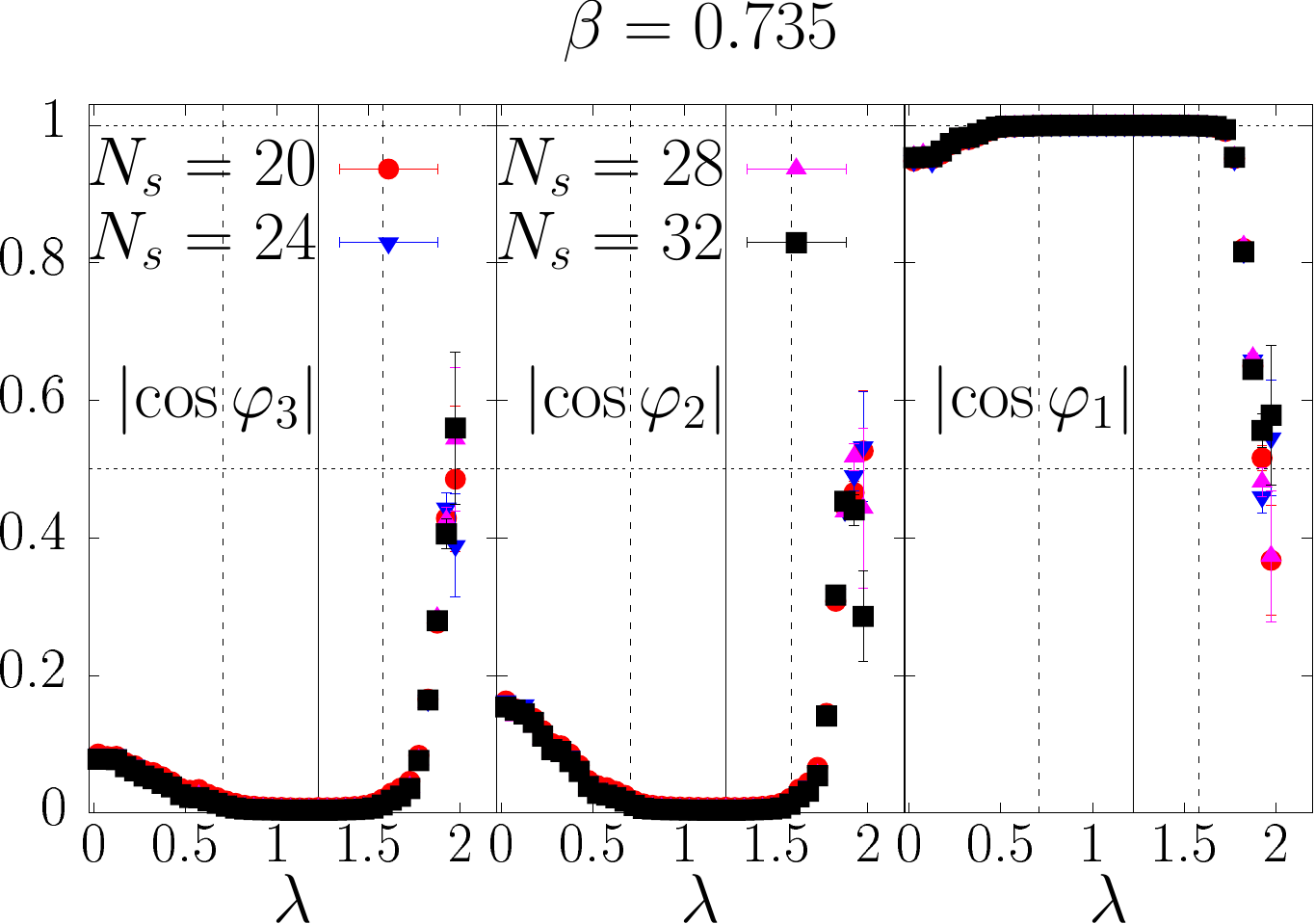}
  \hfil
  \includegraphics[width=0.45\textwidth]{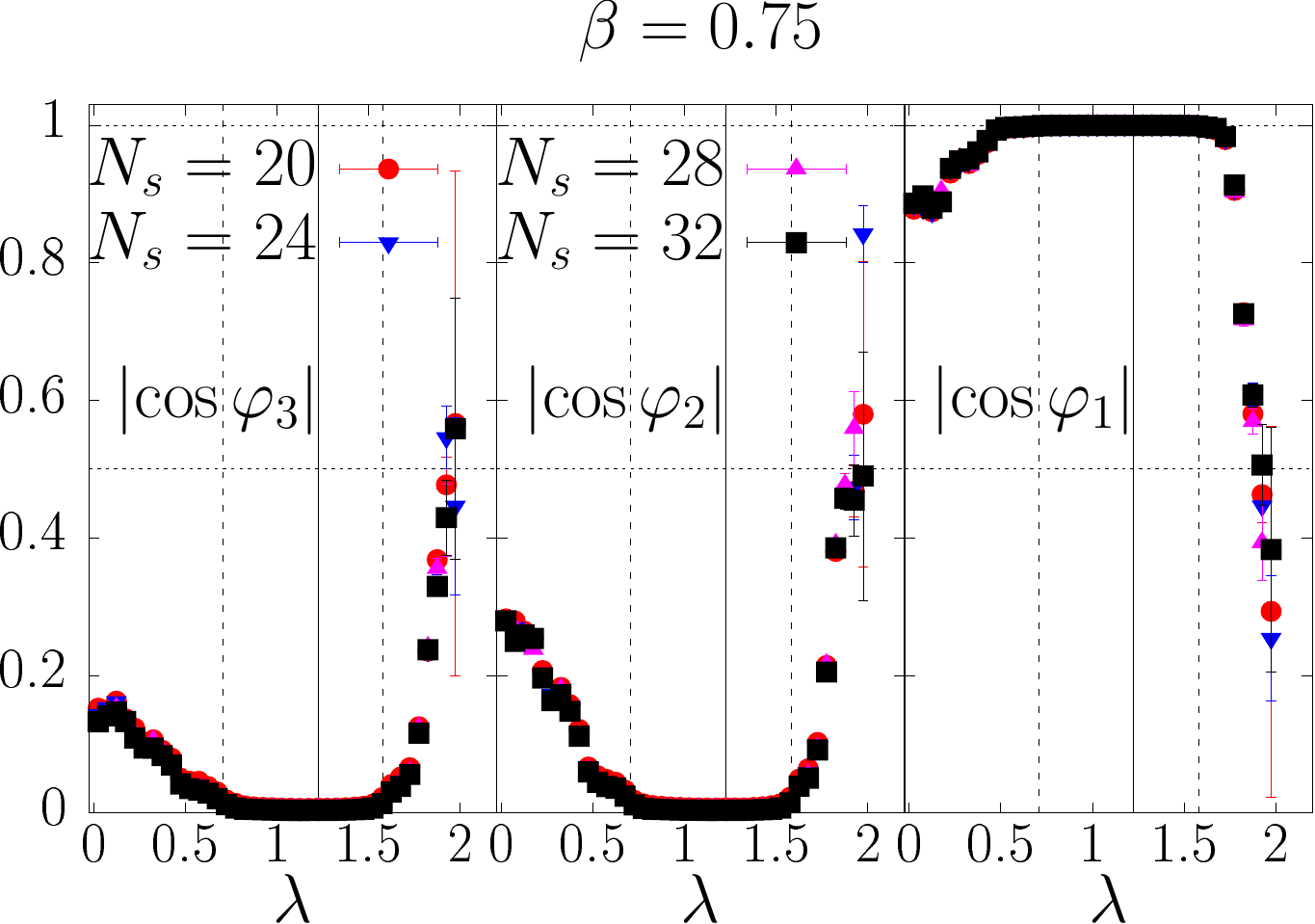}
  
  \caption{Confined (top row) and deconfined (bottom row) phase,
    physical sector: average orientation of the principal axes with
    respect to the time direction.}

  \label{fig:angle}
\end{figure}

\begin{figure}[t]
  \centering
  \includegraphics[width=0.45\textwidth]{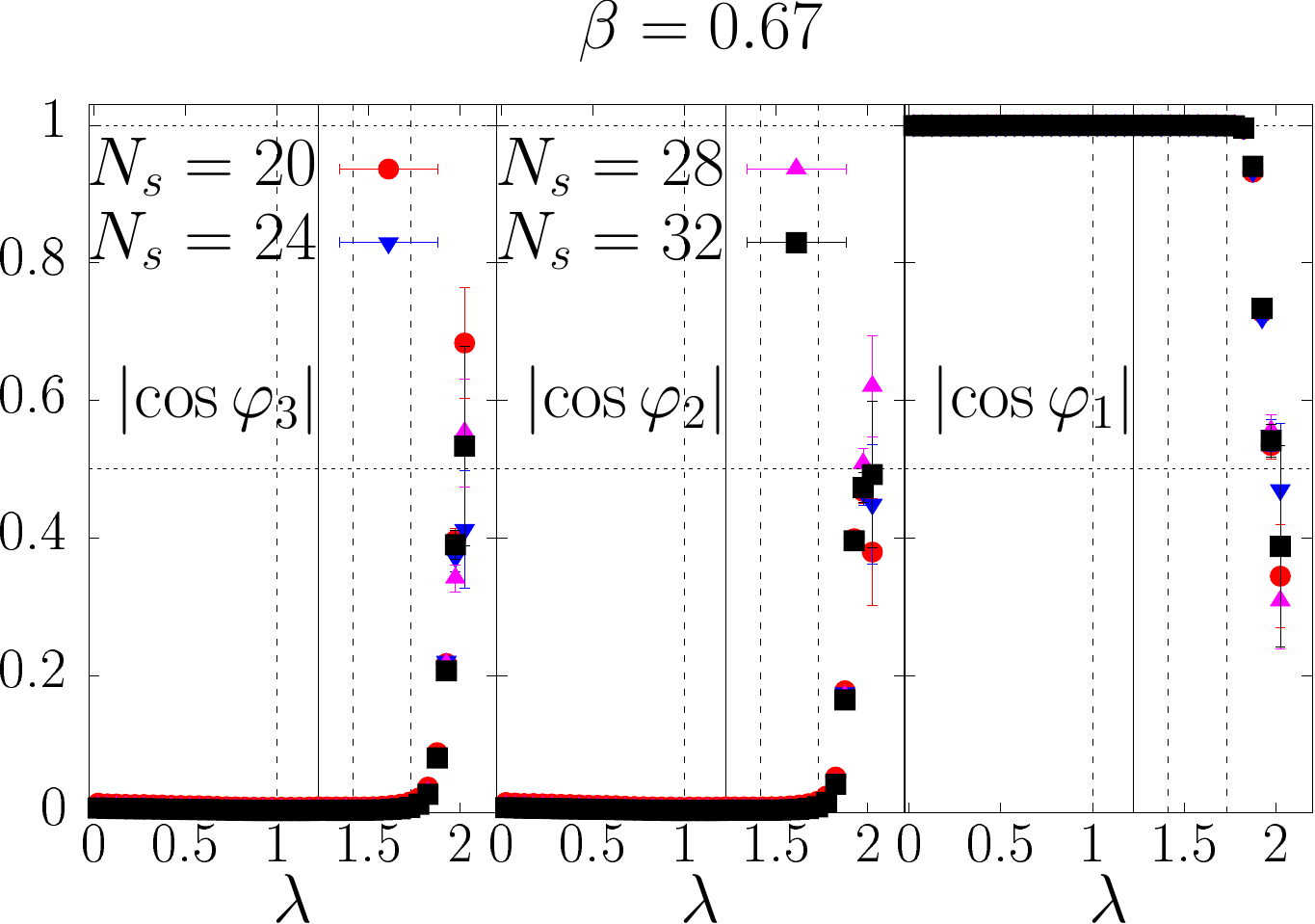}
  \hfil
  \includegraphics[width=0.45\textwidth]{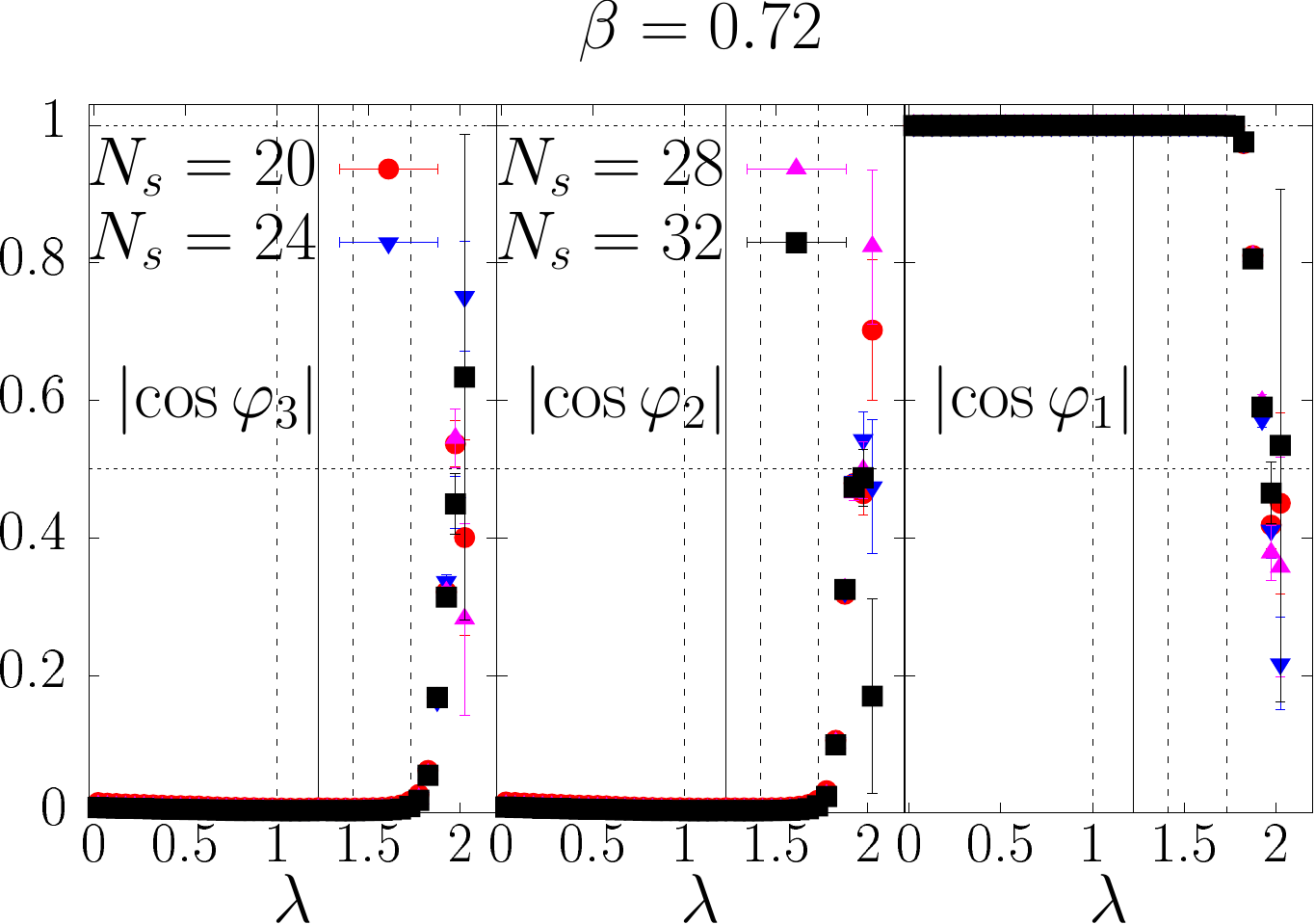}

  \vspace{0.2cm}
  \includegraphics[width=0.45\textwidth]{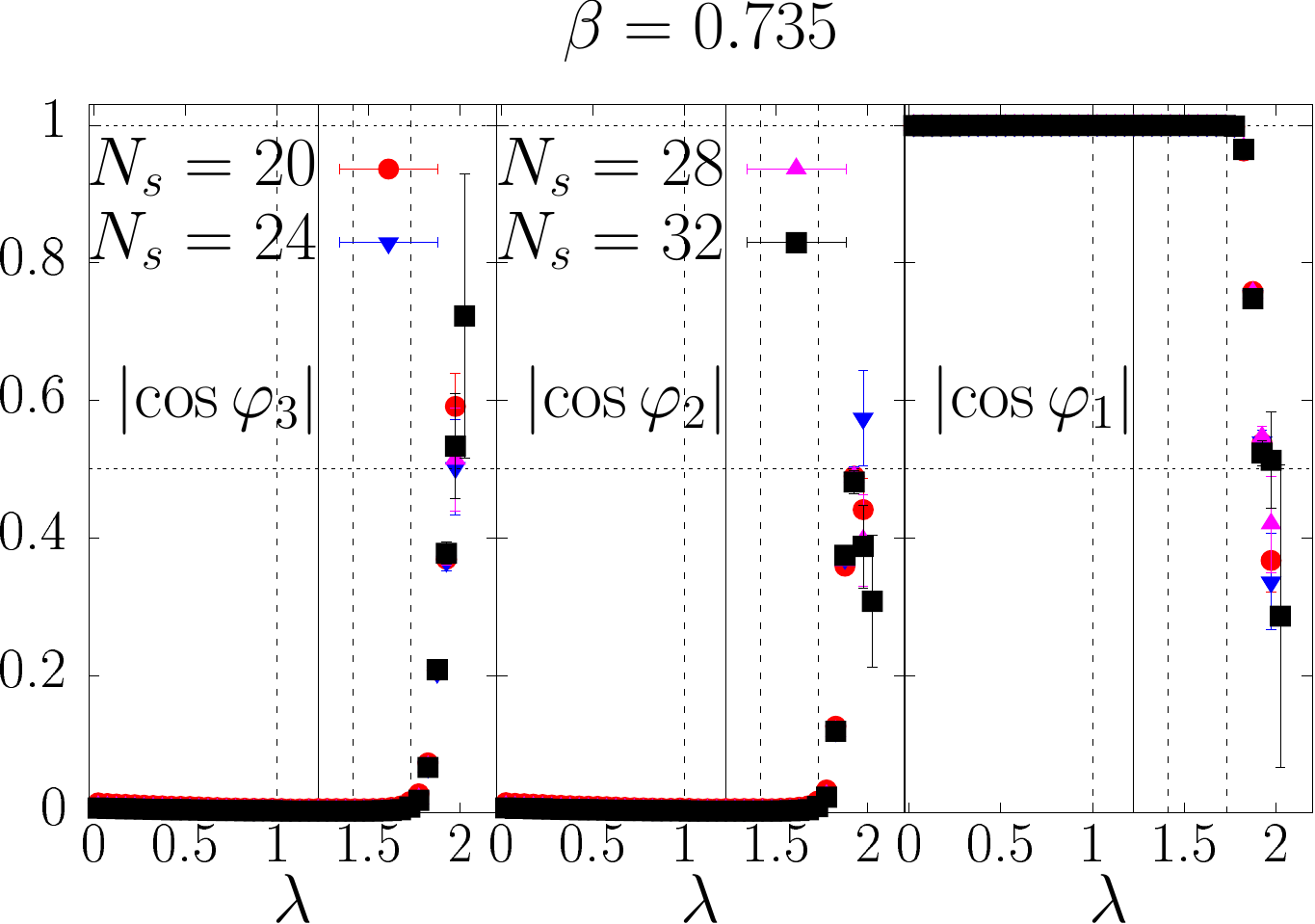}
  \hfil
  \includegraphics[width=0.45\textwidth]{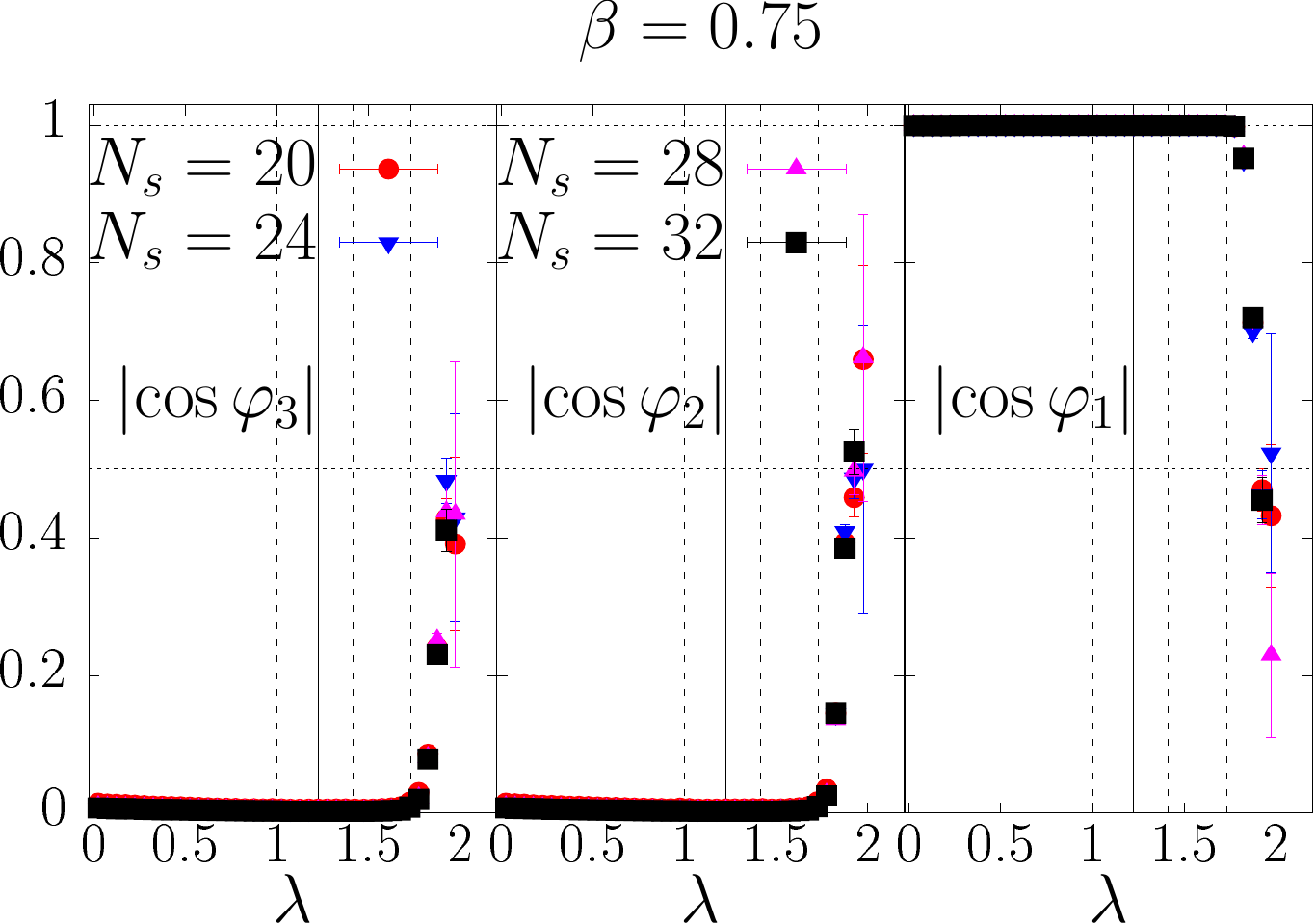}
  
  \caption{Confined (top row) and deconfined (bottom row) phase,
    unphysical sector: average orientation of the principal axes with
    respect to the time direction.}

  \label{fig:angle_ws}
\end{figure}

\paragraph{Mode shape} Information on the shape of the modes is
summarized in Figs.~\ref{fig:shape} and \ref{fig:shape_ws}. In the
physical sector (Fig.~\ref{fig:shape}), low modes become more and more
rod-like/prolate as $\beta$ increases. Bulk modes are instead
slab-like/oblate, as expected for fully delocalized modes. High modes
again tend to become more prolate as $\beta$ increases, but they also
become more spherical as one moves to higher $\lambda$. In the
deconfined phase, this is a clear difference between the low localized
modes and the high localized modes, the former being elongated while
the latter are almost spherical. In the unphysical sector
(Fig.~\ref{fig:shape_ws}), low modes are now slab-like/oblate,
becoming more so as $\beta$ increases. Bulk modes are again slab-like,
fully delocalized, and high modes are again localized and almost
spherical, with little difference between the two phases, and with the
physical sector.

\begin{figure}[t]
  \centering
  \includegraphics[width=0.59\textwidth]{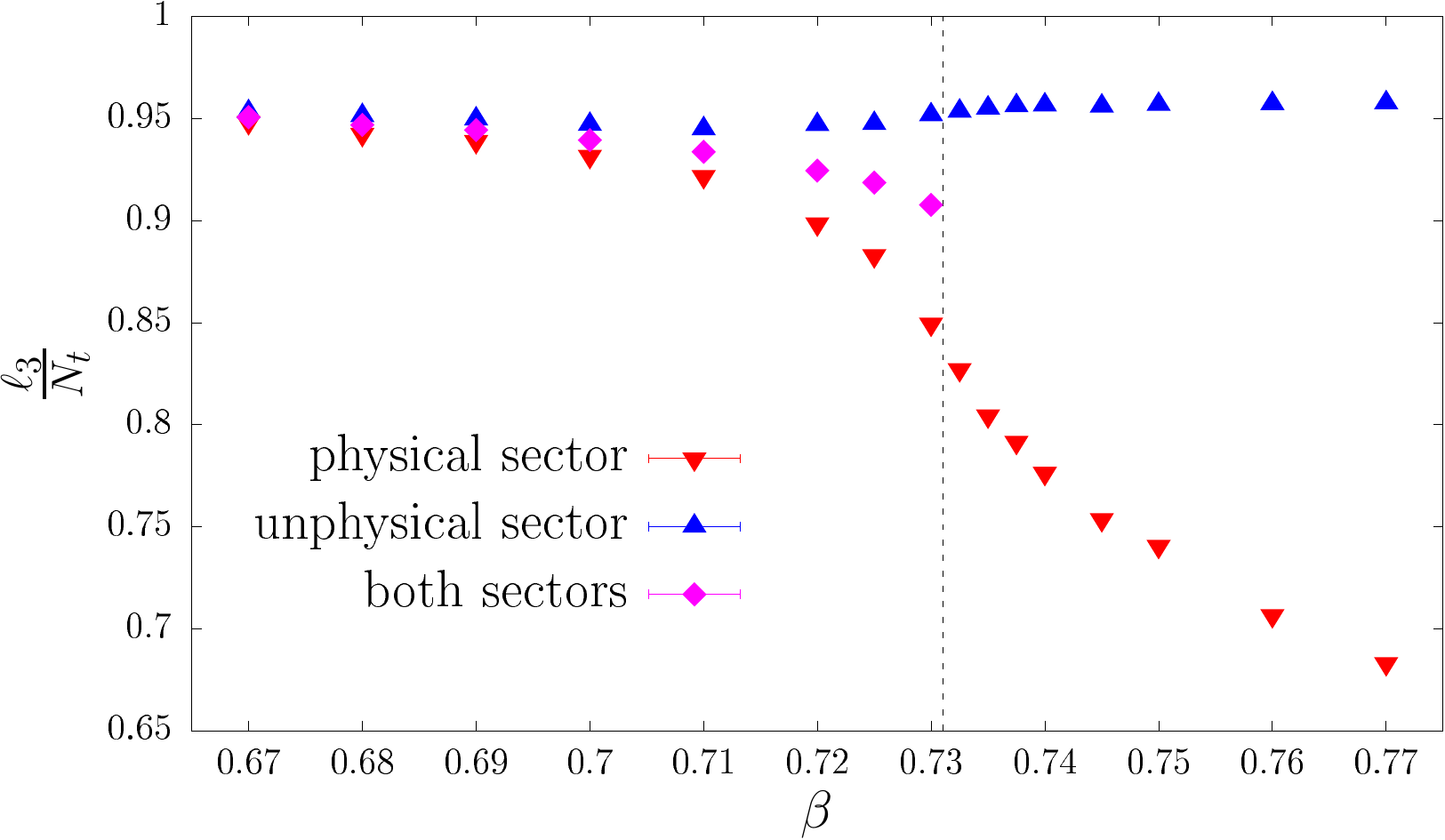}
      
  \vspace{0.35cm}
  \includegraphics[width=0.59\textwidth]{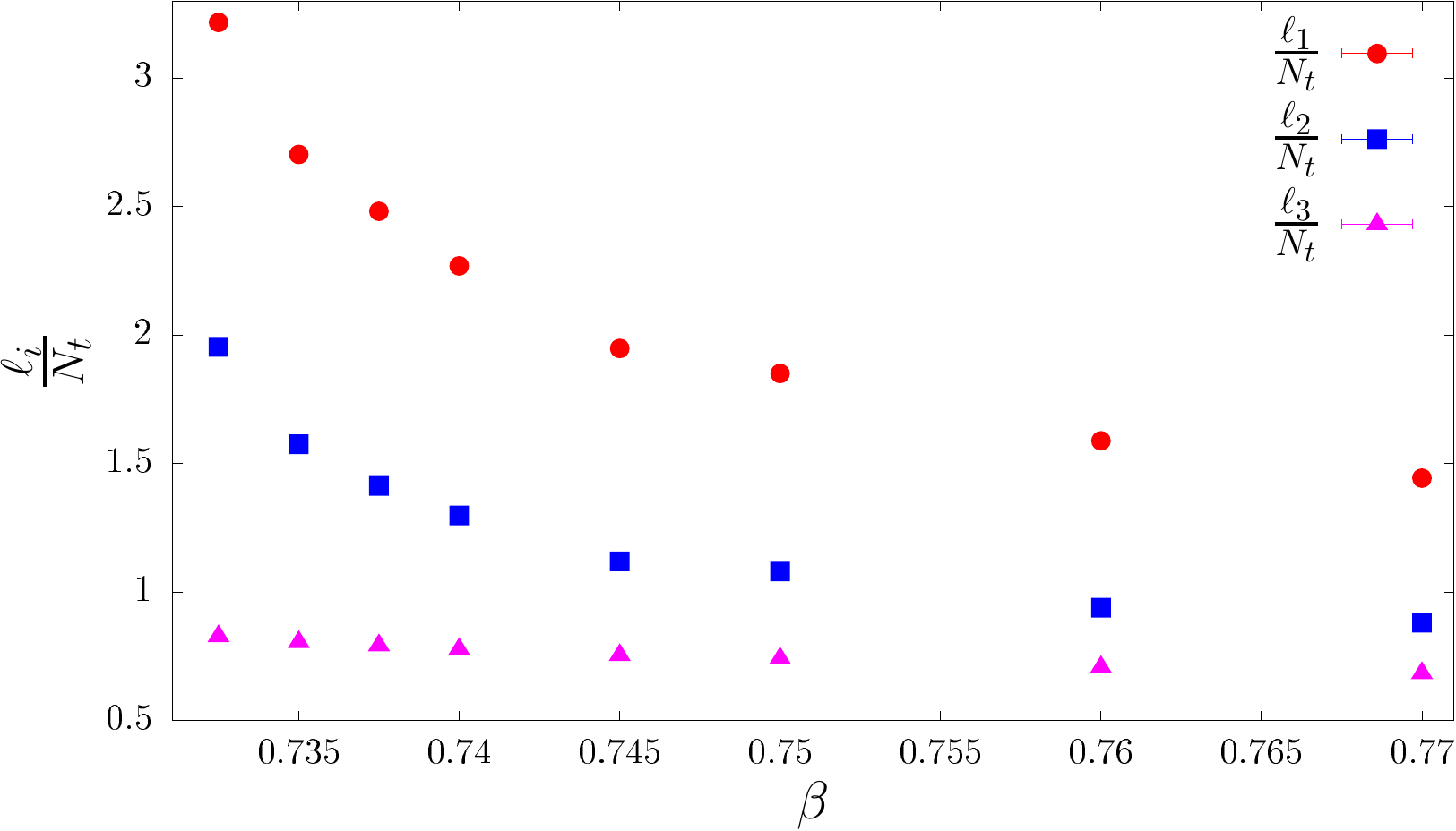}
      
  \caption{Top panel: temporal length scale $\ell_3$ of the lowest
    modes in units of the inverse temperature in all
    phases/sectors. Bottom panel: length scales $\ell_{1,2,3}$ of the
    lowest modes in units of the inverse temperature in the deconfined
    phase, physical sector. $N_s=32$ in both panels.}
  \label{fig:inmom_lengths}
\end{figure}

\paragraph{Mode orientation} The average orientation of the modes with
respect to the temporal direction (see right before
Eq.~\eqref{eq:loc17}) is shown in Figs.~\ref{fig:angle} and
\ref{fig:angle_ws}.  In the physical sector (Fig.~\ref{fig:angle}),
one finds for the low modes in the confined phase
$\langle \cos \varphi_3 \rangle \approx \langle \cos \varphi_2 \rangle
\approx 0$ and $\langle \cos \varphi_1 \rangle \approx 1$, as expected
for delocalized modes. Above the deconfinement transition, the
deviations from 0 and 1 become much larger as the modes become
localized, but they remain clearly away from $\f{1}{2}$. This is
consistent with the fact that they are typically not spherical but
prolate, localized objects. For bulk modes one has
$\langle \cos \varphi_3 \rangle \approx \langle \cos \varphi_2 \rangle
\approx 0$ and $\langle \cos \varphi_1 \rangle \approx 1$ in both
phases, as expected. For high modes
$\langle \cos \varphi_i \rangle \approx \frac{1}{2}$, confirming that
they are approximately spherical.  In the unphysical sector
(Fig.~\ref{fig:angle_ws}), the average orientation of low and bulk
modes is again consistent with what one expects for delocalized modes,
and that of high modes is again consistent with what one expects for
spherical modes.

Overall, the results discussed above are consistent with the picture
obtained from the participation ratio. In addition, they distinguish
between the localized modes found at the low end (when present) and at
the high end of the spectrum, that have different shapes.

\paragraph{Characteristic lengths of low modes }
Finally, in Fig.~\ref{fig:inmom_lengths} we show the length scales
$\ell_i^{\rm phys}=a\ell_i$ of the equivalent cuboid for the lowest
modes ($\lambda\in[0,\Delta\lambda]$) and the largest system size
($N_s=32$). More precisely, in Fig.~\ref{fig:inmom_lengths} (top) we
show the ratio $\ell_3/N_t = \ell_3^{\rm phys} T$, for all $\beta$s
and sectors. This quantity is expected to have a finite thermodynamic
limit, independently of the spatial localization properties of the
modes, as it is expected to correspond essentially to the extension of
the mode in the temporal direction. In Fig.~\ref{fig:inmom_lengths}
(bottom) we show instead
$\ell_{1,2,3}/N_t = \ell_{1,2,3}^{\rm phys} T$ in the physical sector
of the deconfined phase, which are again expected to have a finite
thermodynamic limit due to the localized nature of the modes. The
volume dependence is indeed mild, except for $\ell_{1,2}^{\rm phys}$
closer to $\beta_c$, where finite-size effects are still important. In
particular, $\ell_3^{\rm phys}$ is of the same order of $1/T$ for all
$\beta$. In the deconfined phase, in the physical (resp.\ unphysical)
sector it becomes smaller (resp.\ larger) than in the confined phase,
indicating a reduced (resp.\ increased) extension in the temporal
direction. In the physical sector in the deconfined phase, all the
length scales $\ell _i^{\rm phys}$ are of the order of the inverse
temperature.

\section{Conclusions and outlook}
\label{sec:concl}

The finite-temperature deconfinement transition of gauge theories
leads to the appearance of localized Dirac modes at the low end of the
spectrum in a variety of gauge theories, both in
3+1~\cite{Gockeler:2001hr,Gattringer:2001ia,Gavai:2008xe,GarciaGarcia:2005vj,
  GarciaGarcia:2006gr,Kovacs:2009zj,Bruckmann:2011cc,Kovacs:2012zq,
  Giordano:2013taa,Cossu:2016scb,Holicki:2018sms,Kovacs:2017uiz,Vig:2020pgq,
  Giordano:2016nuu,Bonati:2020lal,Kovacs:2010wx} and
2+1~\cite{Bruckmann:2017ywh,Giordano:2019pvc} dimensions. This has
been put in a relation with the ordering of the Polyakov loop in the
deconfined phase, and the formation of ``islands'' of fluctuations in
the ``sea'' of ordered Polyakov loops, which are expected to be the
localization centers for the low Dirac
modes~\cite{Bruckmann:2011cc,Giordano:2015vla,Giordano:2016cjs,
  Giordano:2016vhx}. This leads one to generally expect localization
of the low Dirac eigenmodes in the high-temperature, deconfined phase
of a gauge theory (in the physical center sector).

In this paper we have studied the localization properties of the
eigenmodes of the staggered Dirac operator in the background of
$\mathbb{Z}_2$ gauge field configurations on the lattice in 2+1
dimensions. This is the simplest gauge theory displaying a deconfining
transition at finite temperature, and so provides the most basic test
of the ``sea/islands'' picture of localization. We have studied this
theory by means of numerical simulations, producing full staggered
spectra for configurations in both center sectors in order to study in
detail the effects of the ordering of the Polyakov loop, $P(\vec{x})$,
throughout the whole spectrum. Our numerical results confirm the
predictions of the sea/islands picture: the low-lying Dirac modes,
which are delocalized in the low-temperature phase, become localized
in the high-temperature phase of the theory in the ``physical'' center
sector selected by external fermion probes, $\la P(\vec{x})\ra > 0$,
displaying a strong correlation with the location of negative Polyakov
loops.  Instead, in the ``unphysical'' center sector selected by
external pseudofermion probes, $\la P(\vec{x})\ra < 0$, the low modes
remain delocalized, and show a positive correlation with the location
of positive Polyakov loops. Our results suggest that localization of
the lowest modes in the physical sector takes place exactly at
deconfinement, although a study on larger volumes is required to make
a conclusive statement.

The available lattice sizes and accumulated statistics do not allow
for a precise determination of the mobility edge separating the
low-lying, localized modes and the bulk, delocalized modes in the
deconfined phase of the theory in the physical sector. A dedicated
study using larger lattices is required to determine the location of
the mobility edge and understand better the nature of the
corresponding Anderson transition, which is expected to be of BKT
type. However, the existence of a mobility edge seems very likely, and
there are indications of non-trivial scaling of eigenmodes in the
bulk, expected for BKT-type Anderson transitions.

A novel result obtained in this paper is that the very high modes,
near the upper end of the spectrum, are localized in both phases of
the theory, irrespectively of the center sector. The localized nature
of these modes can be understood similarly to that of the low modes in
the physical sector at high temperature, if one recalls the
``Dirac-Anderson'' form of the staggered
operator~\cite{Giordano:2016cjs,Giordano:2016vhx}. In the
Dirac-Anderson basis, the staggered operator looks like a set of $N_t$
coupled Anderson models, with on-site ``energies'' provided by the
phase $\phi(\vec{x})=0,\pi$ of the Polyakov loop along the different
``branches'' of the relation\footnote{Here the temporal direction is
  taken to be direction 1, so that the corresponding staggered phase
  entering $\varepsilon_k(\vec{x})$ is just $\eta_1(x)=1$.}
$\varepsilon_k(\vec{x})=\sin\f{(2k+1)\pi + \phi(\vec{x})}{N_t}$,
$k=0,\ldots,N_t-1$. Here for simplicity we take $N_t$ to be a multiple
of 4. If hopping terms are ignored, the ``unperturbed'' eigenstates
are all spatially localized at the spatial points $\vec{x}$ of the
lattice. Moreover, when the phase of the Polyakov loop is $\pi$, two
of the branches give 0 ($k=\f{N_t}{2}-1$ and $k=N_t-1$), and another
two branches give $\pm 1$ ($k=\f{N_t}{4}-1$ and $k=\f{3N_t}{4}-1$) for
the on-site energy of the corresponding Anderson model. When the phase
of the Polyakov loop is $0$, four of the branches give
$\pm \sin\f{\pi}{N_t}$ ($k=0,\f{N_t}{2}-1$ and $k=\f{N_t}{2},N_t-1$),
and other four of the branches give $\pm \cos\f{\pi}{N_t}$
($k=\f{N_t}{4}-1,\f{N_t}{4}$ and $k=\f{3N_t}{4}-1,\f{3N_t}{4}$). When
the effect of the hopping terms is taken into account, the localized
unperturbed modes mix and tend to delocalize, but they have a chance
of remaining localized if they are in a spectral region where the
spectral density remains sufficiently small, so that mixing is
reduced. Such regions can only be outside of the spectral range of the
free 2+1 dimensional Dirac operator in the two center sectors (i.e.,
setting either $U_1(N_t-1,\vec{x})\equiv 1$ or
$U_1(N_t-1,\vec{x})\equiv -1$, $\forall\vec{x}$, and all other link
variables to 1). While for the spectral region near zero this is the
case only at large $\beta$ in the physical sector, for the high end
(and its symmetric reflection in the negative spectrum) this happens
in both phases of the theory. In the deconfined phase in the physical
sector, the rare sites where $\phi=\pi$ are preferred by the localized
modes, while in the unphysical sector only the rare sites with
$\phi=0$ can support localization in the sea of $\phi=\pi$ sites. It
would be interesting to check whether the same situation is found in
physically more relevant gauge theories, e.g., in QCD.

The correlation between low modes and positive Polyakov loops observed
in the unphysical sector in the deconfined phase, while not
explainable in terms of ``energetically'' favorable islands
``attracting'' the mode, is not in contrast with the sea/islands picture
of localization. As pointed out in Ref.~\cite{Giordano:2016cjs}, an
important effect of the ordering of Polyakov loops in the deconfined
phase is that it induces a strong correlation between different time
slices, that leads to the decoupling of the Anderson models mentioned
above. Conversely, the reduced correlation in the confined phase is
important for their mixing, which is needed for the accumulation of
low modes and the consequent spontaneous breaking of chiral symmetry.
In the unphysical sector above $\beta_c$, the correlation between time
slices is locally reduced where the Polyakov loop fluctuates away from
order, which in this case means that it takes the value $+1$. These
locations are favorable for the different Anderson models to mix,
which can lead to a lowering of the eigenvalue, and can explain the
enhancement of the low modes near positive Polyakov loops.

We have also demonstrated that all localized modes display a strong
correlation with the position of negative plaquettes, again in both
phases of the theory and irrespectively of the center sector. This is
not entirely surprising, given the above-mentioned correlation of
localized modes with Polyakov loops, and the correlation expected
between Polyakov loop fluctuations and clusters of negative
plaquettes. Nevertheless, the interplay between Polyakov loops,
negative plaquettes, and localization of Dirac modes certainly
deserves to be studied in more detail, as it may shed light on the
mechanisms of the confinement/deconfinement transition and of
localization of the low modes.

This work provides further confirmation of the close connection
between localization of the low Dirac modes and deconfinement in
finite-temperature gauge theories. Given the extreme simplicity of the
model, there remains little doubt concerning the universality of this
connection. A possible loose end may seem the case of gauge theories
that display a deconfinement transition but whose gauge group has a
trivial center, in which case the finite-temperature transition
clearly cannot be associated with the spontaneous breaking of center
symmetry. Localization of Dirac modes has not been studied yet in such
theories, and so it is not clear whether it is present or not. In this
case it may seem difficult to identify islands of Polyakov loop
fluctuations. However, the sea/islands picture, as can be seen most
easily making use of the Dirac-Anderson form of the staggered
operator, favors localization on sites where the phase of one of the
eigenvalues of the Polyakov loop is close to $\pm\pi$, irrespectively
of whether this corresponds to a center element, or whether a
nontrivial center even exists at all. It may then be worth studying
the $\mathbb{Z}_3$ gauge theory as well, where the center is not
trivial but the Polyakov loop phase cannot get closer to $\pm\pi$ than
$\pm\f{2\pi}{3}$. We hope that these kind of studies can help to shed
more light on the mechanisms responsible for the deconfinement
transition in finite-temperature gauge theories.

\section*{Acknowledgments}
  We thank T.~G.~Kov\'acs for useful discussions and a careful reading
  of the manuscript.  MG was partially supported by the NKFIH grant
  KKP-126769.

\appendix

\section{IPR for degenerate eigenspaces}
\label{sec:appA}

Suppose to have an $n$-fold degenerate eigenvalue of some Hermitian
Hamiltonian $H$, describing a system living on a finite lattice, and
let $\{\psi_1(x),\ldots,\psi_n(x)\}$ be an orthonormal basis of the
corresponding eigenspace, with $x$ denoting the lattice sites. It is
easy to see that the IPR, ${\rm IPR}= \sum_x|\psi(x)|^4$, is not
constant over the vectors of unit norm, $\sum_x|\psi(x)|^2=1$,
belonging to this eigenspace. Since no such vector is {\it a priori}
singled out, it seems more appropriate to assign an average
${\rm IPR}$ to the whole eigenspace by averaging over all possible
vectors of unit norm. The most general such vector can be obtained as
a linear combination of basis vectors as
$\psi_a^U(x)=\sum_{i=1}^n U_{ai}\psi_i(x)$, where $U_{ai}$ are the
entries of a unitary $n\times n$ matrix. Here $a$ can be fixed to any
value $a=1,\ldots, n$, and it is only included for notational
convenience; the final result should be independent of it. One has for
the mode $\psi_a^U$
\begin{equation}
  \label{eq:degipr1}
  {\rm IPR}_a(U) = \sum_x|\psi^U_a(x)|^4
  = \sum_x\sum_{i,j,k,l=1}^n U_{ai}\psi_i(x) U_{aj}^*\psi_j(x)^*
  U_{ak}\psi_k(x) U_{al}^*\psi_l(x)^*\,. 
\end{equation}
The average over all unit-norm vectors is obtained by group
integration over $U$ with the U$(n)$ invariant (Haar) measure, $dU$.
To this end, we will need the following results~\cite{Collins_2003},
\begin{equation}
  \label{eq:degipr2}
  \begin{aligned}
    \int dU\, U_{ij} U^{-1}_{kl} &=
    {\rm Wg}((1),n)\delta_{il}\delta_{jk}\,,\\
    \int dU\, U_{ij} U^{-1}_{kl} U_{mn} U^{-1}_{pq} &= {\rm
      Wg}((1,1),n) (\delta_{il}\delta_{jk}\delta_{mq}\delta_{np} +
    \delta_{iq}\delta_{jp}\delta_{ml}\delta_{nk}) \\ &\phantom{=}+
    {\rm Wg}((2),n) (\delta_{il}\delta_{jp}\delta_{mq}\delta_{nk} +
    \delta_{iq}\delta_{jk}\delta_{ml}\delta_{np})\,,
  \end{aligned}
\end{equation}
where ${\rm Wg}(\cdot,n)$ is the Weingarten function for the unitary
group, and
\begin{equation}
  \label{eq:degipr3}
  {\rm Wg}((1),n) =\f{1}{n}\,, \quad
  {\rm Wg}((1,1),n)  =\f{1}{n^2-1}\,, \quad  {\rm Wg}((2),n)=
  -\f{1}{n(n^2-1)}\,. 
\end{equation}
It is now straightforward to obtain
\begin{equation}
  \label{eq:degipr4}
  \begin{aligned}
    \int dU \, {\rm IPR}_a(U) &= \sum_x\sum_{i,j,k,l=1}^n
    \psi_i(x)\psi_j(x)^* \psi_k(x)\psi_l(x)^*\int dU\,U_{ai}
    U_{ja}^\dag
    U_{ak} U_{la}^\dag\\
    &= \sum_x\sum_{i,j,k,l=1}^n \psi_i(x)\psi_j(x)^*
    \psi_k(x)\psi_l(x)^* \\ &\phantom{\sum_x\sum_{i,j,k,l=1}^n }\times
    \left[\f{1}{n^2-1}(\delta_{ij}\delta_{kl} +
      \delta_{il}\delta_{jk})  - \f{1}{n(n^2-1)}(\delta_{il}\delta_{kj}
      + \delta_{ij}\delta_{kl})
    \right]\\
    & = \f{2}{n(n+1)}\sum_x \left(\sum_i|\psi_i(x)|^2\right)^2 \,.
  \end{aligned}
\end{equation}
This is independent of $a$, and manifestly invariant under a unitary
change of basis, as it should be: indeed,
$\sum_{i=1}^n|\psi_i(x)|^2=E(x,x)$ equals the diagonal element of the
projector $E(x,y)$ over the degenerate eigenspace.

It is useful to notice that for quantities of the form
\begin{equation}
  \label{eq:degipr6}
  O_i =   \sum_x O(x)|\psi_i(x)|^2
\end{equation}
it is sufficient to average the observables over a set of basis
vectors for any orthonormal basis to automatically obtain the average
of the observable over the whole degenerate eigenspace. In fact,
setting
\begin{equation}
  \label{eq:degipr7}
  \begin{aligned}
    O_a(U) &= \sum_x O(x)|\psi_a^U|^2 = \sum_x O(x)\sum_{i,j=1}^n
    U_{ai} U_{aj}^* \psi_i(x) \psi_j(x)^* \,,
  \end{aligned}
\end{equation}
and repeating the same procedure as above, one finds
\begin{equation}
  \label{eq:degipr8}
  \begin{aligned}
    \int dU\,O_a(U) &= \sum_x O(x)\sum_{i,j=1}^n \psi_i(x)\psi_j(x)^*
    \int dU\,U_{ai} U_{aj}^* \\ & = \sum_x O(x)\sum_{i,j=1}^n
    \psi_i(x)\psi_j(x)^*\f{1}{n}\delta_{ij} = \f{1}{n} \sum_x O(x)
    \left(\sum_{i=1}^n|\psi_i(x)|^2\right)\,,
\end{aligned}
\end{equation}
which is manifestly $a$- and basis-independent.

\bibliographystyle{h-physrev-fix}
\bibliography{references_gt}

\end{document}